\RequirePackage{snapshot}
\documentclass[english,prb,reprint]{revtex4-1}
\usepackage[T1]{fontenc} \usepackage[latin9]{inputenc}
\usepackage{color} \usepackage{babel} \usepackage{amsmath}
\usepackage{amssymb} \usepackage{graphicx} \usepackage{esint}

\usepackage[unicode=true,pdfusetitle,
bookmarks=true,bookmarksnumbered=false,bookmarksopen=false,
breaklinks=true,pdfborder={0 0 0},backref=false,
citecolor=blue,linkcolor=blue,colorlinks=true] {hyperref}

\makeatletter
\renewcommand{\Im}{\ensuremath{\mbox{Im}}}
\newcommand{\Tr}{\ensuremath{\mbox{Tr}}}
\newcommand{\pref}[1]{(\ref{#1})} \newcommand{\etal}{\emph{et al.}}

\newcommand{\boldk}{\boldsymbol{k}}
\newcommand{\ket}[1]{\left|#1\right\rangle}
\newcommand{\bra}[1]{\left\langle#1\right|}
\newcommand{\inner}[2]{\left\langle\left . #1\right|#2\right\rangle}

\@ifundefined{textcolor}{} {%
  \definecolor{BLACK}{gray}{0} \definecolor{WHITE}{gray}{1}
  \definecolor{RED}{rgb}{1,0,0} \definecolor{GREEN}{rgb}{0,1,0}
  \definecolor{BLUE}{rgb}{0,0,1} \definecolor{CYAN}{cmyk}{1,0,0,0}
  \definecolor{MAGENTA}{cmyk}{0,1,0,0}
  \definecolor{YELLOW}{cmyk}{0,0,1,0} }

\@ifundefined{showcaptionsetup}{}{%
  \PassOptionsToPackage{caption=false}{subfig}} \usepackage{subfig}
\makeatother

\newcommand{\boldr}{\boldsymbol{r}}

\begin{document}

\title{Microscopic theory to quantify the competing kinetic processes
  in photoexcited surface-coupled quantum dots}

\author{Kuljit S. Virk}

\email{kuljitvirk@gmail.com}

\affiliation{Department of Chemistry, Columbia University, 3000
  Broadway, New York, NY 10027 }

\author{Mark S. Hybertsen} \email{mhyberts@bnl.gov}
\affiliation{Center for Functional Nanomaterials, Brookhaven National
  Laboratory, Upton, New York 11973, USA}

\author{David R. Reichman}

\email{drr2103@columubia.edu}

\affiliation{Department of Chemistry, Columbia University, 3000
  Broadway, New York, NY 10027}

\begin{abstract}
  We present a self-contained theoretical and computational framework
  for dynamics following photoexcitation in quantum dots near planar
  interfaces. A microscopic Hamiltonian parameterized by first
  principles calculations is merged with a reduced density matrix
  formalism that allows for the prediction of time-dependent charge
  and energy transfer processes between the quantum dot and the
  electrode.  While treating charge and energy transfer processes on
  an equal footing, the non-perturbative effects of sudden charge
  transitions on the Fermi sea of the electrode are included.  We
  illustrate the formalism with calculations of an InAs quantum dot
  coupled to the Shockley state on an Au[111] surface, and use it to
  concretely discuss the wide range of kinetics possible in these
  systems and their implications for photovoltaic systems and tunnel
  junction devices.  We discuss the utility of this framework for the
  analysis of recent experiments.
\end{abstract}
\maketitle

\section{Introduction}

Nanostructured materials represent one of the most promising routes
for the creation of novel energy harvesting and optoelectronic
devices. One of the key challenges in this field is navigating the
vast design space to search for an appropriate combination of material
properties for a specific application. At present, the solution to
this fundamental problem can at best be constructed for specific
classes of systems. In the present paper, we tackle this problem for
one of the key configurations that has emerged in the fields of
photovoltaics and nanoelectronics: quantum dots acting as chromophores
in the vicinity of semiconductor and metallic
surfaces~\citep{Tisdale2010,Choi2010a,Kleemans2010}.

We focus on charge kinetics in these systems and construct a
microscopic dynamical theory that is capable of both describing the
observable experimental phenomena, and of making quantitative
predictions for future research. The key challenge addressed in the
paper is the construction of a framework in which a strong connection
is maintained between the microscopic parameterization of the
Hamiltonian, and the description of dynamics within a restricted set of
excited states. Thus key insights from an efficient navigation of the
parameter space can be related to specific physical properties of the
underlying materials in these systems.

Theories based on model
Hamiltonians~\citep{Fainberg2011,Galperin2009,may,Fainberg2007p2316,
  Galperin2005p2329} allow one to construct a picture of the possible
regimes of dynamics, but the connection to realistic modeling of specific materials may be
 hard to establish.  On the other hand, parameter
free \emph{ab-initio} studies~\citep{kanai,Prezhdo200930} can only
address relatively small systems and the key physical features of
quantum dot systems can not be treated. The work presented in this
paper may be viewed as a compromise between these two
extremes. Starting from the fundamental many-body Hamiltonian for the
light-matter interaction, we derive a low energy effective Hamiltonian
in which all matrix elements are calculated from the single particle
wavefunctions of the subsystems, each represented by a
well-established model.  The wavefunctions and other characteristics
of the surface are obtained from an \emph{ab-initio} Density
Functional Theory based calculation.  The frontier states in the
semiconductor quantum dot are obtained from established effective mass
models.  In the region of overlap between these subsystems, these
wavefunctions can be treated on equal footing.  In this way, our low
energy effective Hamiltonian is fully derived, with the addition of no
parameters.  This Hamiltonian is then used to develop a master
equation for the reduced dynamics of the quantum dot.  Simulations
based on the master equation provide quantitative analysis of the
range of possible charge kinetics in these systems with a clear
connection to the constituent surface and quantum dot materials.

By constructing the Hamiltonian in this way, a meaningful analysis of
charge and energy transfer channels in isolation and in competition
with each other emerges. Thus one can pursue important questions about
charge kinetics such as: How does the extinction of optical power by
the quantum dot change in the presence of a surface? How is cooling of
hot carriers affected by the presence of energy transfer to the
surface? How does electrostatic coupling of the quantum dot to a
surface affect the lifetime of exciton, bi-exciton, and other
multi-electron states? How does non-radiative recombination of
excitons in these systems affect current extraction and
photoluminescence? Our framework is capable of answering these
questions concretely.  We illustrate this within a specific system of
InAs quantum dot on a gold surface.
 
% Given the generic nature of the systems studied here, and the
% resurgent interest in them from many different fields of research, a
% large body of literature exists on various topics explored in this
% paper, both in theory and in experiment. However, based on the
% following overview of the existing literature, it appears that a
% unified framework for addressing the issues described above is
% largely missing.

% MSH: Changes to wording already; but really, contents of this
% paragraph repeats the ideas of the paragraph above, with some more
% details.  This material should be merged with the above & then our
% work separated into a new paragraph.
In recent years, several studies of kinetic processes in models for
single molecule junctions under the influence of both applied
electrical bias and optical fields have been published
\cite{kanai,Galperin2005p2329,Galperin2009,Molen2009,petrov:204701,
  Prezhdo200930,Fainberg2011,Fainberg2007p2316,may,bergfield}.  The
studies of asymmetric dipole coupling in the steady state conductance
\cite{Galperin2005p2329,Galperin2009} have revealed photo-induced
current generation, and current induced photo-emission. This scenario,
however, is inapplicable to single interface systems in which only a
photo-generated current can can exist, and interfacial polarization
plays a different role, as described below. Electron dynamics in
molecular chromophores at semiconductor interfaces has been studied
in depth for a small number of systems and ideal situations fully
\emph{ab-initio}\cite{Prezhdo200930}. While these studies are
invaluable for quantitatively settling many questions about the
microscopic processes and their dynamical interplay, they do not
capture the aspects arising from the relatively large sizes of quantum
dots. Furthermore, the treatment of image potential at the interface
remains an external input to any computation based purely on density
functional theory\cite{Chulkov1999330,neaton-hybertsen,thygesen}.

% KSV: The main issue Mark had raised was that the distinction between
% a molecule and a quantum dot was made too sharp. I have changed the
% following paragraph to not talk about this distinction anymore, but
% just talk about polarizability.

The highly polarizable surfaces of the planar electrode and the
quantum dot lead to electrostatic interactions that significantly
affect the quasiparticle and optical bandgaps, tunneling rates, as
well as energy transfer.  For a spherical quantum dot, an important
fundamental effect of the presence of a planar surface is the
formation of a dipole moment and strong corrections to multipole
moments of its exciton states. Thus, the polarizability of electrode
surface becomes a mechanism for non-radiative recombination of
excitons via energy transfer to the electrode. In our numerical
results, we show the effects of this polarizability and quantify the
significance of high order multipole moments of charge distribution on
the non-radiative exciton recombination in the narrow gap InAs quantum
dot. Having a microscopic Hamiltonian in hand further allows us to
compare this exciton decay with the dissociation across the
junction. This type of analysis, for example, is fundamental to
optimization of current extraction in photovoltaic applications of
these systems.

% KSV: This is changed to more unambiguously compare with HOMO-LUMO
Furthermore, the size of a quantum dot also yields many closely spaced
energy levels, which lead to dynamical effects that often do not arise in single molecules, especially if a simplified treatment is limited to just the highest occupied and lowest unoccupied molecular orbital. The spacing and the
number of energy levels qualitatively affects the dynamics of charge
injection, energy exchange, and the recoil of a hole (electron) in
response to the tunneling of an electron (hole) to the electrode.
Understanding the time for build-up of these transitions in relation
to the magnitude of these transition rates is fundamental to
determining the regimes where a Markovian description of charge and
energy transfer breaks down. In experimental terms, it allows us to
understand when to expect deviations from Lorentzian lineshapes in
both linear absorption and non-linear optical spectroscopy of these
systems. The non-Markovian effects originate physically from the
coupling of quantum dot states to the electrode, as has also been
discussed by Fainberg \emph{et al.} \cite{Fainberg2011} in the case of
molecular junctions.  However, it can be affected significantly by the
spacing of energy levels in the quantum dot, which is a complementary
aspect that exists naturally in our work.  Furthermore, the levels
also couple significantly, via the Coulomb interaction, to the
incoherent particle-hole excitations in the Fermi sea of the
electrode, which opens additional channels of energy transfer beyond
those discussed in studies of molecular systems.

With a few notable exceptions\cite{Fainberg2011,Sukharev2010,elste},
studies in molecular transport generally treat the electrodes as
passive Fermionic reservoirs, thus ignoring the scattering of
electrons in the leads as a result
% MSH
of the excitation dependent Coulomb potential of the molecule. Under
certain conditions, discussed in this paper, this potential can
significantly alter the transport and relaxation in tunneling
junctions \cite{PhysRevB.46.15337,Abanin2004,Abanin2005,segal}, as
well as the optical absorption\cite{recent-Despoja2008}. For example,
tunneling of an electron out of the quantum dot yields sudden
transition of its charge, and can cause significant dynamical
fluctuations in the surface charge density in the electrode. Studies
of the effects of this kind have a long history in X-ray emission and
absorption in bulk metals, and optical absorption in doped quantum
wells\cite{Hawrylak,skolnick,schmitt-rink-review} %,PhysRevLett.84.2006,recent-perkais}.
They are well-known to be composed of two competing contributions: the
Mahan exciton (ME)\cite{Mahan1967,Mahan1980,mahanbook,Ohtaka1990}
arising from the attraction of the electron in the Fermi sea to the hole, and the Anderson
orthogonality catastrophe (AOC)\cite{Anderson:1967p2374,Ohtaka1990}
arising from the vanishing overlap between the initial and the final
many-body state. Together they define the phenomenon of the Fermi edge
singularity (FES). A highly relevant example to the systems considered
in this paper is the recent observation of ME in InGaAs/GaAs quantum
dot heterostructures\cite{Kleemans2010}.

The theory formulated in this paper fully accounts for FES phenomenon
self-consistently alongside charge and energy transfer processes.  A
novel aspect of the geometry considered in this paper is that it is
ideal for exploring FES within the hole bands of a p-doped electrode.
This has remained unexplored in FES studies in bulk metals and quantum
wells because the core charge in this case is the much lighter
electron, the motion of which diminishes the FES signature. On the
other hand, the electron localized inside the dot presents no such
problem and new effects arising from scattering in non-parabolic bands
and the much larger sub-band mixing in hole states can be explored. In
addition, identifying systems in which these effects yield important
signatures in optical response and tunneling current is important for
the correct interpretation of experimental data as well as device
engineering.

The FES has also been studied in resonant tunneling
devices\cite{PhysRevB.46.15337}, and was first predicted in these
systems by Matveev and Larkin \cite{PhysRevB.46.15337}. Abanin
\emph{et al}. emphasized the tunability of the FES effect by
engineering the geometric aspects of the system, and elucidated the
novel effects arising within a non-equilibrium electron
gas\cite{Abanin2005}. The optical response we formulate here naturally
leads to an extension of this idea within the interfacial quantum dot
systems where tuning the relative effect of the ME against the AOC can
be achieved by controlling whether the final state of the optical
excitation lies in the Fermi sea or in the capping layer of the
heterostructure.

In a study by Despoja \emph{et al}\cite{recent-Despoja2008} on the
\emph{ab-initio } calculations of the core-hole spectrum of jellium
surfaces the FES was found to be very weak for a core-hole residing
outside the surface. This is due to very strong screening by the free
electron gas in the metal. While this may be expected for the small
overlap between the sharply screened potential inside the metal, and
the extended states of the slab, our numerical calculations show it to
be true even for the sea of Shockley surface states on Au[111]. On the
other hand, we expect the effect to increase dramatically for thin
films supported on an insulating substrate with low dielectric
constant.

Our paper is organized as follows: in
Sec. \ref{sec:Microscopic-Hamiltonian} we derive the microscopic
Hamiltonian.  The detailed expressions for all the required
Hamiltonian matrix elements are provided in Appendix
\ref{sec:matrix-elements}. In Sec.  \ref{sec:Reduced-density-matrix},
we develop our model for the dynamics, specifically a
time-convolutionless master equation for the density matrix of the
quantum dot. From this equation, we obtain the dynamical rates of
charge and energy transfer, as well as the optical response including
the effects of the FES. In Sec. \ref{sec:Applications} we apply our
model to calculations of charge and energy transfer for an InAs
quantum dot on Au[111] surface, and study the effect of the FES as
well as the competition between tunneling, cooling, and non-radiative
exciton decay on the dynamics. In Sec.  \ref{sec:discussion} we
discuss the application of our theory to the modeling and analysis of
experiments, and the possible extension of the theory developed here
to that of a quantum dot array supporting inter-dot charge and energy
transport. We also discuss in this section how the vibrational modes
of the quantum dot, neglected in the explicit analysis, can be
described within the framework presented. In Sec.~\ref{sec:conclusion}
we conclude. Details not found in the text may be found in Appendices
\ref{sec:matrix-elements} and \ref{sec:math}.

\section{\label{sec:Microscopic-Hamiltonian}Microscopic Hamiltonian}
% ---------------------------------------------------------------------
% Figure: Double sided Feynman Diagram
% ---------------------------------------------------------------------
\begin{figure}
  \subfloat[]{\includegraphics[width=2.5in]{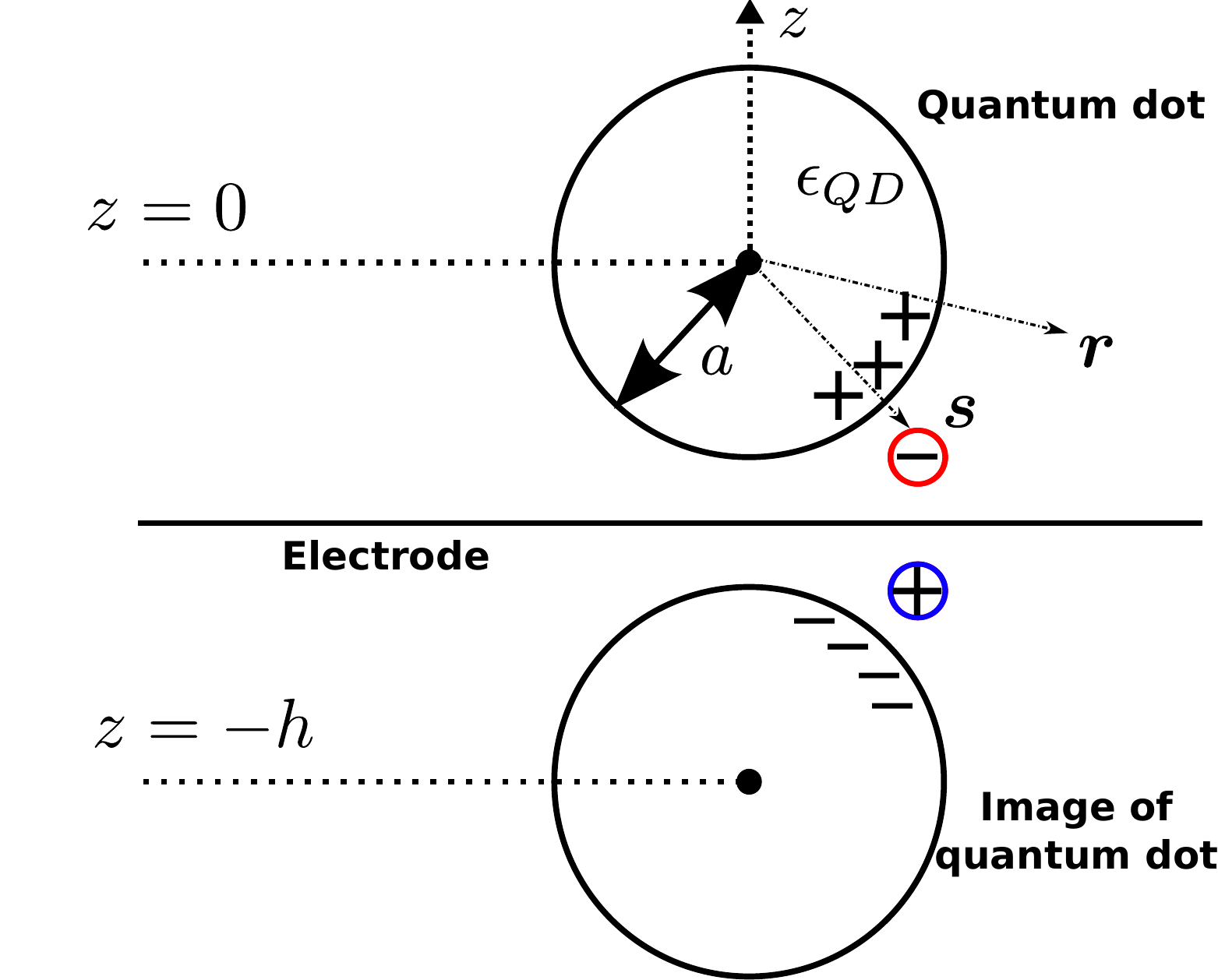}}\hfill
  \subfloat[]{\includegraphics[width=3in]{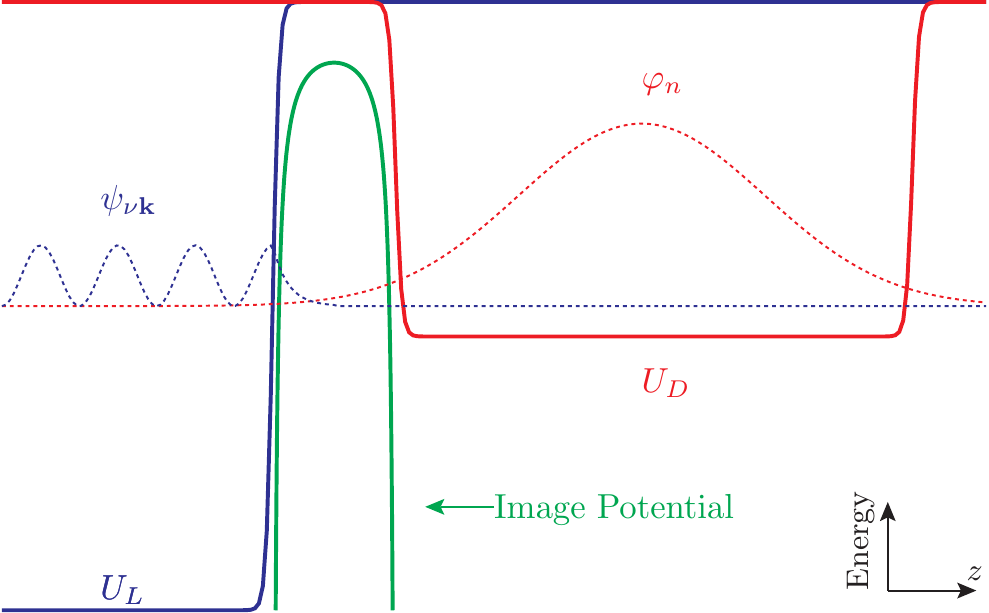}}
  \caption{\label{fig:diagram} (a) The geometry of the system studied
    in this paper showing the important parameters for determining the
    electrostatic potentials. (b) Schematic profile of potentials and
    wavefunctions of the electrode and the quantum dot in a plane
    perpendicular to the electrode surface and passing through the
    center of the dot.}
\end{figure}
% ---------------------------------------------------------------------

We start with the Hamiltonian,
\begin{eqnarray}
  H(t) & = & \int
  d\boldsymbol{r}\,\Psi^{\dagger}(\boldsymbol{r})\left[\frac{1}{2m}\left(\frac{
        \hbar}
      {i}\nabla+e\mathbf{A}(\boldsymbol{r},t)\right)^{2}\right]\Psi(\boldsymbol{r}
  )\label{eq:H fundamental-1}\\
  &  & +\int
  d\boldsymbol{r}\,\Psi^{\dagger}(\boldsymbol{r})\left[U_{D}(\boldsymbol{r})+U_{L}
    (\boldsymbol{r})\right]\Psi(\boldsymbol{r})\nonumber \\
  &  & +\frac{1}{2}\int d\boldsymbol{r}\int
  d\boldsymbol{r}'\,\Psi^{\dagger}(\boldsymbol{r})\,
  \Psi^{\dagger}(\boldsymbol{r}')V(\boldsymbol{r}-\boldsymbol{r}')\Psi(\boldsymbol
  {r}')
  \Psi(\boldsymbol{r}),\nonumber 
\end{eqnarray}
where $\Psi(\boldsymbol{r})$ is the fermionic annihilation field
operator. The interaction of electrons with light occurs through the
vector potential $\mathbf{A}(\boldsymbol{r},t)$ associated with the
optical field. We treat the field as a classical external force in
this work.  Therefore we do not include energy stored in this field in
the above Hamiltonian. The potentials associated with the QD and the
electrode are given by the functions $U_{D}(\boldsymbol{r})$ and
$U_{L}(\boldsymbol{r})$ respectively (see also
Fig.~\ref{fig:diagram}). In the last term of~\eqref{eq:H
  fundamental-1} $V(\boldsymbol{r}-\boldsymbol{r}')$ is the Coulomb
interaction among all fermions. We have neglected the phonons and
electron-phonon interaction in the present model for brevity, but it
can be incorporated straightforwardly in our formalism, and we comment
in Section~\ref{sec:Applications} on how this can be accomplished.

Our approach is akin to the Bardeen approach to
tunneling~\cite{bardeen}.  We make a physically reasonable distinction
between states of the QD and the electrode, and then develop a theory
for the charge and energy exchange between the two sets of states. It
is then convenient to first write the field operator as a sum of field operators
that create/annihilate particles confined to the QD and the electrode,
\begin{equation}
  \Psi(\boldr) = \Psi_D(\boldr)+\Psi_L(\boldr)\label{eq:psidl}.
\end{equation}
With the choice of the single particle basis below, how the states
within the two subsystems can be identified will be detailed. It is
implicit in the formalism of the reduced dynamics of the QD (see
Sec.~\ref{sec:Reduced-density-matrix}) that the electrode states,
which are traced over, are orthogonal to the QD states included in the
dynamics. Any distinction based on exact vanishing of the states
within appropriately defined volumes of the two subsystems would not
yield an orthogonal basis in general~\cite{prange}. However, with the
barrier potential equal to several electron volts, the low energy
states of the QD and those near the Fermi level of the electrode, even
when calculated in isolation of each other, decay exponentially within
the barrier with a characteristic length of about 1-2 \AA. The QD and
electrode states most relevant to the dynamics are then orthogonal to
a good approximation. We exploit this property in the calculations,
but note that the approximation lies in the choice of basis and not in
the formalism.

We substitute~\eqref{eq:psidl} into the Hamiltonian~\eqref{eq:H
  fundamental-1} above, and group the terms as follows,
\begin{widetext}
{\small
\begin{eqnarray}
H_D & = & \int d\boldr \Psi_D^{\dagger}(\boldr)\left[\frac{1}{2m}\left(\frac{
\hbar}
{i}\nabla+e\mathbf{A}(\boldsymbol{r},t)\right)^{2} + U_{D}(\boldr)+U_L(\boldr)\right.
+\left.\int d\boldr' V(\boldr-\boldr')\left\{
\Psi_D^\dagger(\boldr')\Psi_D(\boldr')+\rho_{img}(\boldr')\right\}\right]\Psi_D(\boldr)\label{eq:HD1},\\
H_L & = & \int d\boldr \Psi_L^{\dagger}(\boldr)\left[\frac{1}{2m}\left(\frac{
\hbar}
{i}\nabla+e\mathbf{A}(\boldsymbol{r},t)\right)^{2} + U_{D}(\boldr)+ U_{L}(\boldr)+\int d\boldr' V(\boldr-\boldr')\Psi_L^\dagger(\boldr')\Psi_L(\boldr')\right]\Psi_L(\boldr),\label{eq:HL1}\\
H_{DL}&=&\int d\boldr \Psi_D^{\dagger}(\boldr)\left[\frac{1}{2m}\left(\frac{
\hbar}
{i}\nabla+e\mathbf{A}(\boldsymbol{r},t)\right)^{2}+ U_{D}(\boldr)+U_{L}(\boldr)\right]\Psi_L(\boldr),\label{eq:HDL}\\
V_{DL}&=&\int d\boldr \int d\boldr'\Psi_D^{\dagger}(\boldr)\Psi_D(\boldr)V(\boldr-\boldr')\left[\Psi^\dagger_L(\boldr')\Psi_L(\boldr')-\rho_{img}(\boldr')\right].\label{eq:VDL}
\end{eqnarray}}
\end{widetext}

The term $H_D$ given by~\eqref{eq:HD1} describes the QD states in the
presence of the electrode potential, the Coulomb interaction among
carriers, and the classical electrostatic interaction with the
electrode in the form of an image potential. The latter is included
via the induced density (a real-valued function) $\rho_{img}(\boldr)$,
which is later subtracted in~\eqref{eq:VDL}. We call the basis states
of $H_D$ for $\mathbf{A}=0$ the QD states, and the matrix elements of
$-im^{-1}\hbar\mathbf{A}(\boldr)\cdot\nabla$ between these states
defines the optical excitation of the QD in the presence of an
electrode.

Similarly, $H_L$ in~\eqref{eq:HL1} defines electrode states and their
optical interaction in the presence of the lattice potential $U_L$ and
the Coulomb interaction among carriers. The single particle states of
$H_L$ are calculated with full atomic scale detail, after setting $\mathbf{A}=0$. Then the Coulomb interaction appearing in Eq.~\eqref{eq:HL1} is incorporated into the treatment of the states at the level captured by Density Functional Theory.  In the particular example we use for illustration, only the surface bound states play a direct role
in calculations, while the remaining states are kept formally for
completeness. In the final reduced dynamics of the QD, the sum of the
interaction matrix elements over these states yields well-defined
single and two-particle response functions. The actual problem of
treating the electrode is then reduced to the calculation of these
response functions in the presence of a surface. The surface bound
states are calculated by constructing the surface Green functions from
the Kohn-Sham density functional theory, as discussed at length in
Section~\ref{sec:Applications}.

Next, the term $H_{DL}$ given by~\eqref{eq:HDL} describes
hybridization, including the optically driven excitations in which an
electron or a hole is excited directly into a final state in the
electrode. This term thus describes charge transfer. The last term,
$V_{DL}$, given by~\eqref{eq:VDL}, describes energy transfer mediated
by the Coulomb interaction between the QD and the electrode. The
advantage of adding and subtracting $\rho_{img}(\boldr)$ in the above
expressions is two fold: the QD states include the image potential
non-perturbatively, and the dynamical interaction of these states with
the electrode reduces to \emph{fluctuations} around this density. As
we will see in Section~\ref{sec:et}, this naturally leads to
describing the energy transfer in terms of a dynamical longitudinal
susceptibility of the electrode.

In the above expansion we neglect the term
$\Psi^\dagger_L(\boldr)U_D(\boldr)\Psi_L(\boldr)$ that appears in
$H_L$, which would result in additional renormalization of the
electrode states in response to the presence of the neutral QD. We
disregard this term because the exponential decay of the electrode
wavefunctions in the barrier, combined with the extended nature of
these wavefunctions inside the electrode, makes the effect of this perturbation very
small. On the other hand, the changes induced by the presence of the
electrode has a significant effect on the boundary conditions for the
states localized to the QD, and thus the analogous term
$\Psi_D^\dagger(\boldr)U_L(\boldr)\Psi_D(\boldr)$ is included in the
$H_D$ term. We remark that the neglected term may be included by
calculating the scattering states starting from the eigenstates of
$H_L$~\cite{mahanbook}. This would perhaps be necessary for some
mesoscopic electrodes lying at a very small distance from the QD, or
in cases with small barrier heights such that the electrode
wavefunctions significantly probe $U_D(\boldr)$.

We have also neglected the particular exchange interaction where one
carrier lies in the QD state, and the other in the electrode
state. This interaction should also be exponentially suppressed for
the low energy and well confined states of relevance to this
work. Related to this exchange is the Coulomb-driven tunneling process
which is also neglected in comparison to the contribution by single
particle kinetic and potential energies as well as optical
interactions, $H_{DL}$. The exchange interaction among carriers within
each subsystem is implicit in the above expressions.

Let us now turn to specification of the states comprising $\Psi_D$ and
$\Psi_L$, and begin with a set of single particle basis states. Let
$\ket{g}$ be the ground state of the QD, and $
\varphi_n(\boldsymbol{r})=\left\langle
  \boldsymbol{r}\left|\right.n\right\rangle $
be the single particle excited states satisfying,
\begin{equation}
  \left[-\frac{\hbar^{2}}{2m_{0}}\nabla^{2}+U_{Deff}(\boldsymbol{r}
    )+\Sigma(\boldsymbol{r})-
    \varepsilon_n \right]
  \varphi_{n}(\boldsymbol{r}) = 0.\label{eq:qdstates}
\end{equation}
Here $U_{Deff}$ represents the pseudopotential for a single electron
or a hole added to the neutral ground state of the QD, and we have
introduced the electrostatic self energy of a point charge in the
vicinity of polarizable surfaces such as that of the QD and the electrode. The electrode contribution is given by~\cite{virk-apl} 
\begin{equation}\Sigma_L(\boldsymbol{r})=\int d\boldr'
  V(\boldr-\boldr')\rho_{img}(\boldr').
  \end{equation}
This is a well-known formulation of an exact one-body potential representing the electrostatic energy stored in the polarization reaction field that is induced by a unit charge on dielectric surfaces~\cite{banyai,boettcher,brus:4403}.

We label the solutions to the eigenvalue equation~\eqref{eq:qdstates}
as electron states, $\varphi_e(\boldsymbol{r})$, for the addition of a
single electron to the QD state $e$ above the quasi-particle energy
gap, and hole states, $\varphi_h^*(\boldsymbol{r})$, for the removal
of an electron from a valence state, $h$. While the solutions at
higher energies significantly violate the boundary conditions at the
electrode surface, they are nonetheless useful in forming a convenient
basis set for expanding the multi-particle states. The present paper
discusses only the exciton as the multi-particle excitation, and we
express its wavefunction as,
\begin{equation}
  \Phi_x(\boldsymbol{r}_e,\boldsymbol{r}_h) = \sum_{eh}\varphi_e(\boldsymbol{r}_e)
  \varphi_h^*(\boldsymbol{r}_h)\Phi_{eh;x},\label{eq:exciton}
\end{equation}
where the coefficients $\Phi_{eh;x}$ are determined from a variational
calculation including the Coulomb interaction between the electron,
hole, and their induced surface polarizations (see~\eqref{eq:HD1} and
Sec. \ref{sec:Applications}). This approach has been employed widely in semiconductor optics\cite{Hu-Lindberg-Koch,chuang-schmitt-rink}.

The electrode states diagonalize $H_L$ in ~\eqref{eq:HL1} for
$\mathbf{A}=0$, and we write them as
$\psi_{\nu\boldsymbol{k}}=u_{\nu\boldsymbol{k}}(\boldsymbol{r})e^{i\boldsymbol{k}\cdot\boldsymbol{r}}$. While
it is not essential for the theory presented, we have parameterized
these states by a two-dimensional quasi-momentum $\boldsymbol{k}$,
thus assuming that the electrode has a planar surface.  In a
semi-infinite electrode, $\nu$ may also take continuous values within
regions of the bulk excitations in the projected density of states. In
terms of the (non-orthogonal) basis set above,~\eqref{eq:psidl} can
now be written more precisely as,
\begin{eqnarray}
  \Psi(\boldsymbol{r}) & = & \sum_{n}\varphi_{n}(\boldsymbol{r})c_{n}+\sum_{\nu
    \boldsymbol{k}}u_{\nu\boldsymbol{k}}(\boldsymbol{r})e^{i\boldsymbol{k}\cdot
    \boldsymbol{r}}c_{\nu\boldsymbol{k}},\label{eq:field operator}
\end{eqnarray}
where the first term on the right hand side corresponds to $\Psi_D$
and the second to $\Psi_L$. The operators $c_{n}$ and
$c_{\nu\boldsymbol{k}}$ annihilate particles in states $\varphi_n$ and
$\varphi_{\nu\boldk}$ respectively. The sum over $n$ is truncated to
states that are well localized on the QD in the sense that their
weight in the electrode region is negligible. The remaining states,
which are high in energy and have a significant fraction of their
wavefunction inside the electrode, are then all viewed formally as
part of the states to be traced over in the reduced dynamics.

Separating the optical interaction from the $\mathbf{A}=0$ form of the
terms in~\eqref{eq:HD1}-\eqref{eq:VDL}, we now write the Hamiltonian
as
\begin{equation}
  H  =  H_{D}+H_{L}+H_{T}+V_C+H_{D}^{(r)}+H_{L}^{(r)}+H_{T}^{(r)},\label{eq:full H}
\end{equation}
in which the first two terms are the Hamiltonians for the quantum dot
and the electrode states in the absence of radiation,
\begin{eqnarray}
  H_{D} & = & \sum_{n}\ket{n}\varepsilon_n\bra{n},\label{eq:HD}\\
  H_{L} & = & \sum_{\nu\boldsymbol{k}}\varepsilon_{\nu\boldsymbol{k}}c_{\nu
    \boldsymbol{k}}^{\dagger}c_{\nu\boldsymbol{k}}.\label{eq:Hl}
\end{eqnarray}
Here the $\varepsilon_{n}$ are the energies of the QD states,
including the ground state of the neutral QD, single electron and
hole states, and the neutral exciton states.  The
$\varepsilon_{\nu\boldsymbol{k}}$ are the dispersion of electrode
energies with band index $\nu$. The third term in \pref{eq:full H},
which arises from \eqref{eq:HDL} represents the hybridization between
the QD and the electrode, and we write it as,
\begin{eqnarray}
  H_T &=& \ket{e}\bra{g}\mathcal{T}^{eg}+\ket{h}\bra{g}\mathcal{T}^{hg}\nonumber\\
  &&+\ket{h}\bra{x}\mathcal{T}^{hx}+\ket{e}\bra{x}\mathcal{T}^{ex}+c.c.
\end{eqnarray}
Here we have introduced the \emph{electron transfer operators},
$\mathcal{T}^{ab}$, corresponding to the tunneling induced change in
the charge state of the quantum dot. These formal operators are
introduced to simplify the dynamical model in Sec.
\ref{sec:Reduced-density-matrix} below, and are defined as,
\begin{eqnarray}
  \mathcal{T}^{eg}&=&\sum_{\nu\boldsymbol{k}}T^{eg}_{\nu\boldsymbol{k}}c_{\nu
    \boldsymbol{k}}\label{eq:Teg},
  \\
  \mathcal{T}^{hg}&=&\sum_{\nu\boldsymbol{k}}T^{hg}_{\nu\boldsymbol{k}}c_{\nu\boldsymbol{k}}^\dagger\label{eq:Thg},\\
  \mathcal{T}^{hx}&=&\sum_{\nu\boldsymbol{k}}T^{hx}_{\nu\boldsymbol{k}}
  c^\dagger_{\nu, \boldsymbol{k}}\label{eq:Thx},\\
  \mathcal{T}^{ex}&=&\sum_{\nu\boldsymbol{k}}T^{ex}_{\nu\boldsymbol{k}}c_{\nu\boldsymbol{k}}\label{eq:Tex}.
\end{eqnarray}
The matrix elements $T^{aa'}_{\nu\boldsymbol{k}}$ represent tunneling
amplitudes, and can be computed from the electronic states $\varphi_n$
and $\psi_{\nu\boldk}$ as described in detail in Appendix
\ref{sec:matrix-elements}. Recall that $\varphi_h$ represents a single electron
 orbital in the valence band~\footnote{For a DFT calculation, this would be a Kohn-Sham orbital}, such that $\varphi_h^*$ is the state of the corresponding hole. Thus the matrix element $T^{hg}_{\nu\boldsymbol{k}}$ describes the transfer of an electron from state $h$ in the valence band of the QD to the electrode. We have now defined all the electron transfer operators needed for the formalism below to 
describe addition or removal of electrons from the QD.

Returning to \pref{eq:full H}, the fourth term represents the Coulomb
interaction in \pref{eq:VDL}, which we expand in QD basis states,
\begin{eqnarray}
  V_{C} & = & \sum_{nm}\ket{n}\bra{m}\hat{V}_{nm},\label{eq:V coul final}
\end{eqnarray}
where the operator $\hat{V}_{nm}$ acts on the electrode states, and is
defined as
\begin{eqnarray}
  \hat{V}_{nn'} & = &
  \sum_{\nu,\nu'}\sum_{\boldsymbol{k},\boldsymbol{k}'}V_{n,\nu\boldsymbol{k};n',
    \nu'\boldsymbol{k}'}c_{\nu\boldsymbol{k}}^{\dagger}c_{\nu'\boldsymbol{k}'}
  .\label{eq:Vhat nn' def}
\end{eqnarray}
The matrix elements in this expression follow directly
from~\eqref{eq:VDL}, and are defined in full form in~\pref{eq:eff
  coulomb mat elem}.  For $n\neq m$, the $\hat{V}_{nm}$ represent
quantum fluctuations around the classically induced density of the
electron gas and lead to energy and polarization transfer between the
QD and the electrode. On the other hand, the diagonal terms,
$\hat{V}_{nn}$, cause random fluctuations in the energy level of the QD.  
This may be thought of as a back-action from the excitations in the electrode induced by the QD potential. As was discussed in the
Introduction, this coupling results in the FES and AOC phenomena. 

At this stage we let $\hat{V}_{nm}$ represent all electronic
excitations of the system so that the matrix elements may be taken to
be the bare Coulomb interaction. In actual calculations, however, it
is more convenient to identify a set of elementary excitations of the
electrode strongly coupled to the QD, and then renormalize this
coupling by the interactions among these excitations, and their
interactions with the weakly coupled excitations. We will see
in Sec.~\ref{sec:Reduced-density-matrix} that this can be achieved
essentially by defining a frequency dependent dielectric function for
the electrode surface, in which the relevant interactions are included
by construction.

We now turn to the last three terms of \pref{eq:full H}, which
describe the interaction of the entire system with the external
electromagnetic (EM) field. We first write the matrix elements of the
velocity operator in the standard form as,
\begin{eqnarray}
  \mathbf{v}_{nn'}(t) & = & \frac{e}{2m}\mathbf{A}(t)\delta_{nn'}\label{eq:current
    density 
    operator}\\
  &  & +\frac{\hbar}{2mi}\int
  d\boldsymbol{r}\left[\phi_{n}^{*}(\boldsymbol{r})\nabla\phi_{n'}
    (\boldsymbol{r})-\phi_{n'}(\boldsymbol{r})\nabla\phi_{n}^{*}(\boldsymbol{r}
    )\right],\nonumber 
\end{eqnarray}
where $\phi_{n}$ may be set to a QD state $\varphi_{n}$ or an
electrode state $\psi_{\nu}$. The fundamental optical transition is
the exciton, and we define its matrix element as,
\begin{eqnarray}
  \mathbf{v}_{xg}(t) & = & \sum_{e,h}\mathbf{v}_{eh}(t)\Phi_{eh;x},\label{eq:vxg}
\end{eqnarray}
and write the interaction between the EM field and the QD as,
\begin{eqnarray}
  H_{D}^{(r)}(t) & = &
  e\sum_{x}\mathbf{A}(t)\cdot\mathbf{v}_{xg}(t)\ket{x}\bra{g}+\mbox{c.c}.
  \label{eq:Hr D}
\end{eqnarray}
Similarly the interaction between the electrodes and the EM field is
given by
\begin{eqnarray}
  H_{L}^{(r)}(t) & = & e\sum_{\nu,\nu'}\sum_{\boldsymbol{k},\boldsymbol{k}'}\mathbf{A}(t)\cdot\mathbf{v}_{\nu
    \boldsymbol{k},\nu'\boldsymbol{k}'}(t)c_{\nu\boldsymbol{k}}^{\dagger}c_{
    \nu'\boldsymbol{k}'}+
  \mbox{c.c.}\;\;\label{eq:Hr L}
\end{eqnarray}

An additional light-matter coupling that we do not consider in much detail here,
$H_{T}^{(r)}$, describes the radiation driven charge transfer between
the QD and the electrode. This term introduces an externally
controllable exciton dissociation between the QD and the electrode,
and takes the form,
\begin{eqnarray}
  H_{T}^{(r)}(t) & = &
  e\sum_{e}\ket{e}\bra{g}\sum_{\nu\boldsymbol{k}}\mathbf{A}(t)\cdot
  \mathbf{v}_{e;\nu\boldsymbol{k}}(t)c_{\nu\boldsymbol{k}}\label{eq:Hr LD}\\
  &  &
  -e\sum_{h}\ket{h}\bra{g}\sum_{\nu\boldsymbol{k}}\mathbf{A}(t)\cdot\mathbf{v}_{
    h;\nu
    \boldsymbol{k}}(t)c_{\nu\boldsymbol{k}}^{\dagger}+\mbox{c.c.}\nonumber 
\end{eqnarray}
This completes our construction of the microscopic Hamiltonian, and we
now turn to the description of the charge dynamics of an electrode
coupled quantum dot.

\section{\label{sec:Reduced-density-matrix}Dynamics following
  photo-excitation}

We study dynamics within the restricted Hilbert space of four classes
of states: the ground state $\left|g\right\rangle $, single electron
states, $\left|e\right\rangle $, single hole states
$\left|h\right\rangle $, and the exciton states $\left|x\right\rangle
$. This restriction is only for convenience, and may be lifted by
expanding the set to include bi-exciton states and even larger
multi-electron complexes. Within each class, however, we do allow for
an arbitrary number of states to exist. Since this system is coupled
to the electrodes, the unitary evolution governed by the
Schr\"{o}dinger equation applies to the full density matrix, which we
denote as $\xi(t)$, and it obeys the equation,
\begin{eqnarray}
  \frac{d}{dt}\xi(t) & = & \frac{-i}{\hbar}\left[H(t),\xi(t)\right],
\end{eqnarray}
where $H(t)$ is the Hamiltonian \eqref{eq:full H} discussed in the
previous section. Fundamentally, the equation describes optical
excitations acting as the external force driving the system out of
equilibrium (the last three terms in \eqref{eq:full H}), and the
dynamical couplings between QD and electrodes returning the two
subsystems towards a state of mutual equilibrium (via both $H_{T}$ and
$V$). Assuming that the electrode stays in equilibrium, we now reduce
this equation to the description of excitation, dephasing, and
relaxation of the QD alone.

\subsection{Reduced density matrix dynamics }\label{sec:reduced dm}

Recent experiments have pinpointed subtle FES effects in optical
spectra of quantum dots coupled to quantum
wells~\cite{recent-Latta,Kleemans2010,PhysRevLett.106.107402}.  Thus
it is crucial to include the effects of FES in our theory. In order to
achieve this, a special interaction picture must be constructed to
fully account for the non-perturbative scattering of the electrode
states in response to the electrostatic potential of the QD. Our
approach is motivated by the one body formulation of the X-ray spectra
by Mahan\cite{Mahan1980}, Nozieres and De
Dominicis\cite{NOZIERES1969}, and the analysis of orthogonality
catastrophe by Anderson\cite{Anderson:1967p2374}.  In the original FES
papers, the authors captured how the sudden shift in the Hamiltonian
of electrons forming a Fermi sea affects the photoemission and
absorption spectra in metallic systems\cite{Mahan1967}. Following this
work, investigations into the FES in the photo-excitation of doped
semiconductors were also carried out\cite{Hawrylak}. We also note the
studies of the FES within the context of pump-probe experiments
sensitive to the coherent non-linear optical response of doped
semiconductors\cite{perakis-prb,perakis-review}. All these works focus
on changes in the lineshape of the optical spectrum of a Fermi sea.

Here we explore the consequences of these fundamental effects in the
exciton dissociation across a QD-electrode interface. We formulate a
master equation for an electrode coupled QD, and show how the effects
of FES can be introduced in this theory, and how they can be
re-captured naturally in the time-dependent couplings defining the
resulting equation. Thus the FES becomes an integral part of the
dynamical map that propagates the state of the QD towards
equilibrium. The couplings in which the FES appears take the form of
correlation functions bearing many similarities to the results of
Nozieres and De Dominicis\cite{NOZIERES1969}. However, our equations
address a very different physical scenario, and they are applied
without making any simplifying assumptions on the spatial profiles of
the different electrostatic potentials created by QD states. We also
develop one-body formulas for calculating and interpreting these
correlation functions. We will first discuss the derivation of the
equation of motion for a general electrode, and then specialize to the
case of a Fermionic reservoir.

% An alternative perspective on our approach is the solvent
% reorganization due to large system-bath coupling in the spin-boson
% models. These are often used in chemical physics, and the general
% analysis below reproduces the results of Golosov
% \emph{et. al.}\cite{golosov} in the limit of a two-state system and
% a generic bath. However, we diverge away from the past
% works~\cite{golosov,Galperin2008,Galperin2009,skinner} by
% specializing our formalism to a fermionic bath, and have found this
% route to have crucial differences, which also help simplify the
% equations. In addition the asymmetric treatment of system-bath
% coupling diagonal and off-diagonal in system states can lead to
% various unphysical effects, which are mitigated by a fermionic bath
% and the nature of system bath coupling, but which do plague
% calculations involving a spin-boson model [refs].
\subsubsection{Derivation of the general form}

From the full Hamiltonian defined in \eqref{eq:full H} in which the
Coulomb interaction is given by \eqref{eq:V coul final}, we take three
contributions to construct the interaction picture by defining a
Hamiltonian,
\begin{eqnarray}
  H_{0} & \equiv &
  H_{D}+H_{L}+\sum_{n}\hat{V}_{nn}\ket{n}\bra{n}\label{eq:int-pict},
\end{eqnarray}
where $n$ is a label for the QD states. The last term in the
expression above scatters electrode states via a potential that is
conditioned upon the QD state. This term is also the key to capturing
the FES effects, but it complicates the interaction picture by
yielding a non-perturbative coupling between the system and the bath.

An analogue of this way of partitioning the Hamiltonian has been used
in the chemical physics literature~in the
past\cite{zhang-mukamel,Yang2002,golosov}. Such a reference system
leads to the so called ``modified Redfield''
approaches~\cite{zhang-mukamel,Yang2002}. The crucial difference here
is that our bath is composed of fermionic excitations, which alter
both the physics and the formalism compared to the bosonic vibrational
degrees of freedom at play in the previous work.

% in the paper by Golosov \emph{et. al.} for a generic bosonic
% bath~\cite{golosov}, but with a crucial difference in the form of
% the potentials $\hat{V}_{nn}$. In spin-boson models, such as the
% ones in that paper~\cite{golosov}, $\hat{V}_{nn}$ consists of
% oscillator position operators and therefore is a linear
% superposition of creation and annihilation operators of the bath
% modes.  Thus the finite unitary transformations generated by
% $\hat{V}_{nn}$ contain terms that both destroy and create bososns,
% as well as terms that conserve the particle number. On the other
% hand, for a fermionic bath in which $\hat{V}_{nn}$ represents
% Coulomb interaction, only particle conserving terms appear in such
% transformations. This leads to an important simplification that
% makes charge transfer vanish to the first order in hybridization for
% the fermionic bath (as discussed following Eq. \eqref{eq:G expand}
% below). In contrast particle transfer survives to first order in a
% bosonic bath, in which the special interaction picture
% transformation, which includes system-bath coupling diagonal in
% system states, may become pathological.

To proceed with our analysis, we define the full density matrix within our interaction
picture as,
\begin{equation}
  \tilde{\xi}(t) = e^{iH_0t/\hbar}\xi(t)e^{-iH_0t/\hbar}\label{eq:int-schrod}.
\end{equation}
In the Schr\"{o}dinger picture, the \emph{reduced density matrix}
defined only over the QD states can be obtained by tracing $\xi(t)$
over the electrode degrees of freedom. However, we begin with a
slightly different version of this procedure and relate the reduced
density matrix in the Schr\"{o}dinger picture to the $\tilde{\xi}(t)$
via the equation,
\begin{equation}\label{eq:rho(t)}
  \rho(t) = \Tr_L\left[e^{-iH_0t/\hbar}\tilde{\xi}(t)e^{iH_0t/\hbar}\right].
\end{equation}
The FES arises when many body states are subjected to unitary rotation
by two (or more) different Hamiltonians. The exponentials in the above
formula accomplish this exactly, and allow us to expand the remaining
interactions between QD and the electrode perturbatively. From the
chemical physics perspective, the exponentials take into account the
bath-induced random fluctuations of the QD energy levels, which cause
the phenomenon of pure dephasing (decoherence without population
relaxation)\cite{golosov,skinner}.

Returning to the general derivation, we write the Schr\"{o}dinger
equation within the interaction picture as
\begin{eqnarray}
  \frac{d}{dt}\tilde{\xi}(t) & = &
  -\frac{i}{\hbar}\left[H_{T}(t),\tilde{\xi}(t)\right]-\frac{i}{\hbar}
  \left[V_{C}(t),\tilde{\xi}(t)\right]\label{eq:full EOM}\\
  &  &
  -\frac{i}{\hbar}\left[H_{D}^{(r)}(t),\tilde{\xi}(t)\right]-\frac{i}{\hbar}\left[
    H_{L}^{(r)}(t),
    \tilde{\xi}(t)\right]\nonumber \\
  &  &
  -\frac{i}{\hbar}\left[H_{T}^{(r)}(t),\tilde{\xi}(t)\right]
  % -\frac{i}{\hbar}\left[H_{T}^{(r)}(t),\tilde{\xi}(t)\right]
  .\nonumber 
\end{eqnarray}
The formal solution of this equation in the absence of an external
electromagnetic field may be written as,
\begin{equation}
  \tilde{\xi}(t) = T_{+}\exp\left(-i\int_{0}^{t}dt'\mathcal{J}_{I}
  (t')\right) \xi(0). \label{eq:Exact sol}
\end{equation}
The symbol $T_{+}$ enforces time ordering such that
\begin{eqnarray*}
  T_+\exp\left(-i\int_{0}^{t}dt'\mathcal{J}_{I}
  (t')\right)&=&1-i\int_{0}^{t}dt'\mathcal{J}_{I}(t')\\
  &&-\int_{0}^{t}dt'\int_{0}^{t''}
  dt'\mathcal { J } _ { I }(t')\mathcal{J}_{I}(t'') +\ldots. 
\end{eqnarray*}
The superoperator $\mathcal{J}_{I}(t)$ in~\eqref{eq:Exact sol} acts on
an arbitrary operator $\hat{O}$ as,
%\begin{eqnarray*}
%  \mathcal{J}_{I}(t)\hat{O} & = &
%  -\frac{i}{\hbar}\left[H_{T}(t)+V_{C}(t),\hat{O}\right].
%\end{eqnarray*}
\begin{eqnarray*}
  \mathcal{J}_{I}(t)\hat{O} & = &
  \frac{1}{\hbar}\left[H(t)-H_0,\hat{O}\right].
\end{eqnarray*}
For our discussion below we also define superoperators for the radiative and Coulomb perturbations,
\begin{eqnarray*}
  \mathcal{J}_{T}(t)\hat{O} & = &
  \frac{1}{\hbar}\left[H_{T}(t),\hat{O}\right],\\
  \mathcal{J}_{C}(t)\hat{O} & = &
  \frac{1}{\hbar}\left[V_{C}(t),\hat{O}\right].
\end{eqnarray*}
% MSH This assumption, which really is central, deserves a better
% intro and discussion
To proceed, we assume that $\xi(0) = \rho(0)\otimes\mathfrak{R}$,
return to the Schr\"{o}dinger picture and take the trace of
\eqref{eq:Exact sol} over the electrode states,
% KSV: This is my attempt to make the discussion compact after
% rearranging equations: begin with the symbol "G" right away and
% define it immediately. Then define its ingredients.
\begin{equation}\label{eq:exact rho}
  \rho(t)  = G(t,0)\rho(0),
\end{equation}
where $G(t,0)$ is a propagator for the reduced density matrix,
\begin{eqnarray}
  G(t,0) & = & \left\langle
    e^{-i\mathcal{L}_{0}t}T_{+}\exp\left(-i\int_{0}^{t}dt'\mathcal{J}_{I}(t')\right)\right
  \rangle _{L}.\label{eq:G exact}
\end{eqnarray}
We have introduced $\left\langle \cdot\right\rangle _{L}$ as the trace
over electrodes, including $\mathfrak{R}$.  The superoperator
$e^{-i\mathcal{L}_{0}t}$ is defined by its action on $\hat{O}$ as,
\begin{eqnarray*}
  e^{-i\mathcal{L}_{0}t}\hat{O} & = & e^{-iH_{0}t}\hat{O}e^{iH_{0}t}.
\end{eqnarray*}

% \begin{eqnarray}
%\rho(t) & = \left\langle
%e^{-i\mathcal{L}_{0}t}T_{+}\exp\int_{0}^{t}\mathcal{J}_{I}(t')dt'
%\right\rangle _{L}\rho(0), \label{eq:Exact sol rho}\\ 
%\end{eqnarray}

To manage the subtleties of present choice of interaction picture, we
explicitly find its matrix representation in the vector space of pairs
of QD states by arranging the density matrix elements in a column
vector with some arbitrary but fixed order. We define $G$ as a matrix
acting over vectors in this space with the matrix elements,
\begin{eqnarray*}
  &  & G_{ac;a'c'}(t)= \mbox{Tr}\left[(\ket{a}\bra{c})^\dagger
      e^{-i\mathcal{L}_{0}t}
      \vphantom{\int_{0}^{t}}\right.\nonumber\\
      &&\left.T_{+}\exp\left(-i\int_{0}^{t}dt'
        \mathcal{J}_{I}(t')\right)\mathfrak{R}\otimes\left|a'\right\rangle \left\langle
        c'\right|\right]  
\end{eqnarray*}
We obtain the matrix elements $G_{ac;a'c'}(t)$ by expanding the
evolution operator on the right hand side to second order. Then,
denoting its matrix form as $\mathbf{G}(t)$, we obtain
\begin{eqnarray}
  &  & \mathbf{G}(t)\label{eq:G expand}\\
  & = & \mathbf{D}(t)\left[1-i\mathbf{D}^{-1}(t)\int_{0}^{t}dt'\left\langle
      e^{-i\mathcal{L}_{0}t}
      \mathcal{J}_{I}(t')\right\rangle _{L}\right.\nonumber \\
  &  & -\mathbf{D}^{-1}(t)\left\langle e^{-i\mathcal{L}_{0}t}
    \int_{0}^{t}dt'\int_{0}^{t'}dt''\mathcal{J}_{I}(t')\mathcal{J}_{I}
    (t'')\right\rangle _{L}\nonumber \\
  &  & \left.+\vphantom{-\mathbf{D}^{-1}(t-t_{0})\left\langle
        e^{-i\mathcal{L}_{0}(t-t_{0})}
        \int_{t_{0}}^{t}dt'\int_{t_{0}}^{t'}dt''\mathcal{J}_{I}(t')\mathcal{J}_{I}
        (t'')\right\rangle _{L}}
    \ldots\right].\nonumber 
\end{eqnarray}
Here $\mathbf{D}(t)$ is defined to be a diagonal matrix over the same
space as that of $\mathbf{G}$, and its elements are given by,
\begin{eqnarray}
  D_{ab;a'b'}(t)&=&R_{ab}(t)e^{-i\omega_{ab}t}\delta_{aa'}\delta_{bb'},\label{eq:D}\\
  R_{ab}(t) & = & \mbox{Tr}_{L}\left[e^{iK_{b}t}e^{-iK_{a}t}\mathfrak{R}\right].\label{eq:Dab}
\end{eqnarray}
Here we have defined operators $K_a$ that act only on the electrode degrees of freedom, but
are conditioned on the QD state,
\begin{equation}
  K_a = \frac{1}{\hbar}(H_L + \hat{V}_{aa}).\label{eq:Ka}
\end{equation}
The general expression~\eqref{eq:G expand} for the propagator has also been derived
earlier by Golosov \emph{et. al.}\cite{golosov} for a two-state system
of electronic degrees of freedom coupled to bosonic nuclear
motion. Below, we will specialize this propagator to a bath of
fermions, and provide mathematical details for the important
distinctions with respect to a bosonic reservoir.

To arrive at the equation of motion, we differentiate \eqref{eq:exact rho} with respect to $t$, set
$\rho(0)=\mathbf{G}^{-1}(t,0)\rho(t)$, and use \eqref{eq:G
  expand} expanded to the order shown there. We thus obtain an initial
value problem with the dynamics governed by a time-convolutionless
master equation, which we write in the following form,
\begin{eqnarray}
  \frac{d}{dt}\boldsymbol{\rho}(t) & = & \mathbf{M}(t)
  \boldsymbol{\rho}(t),\label{eq:Reduced EOM}\\
  \boldsymbol{\rho}(0)&=&\boldsymbol{\rho}_0\nonumber,
\end{eqnarray}
where $\boldsymbol{\rho}$ is now a vector obtained by re-arranging the
matrix elements of $\rho(t)$ as mentioned above. We express the
time-dependent mapping in~\eqref{eq:Reduced EOM} as,
\newcommand{\dpdt}{\ensuremath{\dot{\mathbf{P}}}}
%\begin{eqnarray}
%  \mathbf{M}(t) & = & \frac{d}{dt}\log\mathbf{D}(t)\label{eq:Massmatrix}+\mathbf{D}(t)\dot{\mathbf{P}}(t)\mathbf{D}^{-1}(t)
%   +\mathbf{D}(t)\left[ \dot{\mathbf{B}}(t)+\right.\nonumber\\
%   &&\left.\dot{\mathbf{C}}(t)\right]\mathbf{D}^{-1}(t)+\mathbf{D}(t)\mathbf{C}'(t)\mathbf{D}^{-1}(t)
%  ,\nonumber\\\label{eq:M(t)}
%\end{eqnarray}
\begin{eqnarray}
\mathbf{M}(t) & = & \dot{\mathbf{D}}(t) \mathbf{D}^{-1}(t) \label{eq:Massmatrix}\\
 &  & +\mathbf{D}(t)\frac{d}{dt}\left\{ \mathbf{P}(t) + \mathbf{B}(t)+\mathbf{C}(t) - \frac {1} {2} \mathbf{P}^2(t) \right\}
\mathbf{D}^{-1}(t).
\nonumber 
\end{eqnarray}
where $\mathbf{P}=\mathbf{P}_C+\mathbf{P}_T$, and $\dot{X}$ signifies the derivative with respect to time. The form of this expression is motivated by the separate physical
processes that are described by each of its terms, and the general
expressions for these terms in the superoperator form are as follows  (see \eqref{eq:D} above for definition of $\mathbf{D}$)
{\small
\begin{eqnarray}
\mathbf{P}_{j}(t)&=&-i\mathbf{D}^{-1}(t)\int_{0}^{t}dt'\left\langle e^{-i\mathcal{L}_{0}t}\mathcal{J}_{j}(t')\right\rangle _{L},\;\;j=C,T\\
\mathbf{B}(t)&=& -\mathbf{D}^{-1}(t)\int_{0}^{t}dt'\int_{0}^{t'}dt''\left\langle e^{-i\mathcal{L}_{0}t}
    \mathcal{J}_{T}(t')\mathcal{J}_{T}
    (t'')\right\rangle _{L}
    \label{eq:B gen}\\
    \mathbf{C}(t)&=& -\mathbf{D}^{-1}(t)\int_{0}^{t}dt'\int_{0}^{t'}dt''\left\langle e^{-i\mathcal{L}_{0}t}
    \mathcal{J}_{C}(t')\mathcal{J}_{C}
    (t'')\right\rangle _{L}\nonumber \\
    \label{eq:C gen}
    \end{eqnarray}
}
Returning to \eqref{eq:Massmatrix}, the first term of $\mathbf{M}(t)$ describes the
decoherence caused by sudden switching of the QD potential, the general
trends of which will be discussed in the following paragraphs. Note that $\mathbf{D}(t)$ is a diagonal matrix by definition in \eqref{eq:D}.
The second term in \eqref{eq:Massmatrix}, arises from the
first and second order contributions to the propagator, systematically included in \eqref{eq:Reduced EOM} to second order. The first order contribution $\mathbf{P}(t)$ includes both hybridization and Coulomb interaction terms, although for the case of a Fermionic reservoir, only the latter will be non-zero.  The second order contributions $\mathbf{B}(t)$ and $\mathbf{C}(t)$ respectively capture the hybridization and Coulomb interaction terms.

%We have separated the contribution of pure
%charge transfer and pure energy transfer by associating with these
%processes the matrices $\mathbf{B}(t)$ and $\mathbf{C}(t)$
%respectively, which are defined by Eqs.~\eqref{eq:B gen} and \eqref{eq:C gen} and are discussed at length in Sections \ref{sec:ct} and \ref{sec:et} below. The cross terms arising from the product of first order charge and energy transfer terms are contained  in $\mathbf{C}'(t)$ defined in \eqref{eq:Cprime gen}. We will not discuss this term further because it vanishes exactly in the case of a Fermionic reservoir as shown below.

% ----------------------------------------------------------------------------------------------------------
% AOC Discussion
% ----------------------------------------------------------------------------------------------------------
\subsubsection{Discussion and specialization to a Fermionic
  reservoir}~\label{sec:EOMdisc}

Let us now turn to the special interaction picture transformation
defined by~\eqref{eq:int-schrod} and discuss its subtleties. We begin
by taking the matrix elements of this equation between two QD states,
$\ket{a}$ and $\ket{b}$. Due to the fact that
$e^{iH_0t/\hbar}\ket{a}=e^{i\omega_at+iK_at}\ket{a}$, and similarly
for $\ket{b}$, we find
\begin{eqnarray*}
  \rho_{ab}(t) & = &
  \mbox{Tr}_{L}\left[e^{-iK_{a}t}\bra{a}\tilde{\xi}(t)\ket{b}e^{iK_bt}
  \right] e^{-i\left(\omega_{a}-\omega_{b}\right)t}.\label{eq:int to schrod gen}
\end{eqnarray*}
Therefore, unless $\tilde{\xi}(t)$ is in the form of a product of the
density matrices of the two sub-systems, a simple relation does not
exist between the reduced density matrices of the QD in the two
pictures.
% KSV: changed "outer product" to direct product
In the derivation above, we have developed the dynamical equations
under the assumption of a direct product form at $t=0$.  Therefore, it
is instructive to analyze the consequences of
Eq. \eqref{eq:rho(t)} with the product form, \emph{i.e.},
\begin{equation}
  \tilde{\xi}(t) = \tilde{\rho}(t)\otimes\mathfrak{R},
\end{equation}
where $\mathfrak{R}$ is any admissible density operator within the
Hilbert space of the electrode states. The relationship between the two
pictures is more complex than it is conventionally,
\begin{eqnarray*}
  \rho_{ab}(t) & = & R_{ab}(t)\tilde{\rho}_{ab}(t)e^{-i\left(\omega_{a}-\omega_{b}
    \right)t}.
\end{eqnarray*}
Here, by $\tilde{\rho}$ we mean the QD density matrix in the
interaction picture. Thus, in addition to the coherent oscillatory
factors arising from the QD states alone, there is an additional
complex-valued multiplicative factor, $R_{ab}(t)$ (see
\eqref{eq:Dab}), in the transformation from interaction to
Schr\"{o}dinger picture.
% \begin{eqnarray}
%R_{ab}(t) & = & \mbox{Tr}_{L}\left[e^{iK_{b}t}e^{-iK_{a}t}\mathfrak{R}\right].\label{eq:Dab}
%\end{eqnarray}
%
The function $R_{ab}(t)$ is a manifestation of the Anderson
orthogonality catastrophe (AOC), which together with the Mahan exciton
describes the Fermi edge singularity effects~\cite{Ohtaka1990}. To
make the link with AOC more explicit, we pick a basis set
$\{\ket{\psi_\alpha}\}$ for the electrode states in which the operator
$\mathfrak{R}$ is represented by a diagonal matrix with elements
$\mathfrak{R}_{\alpha}$.
Let$\ket{\psi^a_\alpha(t)}=e^{iK_at}\ket{\psi_\alpha}$, and write the
function $R_{ab}(t)$ as
\begin{eqnarray}
  R_{ab}(t) & = & \sum_\alpha
  \mathfrak{R}_{\alpha}\inner{\psi_\alpha^b(t)}{\psi_\alpha^a(t)}.
\end{eqnarray}
As pointed out by Anderson\cite{Anderson:1967p2374}, owing to the
macroscopic size of the electrode, the overlap of the two states
rapidly decays for a finite scattering of single particle states. We
may also view the function $R_{ab}(t)$ as the average decoherence
caused by the electrode states, where the latter act effectively as a
measurement distinguishing between the coherently superimposed states
$\ket{a}$ and $\ket{b}$ of the QD.

The solution to the equations of motion beyond the point at which
$R_{ab}(t)$ vanishes, or crosses zero, can be an extremely poor
approximation to the correct solution. To understand whether this
presents a difficulty in our theory, we note that the leading
contribution to AOC arises from differences in monopole moments of the
initial and final potentials. Thus we expect the decoherence due to
this mechanism to be very weak between QD states of the same net
charge, so that $R_{ab}(t)\approx1$ within each class of states
introduced above. When there is a charge transition, the AOC function
decays as a power law at low temperature, or as an exponential at high
temperatures. In either case, the function does not vanish exactly
within finite time except in the case of strong
coupling~\citep{breuerbook}. To exclude strong coupling from the
present scenario, we note that the coupling in our Hamiltonian is the
Coulomb interaction, which is always screened by the electrode. For
situations like those considered here the magnitude of the coupling is
far from that of the strong coupling regime. In fact, we have verified
this by explicit calculations of $R_{ab}(t)$ for typical values of
matrix elements in our Hamiltonian. Thus we proceed assuming that the
functions $R_{ab}(t)$ decay but do not vanish exactly within the
relevant temporal window of the dynamics.
% ----------------------------------------------------------------------------------------------------------
% End of AOC discussion
% ----------------------------------------------------------------------------------------------------------

We now consider the consequences of specializing the above general
formulation to the case of a Fermionic reservoir, which is the focus
of the present work. We assume that the state of the fermionic bath
representing the electrode is that of a normal metal, and described by
a mixture of states $\Phi_{\alpha}(N)$ where $\alpha$ is a state index
for $N$-particle many body states. Under this assumption, the
annihilation operators entering $H_0$ in \eqref{eq:int-pict} possess
the property,
\begin{equation}
  c_{\nu\boldk}\ket{\Phi_{\alpha}(N)}=\ket{\Phi^\prime(N-1)}\label{eq:restrict},
\end{equation}
for the single particle state $\ket{\nu\boldk}$ having a finite
occupation in the many body state $\ket{\Phi_{\alpha}(N)}$, and
$\ket{\Phi^\prime(N-1)}$ being (an un-normalized) $N-1$ particle
state. For such a state, the first order term in \eqref{eq:G expand}
vanishes whenever it corresponds to hybridization coupling. To see
this, write the trace as a sum over the many-body bases
$\ket{\Phi_{\alpha}(N)}$ and consider the matrix element between QD
states $\ket{a}$ and $\ket{b}$, such that $\ket{b}$ has one extra
electron relative to $\ket{a}$. Then the expectation value, $\langle
e^{-i\mathcal{L}_0t}\mathcal{J}_I(t')\rangle$, in the first order term
consists of two terms. One of these is proportional to the following
sum, {\small
  \begin{eqnarray*}
    &&\sum_{N,\alpha}
    \mathfrak{R}_{\alpha}(N)\bra{\Phi_{\alpha}(N)} e^{-i K_a
      (t-t')}c^\dagger_{\nu\boldk}e^{i K_b (t-t')}\ket{\Phi_{\alpha}(N)},
  \end{eqnarray*}
}%
and the other is obtained by $c^\dagger_{\nu\boldk}\rightarrow
c_{\nu\boldk}$. In the equation above, $\mathfrak{R}_\alpha(N)$ is the
probability of state $\Phi_\alpha(N)$ in a grand canonical
ensemble. Since, the operators $K_a$ and $K_b$ do not change the total
number of particles, the result of the above formula is an overlap
between $N$-particle and $N+1$-particle Fock states. Therefore it
vanishes, and so do all odd order terms in \eqref{eq:G expand}. 

In the language of our formalism, this implies that  $\mathbf{P}_T=0$. We
remark that this is a consequence of \eqref{eq:restrict},
and the fact that hybridization involves an odd number of creation
annihilation operators for electrons. If the electrode were, for
example, in a BCS state then the previous sum over states would also
involve states in coherent superposition of different $N$, resulting
in a non-vanishing expectation value of $c_{\nu\boldk}^\dagger$ in
general.

Furthermore, when considering the contribution of Coulomb coupling via $\mathcal{J}_C$, the
first order term, $\mathbf{P}_C$ in \eqref{eq:G expand} does not vanish. However,
since Coulomb coupling does not change the charge state of the QD, this
term can be understood as a Hartree energy correcting for
the fact that electrode states are defined in the presence of a
neutral QD. Thus for a Fermionic reservoir, $\mathbf{P}_C(t)$ in
\eqref{eq:Massmatrix} is the off-diagonal Coulomb potential matrix. It
is discussed below in Sec. \ref{sec:et} with its definition given by
\eqref{eq:Pmaintext}. We also note that this term does not affect
populations, but describes only the dynamical re-organization energy within the
electrode during coherent oscillations between different charge states
of the QD.

For an electrode at equilibrium, the charge and energy transfer
processes may couple at third order. This coupling would affect the
rates of charge transfer such that an excited electron or a hole in a
state with low escape rate may exchange energy with the electrode and
jump to a state with larger escape rate. This modification of the
charge lifetime of the QD may be expected for closely lying hole
levels. However, since the Coulomb interaction is small due to
screening, and tunneling is exponentially suppressed by increase
in junction width, this regime may be an exception rather than a rule.
Thus we do not pursue it in the analysis below.

We now stay within the confines of a Fermionic reservoir, construct
the expressions for $\mathbf{B}$, $\mathbf{P}_j$, and $\mathbf{C}$
matrices, and discuss their physics. From the physical insight gained
into these matrices, we are also able to obtain useful approximations
that simplify their numerical evaluation.

\subsection{Charge transfer rates}\label{sec:ct}

We first write the matrix elements of $\mathbf{B}(t)$ in terms of the
hybridization operators to facilitate the connection with electrode
correlations, and then evaluate the matrix elements as shown
below. The matrix elements $B_{ac;a'c'}(t)$ follow from the general expression \eqref{eq:B gen} 
specialized to a Fermionic reservoir, in which the term $\mathbf{P}_T=0$ as discussed above. Thus we obtain, 
\begin{eqnarray}
  &  & B_{ac;a'c'}(t)\label{eq:B jj}\\
  & = & 
  -D_{ac}^{-1}(t)e^{-i\omega_{ac}t}\left[\frac{1}{\hbar^2}\int_{0}^{t}dt_{1}\int_{0}^{t_{1}}dt_{2}\right.
  \nonumber \\
  &  & \sum_{b}\left\{\delta_{cc'}\left\langle
    \mathcal{T}_{0}^{ca}(t)\,\,\mathcal{T}^{ab}(t_{1})
    \mathcal{T}^{ba'}(t_{2})\mathfrak{R}\right\rangle e^{i(\omega_{ab}t_{1}+
    \omega_{ba'}t_{2})}\right.\nonumber \\
  &  & \left.+\delta_{aa'}\left\langle
    \mathcal{T}^{c'b}(t_{2})\mathcal{T}^{bc}(t_{1})\,\,
    \mathcal{T}_{0}^{ca}(t)\mathfrak{R}\right\rangle
  e^{i(\omega_{c'b}t_{2}+\omega_{bc}t_{1})}\right\}
  \nonumber \\
  &  & -\frac{1}{\hbar^2}\int_{0}^{t}dt_{1}\int_{0}^{t}dt_{2}\nonumber \\
  &  & \left.\left\langle
      \mathcal{T}^{c'c}(t_{2})\,\,\mathcal{T}_{0}^{ca}(t)\,\,\mathcal{T}^{aa'}
      (t_{1})\mathfrak{R}\right\rangle
    e^{i(\omega_{aa'}t_{1}+\omega_{c'c}t_{2})}\right],\nonumber 
\end{eqnarray}
where $\omega_{ac}=\omega_a-\omega_c$. The matrix element $B_{ac;a'c'}(t)$ describes scattering from the
``initial state'' $\rho_{a'c'}$ to the ``final state''
$\rho_{ac}$. Here the phase factors due to the QD energy levels are shown explicitly and the time dependent hybridization operators reflect the action of $K_a$ in \eqref{eq:Ka}.  In addition, we have introduced what we call the \emph{pure dephasing operator},
\begin{equation}
  \mathcal{T}_0^{ca}(t)=e^{iK_ct}e^{-iK_at},\label{eq:pdo}
\end{equation}%
which does not generate any charge transfer, and only affects coherences 
by accounting for the FES effects for oscillations between states of the QD with different effective Coulomb potentials. We remark that it is
straightforward to verify that the matrix elements, $B_{ac;a'c'}(t)$ obey the sum rule,
\begin{eqnarray}
  B_{mm;mm}(t) & = & -\sum_{n\neq m}B_{nn;mm}(t),\label{eq:Bsum rule}
\end{eqnarray}
which is quite general, and in turn ensures that the sum of all the rates for the
population to relax from a state $\left|m\right\rangle $ equals the
total decay rate of this state.

\begin{figure}
  \includegraphics[width=3in]{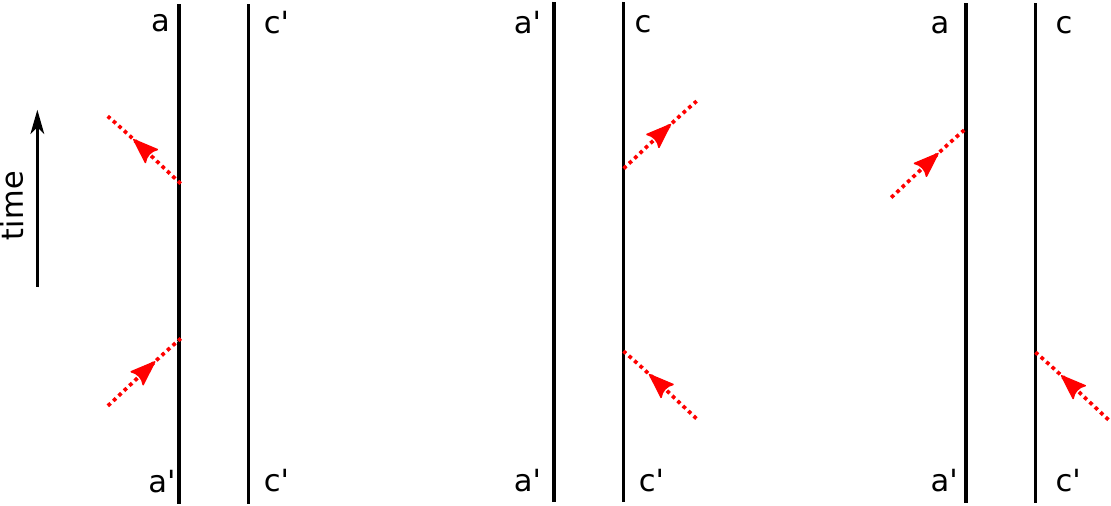}
  \caption{\label{fig:Double-sided-Feynman-1}Double sided Feynman
    diagrams showing the effect of the three terms in the dynamical
    hybridization couplings.  The red arrows represent the addition
    and removal of electrons from the QD respectively, and the two
    vertical lines represent the ``ket'' on the left and ``bra''on the
    right.}
\end{figure}

Returning to~\eqref{eq:B jj}, we note that since there are no hybridization operators of the form
$\mathcal{T}^{cc}(t)$ in the entire Hamiltonian, a given density
matrix element is acted upon by either the first two terms of that equation or the
third but not both.  However, the three terms are not independent, and
satisfy sum rules due to the conservation of total particle number by
the underlying Hamiltonian.  These terms may also be interpreted as
generalized scattering ``in'' and ``out'' rates for populations and
coherences. We depict the effect of these rates on the density matrix
in Fig. \ref{fig:Double-sided-Feynman-1} using double sided Feynman
diagrams. As shown there graphically, the first two terms couple only
coherences to populations, while the third term provides an additional
pathway for coupling populations that differ by one electron. The same
effect would occur at a higher order in the form of the third
diagram. Note that $\mathcal{T}_0(t)$ does not have a representation
in terms of these double sided Feynman diagrams because it does not
change the state of the QD.

% ---------------------------------------------------------------------
% Figure: Double sided Feynman Diagram
% ---------------------------------------------------------------------
% ---------------------------------------------------------------------

From the definition \eqref{eq:Teg}-\eqref{eq:Tex} of the hybridization
operators, and for an electrode consisting of an electron reservoir in
equilibrium, we find
\begin{eqnarray}
  & & B_{ac;a'c'}(t) \label{eq:B ac dc}\\
  & = &
  -D_{ac}^{-1}(t)e^{-i\omega_{ac}t}e^{i(\omega_{aa'}-\omega_{cc'})t}\frac{1}{\hbar^2}\int_0^t
  dt'\int_{-
    \infty}^{+\infty}\frac{d\omega}{2\pi} \left[\right.\nonumber\\
  &  &
  \phantom{+}\gamma_{c'c;aa'}(\omega,t';t)S(\omega-\omega_{a'a},t')\nonumber
  \\
  &  & +\gamma_{a'a;cc'}(\omega,t';t)S(\omega-\omega_{c'c},t')\nonumber \\
  &  &
  -\delta_{a'a}\sum_{b}\lambda^{ac}_{cb;bc'}(\omega,t';t)S(\omega-\omega_{cb},
  t')\nonumber \\
  &  &
  -\delta_{c'c}\sum_{b}\lambda^{ca}_{ab;ba'}(\omega,t';t)S(\omega-\omega_{ab},
  t')\left.\right]\nonumber 
\end{eqnarray}
 The derivation of this formula is provided in Appendix~\ref{sec:math},
and we have defined here a function,
\begin{eqnarray*}
  S(\omega,t) & = & -\frac{\sin(\omega
    t)}{\omega}+\frac{2i\sin^{2}[\omega\frac{t}{2}]}{\omega},
\end{eqnarray*} 
which accounts for the initial condition defined at a finite time, and
allows us to work with Fourier transforms with respect to the initial
time at $-\infty$ (see Appendix \eqref{sec:math}). The functions
$\gamma_{c'c;aa'}(\omega,t';t)$ and
$\lambda^{ca}_{ab;ba'}(\omega,t';t)$ are a generalization of the
spectral functions and defined as Fourier transforms of causal
response functions,
\begin{eqnarray*}
  \gamma_{c'c;aa'}(\omega,t';t) & = & -2\Im\int_{-\infty}^{+\infty}d
  t''\Gamma_{c'c;aa'}(t'',t';t)e^{i\omega t''},\\
%
  % KSV: Typo corrected: subscript is ab;ba'
%
  \lambda^{ca}_{ab;ba'}(\omega,t';t) & = & -2\Im\int_{-\infty}^{+\infty}d
  t''\Lambda^{ca}_{ab;ba'}(t'',t';t)e^{i\omega t''},
\end{eqnarray*}
so that the Fourier transform integrals can extend over the entire
real axis. The integrands in the previous expressions are the
correlation functions appearing in \eqref{eq:B ac dc}; the superscript
in $\Lambda$ identifies the indices on the pure dephasing operator
\eqref{eq:pdo}, while these indices are implied by subscripts in the
function $\Gamma$. The function $\Lambda$ captures the first two
correlation functions in \eqref{eq:B jj}, while $\Gamma$ captures the
last one, and they are defined as,
\begin{eqnarray}
  \Lambda^{ca}_{ab;ba'}(t'',t';t) & = & -i\Theta(t'-t'')\left\langle
    \mathcal{T}_0^{ca}(t)\mathcal{T}^{ab}(t')\mathcal{T}^{ba'}(t'')\mathfrak{R}
  \right\rangle, \nonumber \\
  \Gamma_{c'c;aa'}(t'',t';t) & = & -i\Theta(t'-t'')\left\langle
    \mathcal{T}^{c'c}(t')\mathcal{T}_0^{ca}(t)\mathcal{T}^{aa'}(t'')\mathfrak{R}
  \right\rangle. \nonumber\\ \label{eq:Lambda-Gamma}
\end{eqnarray}

In contrast to conventional correlation functions for a single particle propagation, these two correlation functions contain three time arguments. Their dependence on the third argument arises from the operator $\mathcal{T}^{ca}_0(t)$, which differs from unity only  when $a\neq c$ in general. Furthermore, physically, it is only significantly different from unity when the difference between the potentials of the two states is large enough to cause significant AOC. Thus the third argument describes the shakeup of the final electrode states when the QD oscillates between two different charge states. This is fundamentally different than the processes described by the first two arguments.  Specifically, the time difference $t''-t'$ relates to the particle absorption/emission spectrum of the electrode in the presence of a QD.  The average time $(t''+t')/2$ relates to the memory of the initial state potential of the QD. In our notation, we use a semi-colon to set apart the two different kinds of time arguments.

The full mathematical analysis of correlation functions Eq. (50) is outlined in Appendix~\ref{sec:math}.  To gain physical insight into the results therein, we focus on the process of electron transfer from the QD to the electrodes. Then the pertinent correlation function is denoted by the superscript ``>'' and defined as,
%-----------------------------------------------
\begin{widetext}
\begin{eqnarray}
  \Gamma_{c'c;aa'}^{>}(t'',t';t)
  &=&
  \frac{-i}{Z}\Theta(t'-t'')\sum_{\nu\nu'\boldsymbol{k}\boldsymbol{k}'}T^{c'c}_{\nu\boldk}T^{aa'}_{\nu'\boldk'}
  \left\langle e^{iK_{c'}t'}c_{\nu\boldsymbol{k}}e^{-iK_{c}(t'-t)}e^{iK_{a}(t''-t)}c_{\nu'\boldsymbol{k}'}^{\dagger}e^{-iK_{a'}t''}e^{-\beta
      H_{L}}\right\rangle,\label{eq:Gamma-green-func}
\end{eqnarray}
\end{widetext}
%----------------------------------------------
where we have made the physical assumption that $\mathfrak{R}=\exp(-\beta H_L)/Z$ with the usual partition function $Z=\left<\exp(-\beta H_L)\right>$. 

The correlation
function under the sum on the right hand side of
\eqref{eq:Gamma-green-func} may be interpreted as a
thermal average of overlaps between two electrode states evolving
under different potentials.  To see this, take a many body electrode state with $N$
electrons, $\left|\Phi(N)\right\rangle $, which would appear as a ket when computing the expectation values in~\eqref{eq:Gamma-green-func}.
First consider the case where
$\mathcal{T}_0^{ca}(t)\approx 1$, which occurs when the states
$\ket{a}$ and $\ket{c}$ have the same charge. Then all the correlation
functions above reduce to functions with two essential time arguments. The state $\left|\Phi(N)\right>$ initially evolves under the
potential $\hat{V}_{a'a'}$ until time $t''$, and at this time an
electron is injected from the QD into a single particle state of the electrode with in-plane momentum
$\boldsymbol{k}'$. Within this scenario, the QD can either be in a negatively charged
state or in an exciton state immediately before $t''$; it cannot be in the ground state. After
tunneling, the QD potential \emph{switches} to $\hat{V}_{aa}$ and the
state is evolved \emph{back} to the initial time ($t=0$). Let us call
this state $\left|\Phi''(N+1)\right\rangle $ to denote the fact that
the state corresponds to time $t''$. Similarly, the bra form of 
$\left|\Phi(N)\right\rangle $ evolves under $\hat{V}_{c'c'}$ until
time $t'$, when an electron is added to it in the single
particle level $\boldsymbol{k}$, and the resulting $N+1$ electron
state is evolved back to the initial time under the influence of the
QD potential $\hat{V}_{cc}$. We label the resulting state
$\left|\Phi'(N+1)\right\rangle $.

With these definitions, we see that the correlation function
in~\eqref{eq:Gamma-greater-matrix-main-text} is equal to the overlap
$\left\langle \Phi'(N+1)\left|\right.\Phi''(N+1)\right\rangle $,
the trace in that expression being a thermal average over all the many
body states, and the summation over $\boldsymbol{k}$ and
$\boldsymbol{k}'$ is a summation over all possible in-plane momenta
into which the electron can be injected. The weights for the latter
summation are the tunneling amplitudes, which give the probability of
such an injection to occur. We may interpret in the same way the
correlation functions in which $c_{\boldsymbol{k}'}^{\dagger}$ is to
the left of $c_{\boldsymbol{k}}$. As they describe the evolution with
an electron removed from the initial state, the overlap is
$\left\langle \Phi'(N-1)\left|\right.\Phi''(N-1)\right\rangle $.

It is now clear that if $\hat{V}_{c'c'}$
and $\hat{V}_{a'a'}$ are weak (for example, if their source has a
vanishing charge), the propagation of $\left|\Phi(N)\right\rangle $
until the time of adding the electron is identical in both
$\left|\Phi'(N+1)\right\rangle $ and
$\left|\Phi''(N+1)\right\rangle $. Therefore, the origin of time
loses significance and the resulting correlation becomes a function
only of the \emph{time difference} $t'-t''$. It thus reduces to a correlation function 
for a system in equilibrium or in a steady state, where the effects of initial conditions have decayed. Therefore, the dependence of
the generalized spectral functions on $t'$ in~\eqref{eq:B ac dc} is in
proportion to the memory of the QD potential in the initial state.  Similarly, it
follows from \eqref{eq:Lambda-Gamma} that the dependence on $t$ is in
proportion to the difference between potentials of the superimposed
final states at time $t$.

It follows from above that if only the monopole contribution is significant, the
correlation function would depend only on the time difference, $t'-t''$ so long as
the initial and final state of the QD have the same charge. The Fourier transform of 
the correlation function with respect to this time difference then defines a conventional spectral
function describing single particle absorption or emission as a function of energy.
When coupling to the dipole and higher order potentials is strong enough,
coherent oscillation between the states of different multipoles would, in general, lead to this 
spectral function evolving as a function of the average time $(t'+t'')/2$, thus exhibiting non-Markovian behavior.

%-----------------------------------------------
We now summarize the mathematical expressions that can be evaluated with our physical model.  In particular, we specialize to the case of quadratic Hamiltonians, and follow Blankenbecler \emph{et al}~\citep{bss} and Hirsch~\citep{hirsch} to develop explicit expressions for these correlation functions in terms of a matrix inverse and
matrix multiplication in the single particle basis for the electrode states. Thus for the
correlation function in which an electron tunnels from the QD to the
electrode, the analysis presented in Appendix~\ref{sec:math} yields,
\begin{eqnarray}
  & &\Gamma_{c'c;aa'}^{>}(t'',t';t)=\label{eq:Gamma-greater-matrix-main-text}\\
  &&-i\Theta(t'-t'')r_{c'c;aa'}(t'',t',t)\sum_{\nu,\nu'\boldsymbol{k}\boldsymbol{k}'}T^{c'c}_{\nu\boldk}T^{aa'}_{\nu'\boldk'}\nonumber\\
  &&\left\langle \nu\boldsymbol{k}\right|\mathbf{\bar{N}}_{c'c'}(-t',0)
  \left[\mathbf{I}+\boldsymbol{\Sigma}^{(1)ca}_{c'c;aa'}(t',t'';t)\mathbf{\bar{N}}_{c'c'
    }(-t',0)\right]^{-1}\nonumber\\
  &&\left[e^{-iK_{c'}t'}e^{iK_{a'}t''}\right]\left|\nu'\boldsymbol{k}'\right\rangle 
  ,\nonumber 
\end{eqnarray}
and
\begin{eqnarray}
  &&\Lambda^{>ca}_{ab;ba'}(t'',t';t)=\label{eq:Lambda-greater-matrix-main-text}
  \\
  &&
  -i\Theta(t'-t'')r_{ca;ba'}(t,t'',t')\sum_{\nu,\nu'\boldsymbol{k}\boldsymbol{k}'}T^{ab}_{\nu\boldk}
  T^{ba'}_{\nu'\boldk'}\nonumber\\
  &&\left\langle \nu\boldsymbol{k}\right|\mathbf{\bar{N}}_{ac}(t-t',-t)
  \left[\mathbf{I}+\boldsymbol{\Sigma}^{(2)ac}_{ab;ba'}(t',t'';t)\mathbf{\bar{N}}_{ac}
    (t-t',-t)\right]^{-1}\nonumber\\
  &&\left[e^{-iK_{c'}t'}e^{iK_{a'}t''}\right]\left|\nu'\boldsymbol{k}'\right\rangle.
  \nonumber
\end{eqnarray}
The superscript ``>'' in these definitions indicates that the
correlation functions correspond to the propagation of a state in
which an electron is added to the electrode (see
Appendix~\ref{sec:math}). The pairs of indices on the correlation function are constrained by the type of correlation function, ``lesser'' and ``greater'', in accordance with~\eqref{eq:Gamma-cases}. 

In these expressions, the evaluation of the correlation functions has been divided into two parts.  First, we have defined a more general function to capture the AOC effects for four potentials,
%We have also defined the functions $r_{ab;b'a'}(t'',t';t)$, which describe AOC effects but with possibly
%four rather than two different potentials,
\begin{eqnarray}
  &  & r_{c'c;aa'}(t'',t';t)\label{eq:r function}\\
  & = & \frac{1}{Z}\left\langle
    e^{iK_{c'}t'}e^{-iK_{c}(t'-t)}e^{iK_{a}(t''-t)}e^{-iK_{a'}t''}e^{-\beta
      H_{L}}\right\rangle\nonumber.
\end{eqnarray}
Second, the part that explicitly depends on the hybridization functions is now reduced to the proper combination of matrices that represent each indicated operator in the natural single particle basis for the electrodes, $\left|\nu \boldsymbol{k}\right>$.  The matrix 
$\mathbf{\bar{N}}_{ac}(t,\tau)$ is defined to be,
\begin{equation}
  \mathbf{\bar{N}}_{ac}(t,\tau)\equiv e^{-iK_{a}t}e^{-iK_{c}\tau}\left[1+e^{-\beta H_L}\right]^{-1}e^{iK_{c}
    \tau } e^ { iK_{a}t}\label{eq:Wac(t)},
\end{equation}
In the $\left|\nu\boldsymbol{k}\right>$ basis, the central term is simply the statistical weight for empty states
\begin{equation}\label{eq:W0}
 \frac{\delta_{\nu\boldsymbol{k},\nu'\boldsymbol{k}'}}{1+
    \exp\left\{ -\left(E_{\nu\boldsymbol{k}}-\mu\right)/k_{B}T\right\} }.
\end{equation}
Thus, the process for electron transfer into the electrode is proportional to the empty states available.  However, the $\mathbf{\bar{N}}_{ac}(t,\tau)$ includes the time-dependent shake-up processes induced by the potential of the QD states $\left|a\right\rangle $ and $\left|c\right\rangle $.  In turn, this influences the distribution of states available to receive the electron, opening additional states below or blocking states above the Fermi level.  

The matrices $\boldsymbol{\Sigma}^{(j)}$, $j=1,2$,
in~\eqref{eq:Gamma-greater-matrix-main-text} and
\eqref{eq:Lambda-greater-matrix-main-text} are common to both emission
and absorption, and are defined as,
\begin{eqnarray}
  \boldsymbol{\Sigma}^{(1)bb'}_{c'c;aa'}(t',t'';t)&=&e^{-iK_{c'}t'}e^{iK_{a'}t''}
  e^{-iK_{a}t''}\mathcal{T}_0^{bb'}(t)e^{iK_{c}t'}-
  \mathbf{I},\nonumber\\
  \label{eq:sigma1}\\
  \boldsymbol{\Sigma}^{(2)bb'}_{c'c;aa'}(t',t'';t)&=&e^{-iK_{c'}t'}\mathcal{T}_0^{
    bb'}(t)e^{iK_{a'}t''}e^{-iK_{a}t''}e^{iK_{c}t'}-\mathbf{I}.\nonumber\\
  \label{eq:sigma2}
\end{eqnarray}
For the correlation corresponding to an electron tunneling from the
electrode to the QD, expressions similar to
\eqref{eq:Gamma-greater-matrix-main-text} and
\eqref{eq:Lambda-greater-matrix-main-text} exist (see Eqs.
\ref{eq:Gamma-lesser-matrix} and \ref{eq:Lambda-lesser-matrix}). In those correlation functions,  $\mathbf{\bar{N}}$ is replaced by the
compliment, $\mathbf{N}=\mathbf{I}-\mathbf{\bar{N}}$.
%In the correlation function for the emission of an
%electron from the electrode, $\mathbf{\bar{N}}$ is replaced by the
%compliment, $\mathbf{N}=\mathbf{I}-\mathbf{\bar{N}}$.

% ---------------------------------------------------------------------
% Figure: Illustration of tunneling correlation
% ---------------------------------------------------------------------
% \begin{figure*}
%   \subfloat[]{\includegraphics[height=2in]{correlation-drawing}
% }\hfill{}\subfloat[]{\includegraphics[height=2in]{correlation-drawing-1}
% }
%   \caption{\label{fig:correlation illustration}Pictorial
%   illustration of the process described by the correlation function
%   $\Gamma_{abb'a'}(t_{1},t_{2})$}
% \end{figure*}
% ---------------------------------------------------------------------

\subsection{Energy transfer rates}\label{sec:et}

%We now turn to the expression for the matrices $\mathbf{P}_C(t)$ and
%$\mathbf{C}(t)$, which are the first and the second order terms,
%respectively, in the cumulant expansion of the propagator \eqref{eq:G
%  expand}. As shown in \eqref{eq:first order app}, the first order
%term is given by,
%\begin{eqnarray}
%  &&\frac{d}{dt}P_{Cac;a'c'}(t)=-i\frac{\delta_{cc'}\Delta_{aa'}(t)-\delta_{aa'
%    } \Delta_ { c'c } (t)}{D_{a'c'}(t)}\nonumber\\
%  &&-i\frac{\dot{D}_{ac}(t)}{D_{a'c'}(t)D_{ac}(t)}\int_0^t
%  d\tau(\delta_{cc'}\Delta_{aa'}(\tau)-\delta_{aa'}\Delta_{c'c}(\tau))\nonumber\\
%  &&+\delta P_{Cdipole}(t).\label{eq:Pmaintext}
%\end{eqnarray}
We now turn to the expression for the matrices  $\mathbf{P}(t)$ and $\mathbf{C}(t)$, which are
 the first and the second order terms, respectively, that enter in the cumulant expansion of the
propagator \eqref{eq:G expand}.  The first order term deriving from the Coulomb interaction, Eq. (40), and specialized to the Fermionic reservoir, reads
\begin{eqnarray}
P_{ac;a'c'}(t) = -iD_{ac}^{-1}(t)\frac {e^{-i\omega_{ac}t}} {\hbar} \int_0^t dt'
  \nonumber \\
\left [ \delta_{cc'}\left\langle
\mathcal{T}_0^{ca}(t)\hat{V}_{aa'} (t')\mathfrak{R}\right\rangle e^{-i\omega_{aa'}t'} \right . \nonumber \\
\left . -\delta_{aa'}\left\langle
\hat{V}_{c'c}(t')\mathcal{T}_0^{ca}(t)\mathfrak{R}\right\rangle
e^{-i\omega_{c'c}t'}  \right ]\label{eq:Pmaintext}.
\end{eqnarray}
Recall that the diagonal terms (\emph e.g. $\hat{V}_{aa}$) do not appear in the $\hat{V}_C$ operator, having been put in $H_0$.
As shown in~\eqref{eq:first order app} the above expression for $P_{ac;a'c'}$ may be expressed in terms of a \emph{Hartree energy matrix},
\begin{equation}
  \Delta_{aa'}(\tau) =  (1-\delta_{aa'})\langle
  \hat{V}_{aa'}\mathfrak{R}^{(aa')}(\tau)\rangle,\label{eq:hartree-energy}
\end{equation}
the time-dependence of which arises from the evolution of
the electrode state under the influence of the average of the
potentials of the QD states $\ket{a}$ and $\ket{a'}$, which we write as
\[\mathfrak{R}^{(aa')}(t)\equiv
e^{i\frac{t}{2}(K_a+K_{a'})}\mathfrak{R}e^{-i\frac{t}{2}(K_a+K_{a'})}.\]
By writing the time-dependence in this manner, we fully capture the
monopole contributions to pure dephasing (see Appendix
\ref{sec:energytransfer}). The effects of first and higher moments of this density can be
ignored due to the Friedel sum rule\citep{mahanbook}, 
and have been verified by our calculations to be small. We remark that this approximation is a direct
consequence of the charge conservation implied by the off-diagonal
components of the Coulomb matrix in \eqref{eq:V coul final}. Furthermore, the Hartree correction arises only
in the coupling of coherences to each other and to populations due to the fact that it is of first order and that the diagonal elements of its underlying Coulomb potential matrix have been removed.

%This is so because the matrix
%$\hat{V}_{aa'}$ connects only those states that have an equal net
%charge. Our approximation leaves out the second and higher order
%multipole moments of the transition density of QD states,
%$\varphi^*_{a}(\boldsymbol{r})\varphi_{a'}(\boldsymbol{r})$. As a
%consequence of the Friedel sum rule\citep{mahanbook} the orthogonality
%effects are proportional to the total charge of the QD potential, and
%therefore to the integral of the transition density over all
%space. 

We now turn to the second order term, denoted by the matrix
$\mathbf{C}(t)$, which generates energy transfer between the electrode
and the QD. The full expression for $\mathbf{C}(t)$ is identical in form to that for $\mathbf{B}$, except for the appearance of the Coulomb interaction operators $\hat{V}_{ac}(t)$ instead of the hybridization operators $\mathcal{T}_{ac}(t)$. 
However, instead of starting with the full form, we derive the expression for $\mathbf{C}$ by neglecting the effects of AOC 
altogether~\footnote{Formally, we replace the correlation functions by the factorized form in which the pure dephasing operator is factored out as an expectation value.}. 
These effects can occur only at third order in the Coulomb
interaction between the QD and the electrode.  Furthermore, retaining
these effects amounts to calculating the dielectric function of an
electrode driven out of equilibrium by the fluctuations in the QD
potential. We expect this to be a very weak effect due to the
screening of this potential, and the macroscopic size of the
electrode. Furthermore, since the Coulomb interaction does not change
the charge of the QD states, changes in the potential induced by
Coulomb processes are much weaker than hybridization. The neglect of AOC in energy transfer
simplifies the expression for $\mathbf{P}_C(t)$ by reducing it to $\dot{\mathbf{P}}_C(t)=\left\langle \mathcal{J}_{C}(t)\right\rangle _{L}$.  This also allows us to write the combination $\mathbf{C}-\mathbf{P}^2/2$ in~\eqref{eq:Massmatrix} in terms of the fluctuation of potentials around their average values, with~\eqref{eq:C gen} giving the general expression for $\mathbf{C}(t)$

Thus for states $\left|a\right\rangle $ and $\left|a'\right\rangle $
carrying an equal charge, and $\left|c'\right\rangle $ and
$\left|c\right\rangle $ carrying an equal but not necessarily the same
charge as $\left|a\right\rangle $, we obtain,
\begin{eqnarray}
  &  & C_{ac;a'c'}(t)\label{eq:C matrix}\\
  & = & -\frac{1}{\hbar^2}e^{-i\omega_{ac}t}\int_{0}^{t}dt_{1}\int_{0}^{t_{1}}dt_{2}\left[\sum_{b}\right.\nonumber \\
  &  & \phantom{+}\delta_{c'c}\left\langle
    \Delta\hat{V}_{ab}(t_{1})\Delta\hat{V}_{ba'}(t_{2})
    \mathfrak{R}\right\rangle e^{i(\omega_{ab}t_{1}+\omega_{ba'}t_{2})}\nonumber \\
  &  & \left.+\delta_{aa'}\left\langle
    \Delta\hat{V}_{c'b}(t_{2})\Delta\hat{V}_{bc}(t_{1})\mathfrak{R}
  \right\rangle e^{i(\omega_{bc}t_{1}+\omega_{c'b}t_{2})}\right]\nonumber \\
  &  & -\frac{1}{\hbar^2}\int_{0}^{t}dt_{1}\int_{0}^{t}dt_{2}\nonumber \\
  &  & \left\langle
    \Delta\hat{V}_{c'c}(t_{2})\Delta\hat{V}_{aa'}(t_{1})\mathfrak{R}\right\rangle 
  e^{i(\omega_{aa'}t_{1}+\omega_{c'c}t_{2})}.\nonumber 
\end{eqnarray}

The three terms in this expression act on the density matrix in a way
similar to the corresponding three terms of the tunneling process.
Here we have defined the operator for the fluctuation of
$\hat{V}_{ab}$ around its mean value as,
\begin{eqnarray*}
  \Delta\hat{V}_{ab}(t) & = & e^{iH_{L}t/\hbar}\hat{V}_{ab}e^{-iH_{L}t/\hbar}-\left\langle
    \hat{V}_{ab}
    \mathfrak{R}\right\rangle .
\end{eqnarray*}
Note that we have neglected the effects of AOC due to the QD potential
already in this definition; the subscripts on these operators only
identify the matrix elements when \eqref{eq:Vhat nn' def} is
substituted for performing calculations. We remark that the matrix elements $C_{ac;a'c'}(t)$ also obey the sum rule analogous to~\eqref{eq:Bsum rule}
\begin{eqnarray}
  C_{mm;mm}(t) & = & -\sum_{n\neq m}C_{nn;mm}(t).
\end{eqnarray}

Following the derivation of $\gamma_{c'c;aa'}(\omega,t';t)$ above, we
now define the functions
\begin{eqnarray}
  \chi_{c'c;aa'}(\omega) & = &
  -2\Im\left(1-\delta_{c'c}\right)\left(1-\delta_{a'a}\right)
  \sum_{\boldsymbol{q}}\label{eq:Xkk;}\\
  &  &
  V_{c',\nu;c,\nu'}(-\boldsymbol{q})X_{\nu\nu';\mu\mu'}(\boldsymbol{q},\omega)V_{a
    ,\mu;a',
    \mu'}(\boldsymbol{q}),\nonumber 
\end{eqnarray}
where $X_{\nu\nu';\mu\mu'}(\boldsymbol{q},\omega)$ is the
\emph{renormalized} density-density correlation function, which is
generalized to include inter-subband transitions. By renormalization,
we mean that a summation over an infinite series of particle-hole
excitations is performed between the times of the two
interactions. Thus if $X_{\nu\nu';\mu\mu'}^{0}(\boldsymbol{q},\omega)$
is the response function of a non-interacting gas, and corresponds to
a single particle-hole excitation, then
\begin{eqnarray}
  X_{\nu\nu';\mu\mu'}(\boldsymbol{q},\omega) & = &
  X_{\nu\nu';\alpha\alpha'}^{0}(\boldsymbol{q},
  \omega)\nonumber\\
  & - &
  X_{\nu\nu';\alpha\alpha'}^{0}(\boldsymbol{q},\omega)\mathcal{V}_{
    \alpha\alpha';\beta\beta'}
  (\boldsymbol{q})X_{\beta\beta';\alpha\alpha'}(\boldsymbol{q},\omega)\nonumber,\\
  \label{eq:chi-2d}
\end{eqnarray}
where $\mathcal{V}_{\alpha\alpha';\beta\beta'}(\boldsymbol{q})$ is the
effective Coulomb interaction describing the momentum exchange
$\boldsymbol{q}$ between particles scattering from bands
$\alpha',\beta'$ and into the bands $\alpha,\beta$. This matrix is an
effective interaction because it accounts for the static screening by
the bulk substrate beneath the surface. When this substrate is a
metal, the large plasmon frequency ensures that the bulk response may
be considered instantaneous, which creates a static screening of the
Coulomb potential of a surface electron, and the result defines the
two-particle interaction for the surface modes. Similarly, the
effective bulk dielectric function is also static for insulators or
wide-gap semiconductors. The dynamical screening by the bulk may be
important only for narrow-gap materials.

Following the mathematical steps outlined in Appendix~\ref{sec:math}
we obtain,
\begin{eqnarray}
  &  & \frac{d}{dt}C_{ac;a'c'}(t)\label{eq:dC/dt}\\
  & = &
  \frac{1}{2\hbar^2}e^{-i\omega_{ac}t}e^{i(\omega_{aa'}-\omega_{cc'})t}\int_{-\infty}^{+\infty}
  d\omega\nonumber \\
  &  & \phantom{+}\chi_{c'c;aa'}(\omega)S(\omega-\omega_{cc'},t)\nonumber \\
  &  & +\chi_{a'a;cc'}(\omega)S(\omega-\omega_{aa'},t)\nonumber \\
  &  &
  -\delta_{a'a}\sum_{b}\chi_{c'b;bc}(\omega)S(\omega-\omega_{bc'},t)\nonumber \\
  &  &
  -\delta_{c'c}\sum_{b}\chi_{a'b;ba}(\omega)S(\omega-\omega_{ba'},t).\nonumber 
\end{eqnarray}

This expression forms the basis of our calculations of energy transfer
driven by Coulomb interaction between the QD and the electrode.  We
remark that the electron susceptibility in this expression may be
calculated to any level of sophistication within the computational
constraints. Note that we did not include electron-electron
interaction within the electrode \emph{explicitly} in the Hamiltonian
\eqref{eq:full H}. These interactions enter our theory via the
correlation functions, which depend on elementary excitations that
transfer energy between the electrode and the QD. Thus they can
instead be taken into account fully by constructing appropriate
dynamical dielectric functions and single particle Green functions.

The analysis presented so far applies only to the intrinsic couplings
in the system, which lead the sub-systems to mutual
equilibrium. The radiative interactions that drive these sub-systems out of equilibrium
can be considered with the additional light-matter coupling shown in \eqref{eq:Hr D}-\eqref{eq:Hr LD}.
The full development of optical response of these systems can be developed based on
the above formalism. By computing the linear and
non-linear optical response of surface coupled quantum dots in this way, one can develop ways in
which advanced spectroscopic techniques can yield experimental
measurements of the sub-system couplings. However, this is beyond the scope of the present paper.

\section{\label{sec:Applications}Application}

We now apply our model to the system shown schematically in
Fig.~\ref{fig:diagram} in the beginning of the paper. The goal of this
section is to illustrate the calculations of the main quantitites in
the formalism in the order it has been presented above. We first
calculate the microscopic Hamiltonian from a model of InAs QD, and an
\emph{ab-initio} model of a Au[111] electrode. We use the electronic
states computed from these models to obtain couplings between the
subsystems. We then illustrate the use of the results of these
calculations as input to our dynamical theory. In particular, we
calculate the charge and energy transfer rates, and the non-radiative
exciton recombination rates. In these calculations, we show the impact
of FES and image effects on the dynamical charge and energy transfer
couplings.

\subsection{Quantum Dot states}\label{sec:qd-states}

The QD is modeled as a dielectric sphere of InAs with radius, $a=2.0$ nm,
and dielectric constant $\epsilon_{QD}=15.0$. We set the distance
between the QD center and the image plane of the electrode as
$h/2=2.625\;$nm.  Following Chulkov\etal \citep{Chulkov1999330} we set
the location of the image plane plane to be 2\;\AA\;above the top
atomic layer of the electrode.   To model the QD states, we use
the effective mass approximation, in which we include only the lowest
conduction and heavy-hole bands of bulk InAs. Thus we
use~\eqref{eq:qdstates} in the form
\begin{equation}
  \left[-\frac{\hbar^{2}}{2m_0}\nabla\cdot\frac{1}{m_j(\boldsymbol{r})}\nabla
    +V(\boldsymbol{r})+\Sigma(\boldsymbol{r})-\varepsilon_n\right]\varphi_n(\boldsymbol{r})=0,\label{eq:schrod}
\end{equation}
where $m_j$ is piecewise continuous with $m_j=1$ in the space
between the QD and the electrode. Inside the geometric boundary of the
QD, it is equal to the effective mass of the conduction band or a
heavy-hole band for electron ($j=e$), or hole ($j=h$) states
respectively. Inside the QD, we set $m_e=0.03$, and account for the
anisotropy of the InAs heavy hole bands by setting $m_h=0.52$ parallel
to the electrode surface and $m_h=0.33$ perpendicular to it. Note that
this type of modeling implicitly sets the smallest spatial scale to be
the lattice spacing of the QD such that its boundary is a shell of
zero thickness.

The potential $V(\boldsymbol{r})$ is a square well potential of a
spherical QD with $V=0$ outside the QD and $V=-5.0$ eV inside. The
value inside is typical of the workfunctions of semiconductor
materials, and in our discussion of dynamics below, we explore how the
charge transfer rates depend on the alignment of the square well
potential with the Fermi level of the electrode.  The self energy,
$\Sigma$, describes the electrostatic reaction field due to the
polarization of the electrode and the QD surfaces, thus taking into
account the image effects of both surfaces. An exact calculation of
this electrostatic potential is used, the details of which can be found
in our recent publication~\cite{virk-apl}. Since our smallest spatial
scale is larger than the lattice constant, we interpolate the
electrostatic image potential of the QD across this width.  Another
essential assumption of this electrostatics calculation is the uniform
dielectric constant inside the QD, which is well-justified by several
\emph{ab-initio} calculations published in recent
years~\citep{wang-zunger,delerue-dielectric}.

Using cylindrical symmetry, we reduce the problem to two spatial
dimensions, normal and parallel to the electrode surface, and
solve~\eqref{eq:schrod} numerically over a two-dimensional grid of
$(r,z)$ points using the method of finite differences. We use the the
so-called ghost fluid method \cite{Liu2000} to subsume the mass
discontinuity~\cite{virk-apl}.

\begin{figure}[ht]
  \includegraphics[width=1.75in]{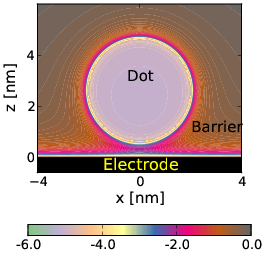}%
  \includegraphics{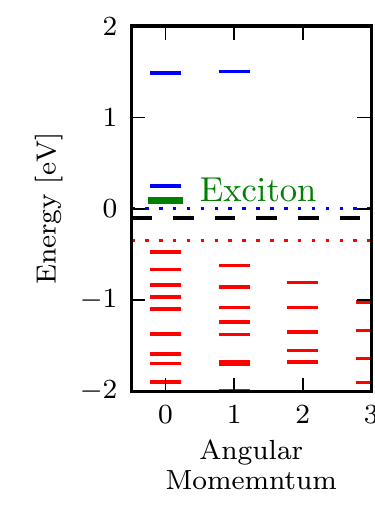}
  \caption{\label{fig:potential-energy}Left: Potential energy of an
    electron in the equatorial plane outside the electrode. Right:
    Energy levels of electron, hole and exciton states. The exciton
    levels is placed such that the hole of the exciton resides in the
    top hole level. The dashed dotted lines refer to conduction (top)
    and valence (bottom) band edges. The thick dashed line is the
    Fermi level of the electrode.}
\end{figure}

Our calculated total potential for the electron in the geometry
described above is shown in Fig.~\ref{fig:potential-energy}. We see
from this figure that the image attraction by both the QD and the
electrode lowers the potential significantly in the narrow
\emph{tunnel junction} around the surface normal passing through the
QD center. This plays an important role in increasing the rate of
electron tunneling. On the other hand tunneling is affected little for
the states whose symmetry places a node in the wavefunction within
this junction.

We calculate the exciton state by employing our exact solution for the
electrostatic polarization to determine the matrix elements of the
effective two-particle interaction potential,
$V(\boldsymbol{r}_e;\boldsymbol{r}_h)$, in which a full account is
taken of the interaction of a hole with the volume density of the
electron, and the surface charge it induces~\cite{virk-apl}. We then
expand $V(\boldsymbol{r}_e;\boldsymbol{r}_h)$ in the basis of product
states lying below an energy cutoff, and increase this cutoff till the
binding energy of the lowest exciton state converges to within 1
meV. The effect of electron-hole correlation on the resulting state is
subtle but important. While the lowest exciton wavefunction is
dominated by the product of the lowest electron and hole states, the
mixing of states introduces corrections that can have noticeable
effects on decay rates, as discussed below.

The right hand side panel in Fig.~ \ref{fig:potential-energy} shows
the energy levels obtained after the exact diagonalization just
described. The exciton level indicated on the figure is shown relative
to the hole level of the QD. One may view this as representing the
electron level correlated to a hole in the top valence level.

% ------------------------------------------------------------------------
% Figure: Potential energy
% ------------------------------------------------------------------------

\subsection{Surface states of the
  electrode}\label{sec:electrode-states}

\begin{figure}[htb]
  \subfloat[]{\includegraphics[width=3in]{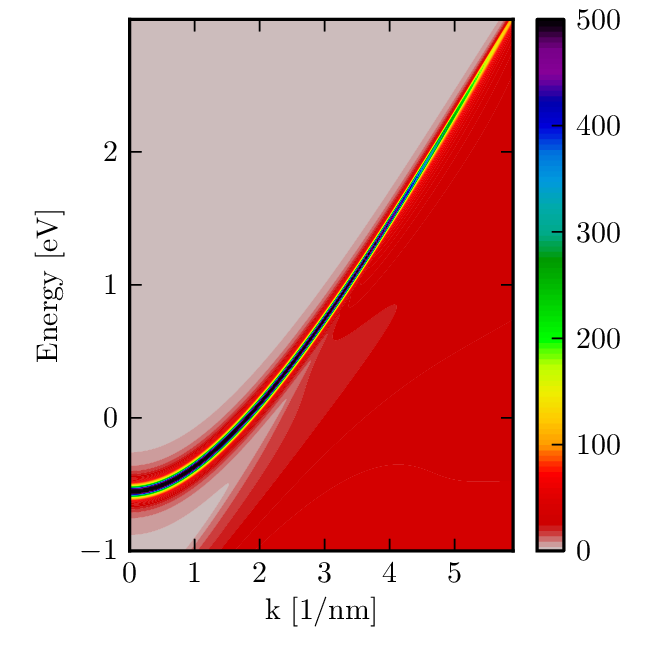}}

  \subfloat[]{\includegraphics{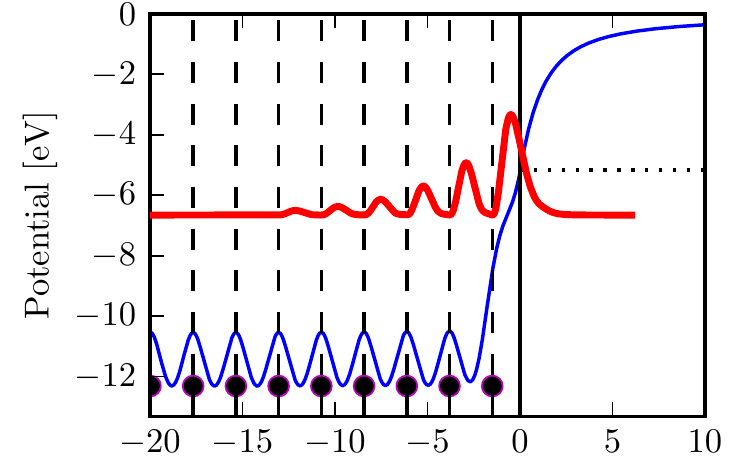}}
  \caption{\label{fig:electrode}(a) Density of states of Au[111]
    surface close to the $\Gamma$ point (a).  (b) The averaged
    potential along the surface normal. The superimposed red curve is
    the Shockley state. The atomic planes are indicated as dashed
    lines, while the image plane is shown as the solid line.}
\end{figure}

To compute the electrode wavefunctions at the surface, we first
perform an \emph{ab-initio} calculation (using the SIESTA
program~\cite{siesta}) for bulk Au, followed by a supercell calculation with 24 mono-layers of Au oriented along the [111] direction separated by the equivalent of 36 layers of vacuum.  The surface unit cell is 1x1, with no reconstruction.
%calculation for a
%unit cell with 12 layers of Au on each end along the [111] direction,
%and separated by vacuum with the distance equal to 36 Au layers. 
The TRANSIESTA~\cite{transiesta} code is then used to generate the Green
functions for the surface Au layers via recursive
decoupling~\cite{williams,dattabook} from the bulk layers. From these
Green functions, we determine the projected density of states (PDOS)
in which the Shockley band~\cite{shockleystate} is identified as shown
in~Fig.\ref{fig:electrode}(a).

From the localized basis orbitals generated by SIESTA, and the
eigenvectors at the poles of the surface Green function, we calculate
the wavefunctions for the surface states along this band. The
eigenvectors are taken along the peak of the PDOS of the surface band.
Fig.\ref{fig:electrode}(b) shows the state at the bottom of the band
superimposed on the atomic layer positions of Au[111]. Since the
dispersion of this state is predominantly parabolic, its exponential
decay remains essentially constant along the band. This is because the
rise in total energy of a parabolic band is cancelled by the rise in
its kinetic energy parallel to the surface, thereby leaving the
wavenumber along the surface normal fixed to its value at the bottom
of the band.

% ------------------------------------------------------------------------
% Figure: Electrode wavefunction and dos and potential
% ------------------------------------------------------------------------
% ------------------------------------------------------------------------

\subsection{\label{sec:couplings}Couplings}

From these wavefunctions, and those of the QD, we determine the
hybridization matrix elements $H_T$ from a direct application of
\pref{eq:hybridization}. To calculate the Coulomb interaction matrix
elements, we employ the procedure described by Eqs.(\ref{eq:V and
  W}-\ref{eq:eff coulomb mat elem}). The main input to this procedure
is the effective potential in which corrections due to the surface
dielectric function of the electrode are applied. In order to do that,
we follow Pitarke \etal~\cite{silkin-asp1} and divide the surface into two linear response systems. The first is
the semi-infinite bulk, and the second is the two-dimensional electron
gas formed within the Shockley band. We let $\chi_b$ and $\chi_s$ be
the susceptibilities of the two systems.

We define $\varepsilon^{-1}_b$ to be the dielectric function
of the bulk, and follow Newns' work~\citep{Newns1970} in calculating
it. Newns' calculation is based on the random phase approximation
(RPA) within a jellium model of the electron gas, and the potential we
calculate thus represents screening by the semi-infinite bulk
only. Due to the large plasmon frequency of the bulk, we
replace $\varepsilon_b$ by its static limit. We include the effects of the surface states 
based on the work by Pitarke \emph{et al}~\cite{silkin-asp1} and Silkin \emph{et al}~\cite{silkin-asp}. In their simplified model,
the surface states comprise a 2D electron gas lying in a plane a small distance $|z_p|$ above the
interface of the semi-infinite jellium representing the bulk substrate~\cite{silkin-asp1}. This separation ensures charge neutrality in the interior of the jellium medium,~\cite{Newns1970} and
in the present case we place the 2D plane at $z=0$, and set $z_p=-3\pi/8k_F\approx2$\AA~\cite{silkin-asp}. 

Given $V_0(q,z)$ as the planar Fourier transform of a bare
 potential, the potential screened by the jellium plane is given by~\cite{Newns1970},
\begin{eqnarray}
  W(q,z)&=&\left[V_0(q,z) - V_0(q,z)
  \frac{1-\varepsilon^{-1}_b(q)}{1+\varepsilon^{-1}_b(q)}\right]\Theta(z-z_p)\nonumber\\
&+&V_0(q,z_p)\frac{2\varepsilon^{-1}(q,z-z_p)}{\varepsilon_b^{-1}(q)+1}\Theta(z_p-z)\label{eq:Wqz},
\end{eqnarray}
where  $\varepsilon^{-1}(q,z-z_p)$ is obtained by multiplying the 
pre-factor in formula (62) of Newns' paper~\cite{Newns1970} by $-q$,
and $z_p$ is the coordinate of the receded plane.  When $V_0(q,z)$ corresponds to 
the potential of the charged QD, $W(q,z)$ enters the Hamiltonian matrix via \eqref{eq:Vhat nn' def}
and accounts for the substrate response. The response of
the surface gas appears explicitly in the dynamical rate expressions, and 
this asymmetry in our treatment arises because the dynamics depends only on
frequencies that are much smaller than the bulk plasmon
frequency. This allows us to use only the static response of the substrate, but since the surface acoustic plasmon branch extends to zero frequency, the full dynamical response of the 2D surface gas must be included. Furthermore, the coupling of the
QD potential is also much larger to the electron hole excitations in
the surface gas than it is in the bulk, since the potential is screened
fully within a few atomic layers below the surface. The dominant
coupling to surface states is also verified by the \emph{ab-initio}
calculation of electrode states as described above.

Thus, having made the choice to let $W(q,z)$ represent the entire bulk
response, we now turn to the susceptibility  $\chi_s$  of the 2D surface electron gas. In the present implementation approximating the general picture developed in Sec.~\ref{sec:et}, the energy transfer rates originate from the response of the 2D
electron gas. We calculate the non-interacting $\chi^0_s$ using Newns' approach~\cite{Newns1970}, and use RPA to construct the interacting response $\chi_s$ as in~\eqref{eq:chi-2d}. 
In the RPA calculation, the electron-electron interaction within the 2D gas is screened
by the substrate, but since the screening is only partial, collective excitations in the 2D electron gas can exist in the form of surface acoustic plasmons~\cite{silkin-asp1}. We find the effective screened
interaction using Newns' work, 
\begin{equation}
v_{\mbox{\scriptsize eff}}(q)=v(q)-v(q)e^{-2q|z_p|}\frac{1-\varepsilon_b^{-1}(q)}{1+\varepsilon_b^{-1}(q)}.
\end{equation}
Here the first term is the Fourier transform of the bare interaction within the 2D plane representing the  surface electrons, and the second term is the interaction with the image charge in the substrate located in the plane a distance $2|z_p|$ below the surface electron plane. Substituting $v_{\mbox{\scriptsize eff}}(q)$ into the RPA summation for the 2D response, we obtain the screened susceptibility, $\chi_s(q,\omega)$, as
\[
\chi_s(q,\omega) = \frac{\chi_s^0(q,\omega)}{1-v_{\mbox{{\scriptsize eff}}}(q)\chi_s^0(q,\omega)}.
\]

\begin{figure}[ht]
  \includegraphics[width=3.1in]{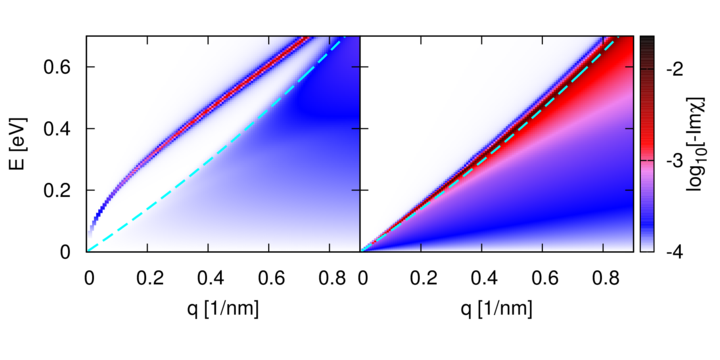}
  \caption{\label{fig:chi-2d} $\log_{10}\left[-\Im\chi\right]$ in the
    absence of the semi-infinite bulk (left), and in the presence of
    static screening by the bulk (right). Note the qualitative change
    in the dispersion of the acoustic plasmon branch, which in the
    presence of the bulk lies close to the incoherent electron-hole
    continuum, the upper boundary of which is marked by the thin
    dashed line. Thus the spectral weight of the plasmon contributes
    to the pair continuum.}
\end{figure}

In the left panel of Fig.~\ref{fig:chi-2d}, we have plotted a
two-dimensional color map of the function $\Im\chi_s(q,\omega)$ in
which the bulk screening is neglected. We have verified that the
acoustic plasmon branch in the absence of bulk substrate follows the
expected trend, $\sqrt{e^2 n q/(2M\varepsilon_0)}$, which holds at low
energies for a 2D gas of quasi-electrons of mass $M$ and density
$n$. By including the screening due to the bulk substrate, we obtain
the plot shown in the right panel. The plasmon branch in the result is
significantly different as it has now acquired a linear profile
that straddles along the incoherent pair continuum. Thus the bulk
substrate shifts the spectral weight of the plasmon branch to a lower
frequency and spreads it into the pair continuum.

% -------------------------------------------------------------------------------
% Chi
% -------------------------------------------------------------------------------
% -------------------------------------------------------------------------------

Let us briefly comment on when this shift can become
important. Consider the processes in which a state $n$ relaxes to a
lower energy state $m$ via the Coulomb interaction, in which the
coupling potential is formed by the transition density
$\varphi_m^*(\boldsymbol{r})\varphi_n(\boldsymbol{r})$. From
\eqref{eq:Xkk;} and energy conservation, we expect that the region in
the $(q,E)$ plane that couples to this process must lie within the
Fourier-Bessel transform of this potential along the $q$ axis, and in
the range of energy differences $E_n-E_m$ along the $E$
axis. Wavefunctions inside a QD of radius 2 nm would generally yield
the peak of the Fourier-Bessel transform to be around $q \approx
0.5\mbox{nm}^{-1}$. The plot in Fig.~\ref{fig:chi-2d} then implies
that $E_n-E_m$ must be 0.4-0.5 eV apart for the states $n$ and $m$ to
experience qualitative change in their coupled dynamics. As shown by
our calculations below, this energy difference is much higher than the
inverse rates implied by the energy transfer matrix $\mathbf{C}(t)$
when the bulk substrate included. On the other hand, in the absence of
the substrate, it is possible to approach the regime where these
changes may significantly modify the time-dependence of energy
transport between the QD and the electrode. We now turn to the effects
of charge and energy transfer on the exciton and hole populations with
the initial state in which the QD is prepared in the exciton state by
photo-excitation.

\subsection{Exciton dissociation and Fermi edge singularity}

In this subsection, we discuss the charge transfer of the electron to the
electrode, which leaves a positively charged QD containing one
hole. Since the charge state of the QD changes from neutral to
positive, we expect to see the effects of the FES in the tunneling
rate.  With reference to the full rate expression in~\eqref{eq:Massmatrix}, we focus on $M_{hh;xx}$ so $D_{hh}$ and $D_{xx}$ enter, both of which are simply unity, and the rate is simply,
\[b_{x}(t)\equiv\sum_h\dot{B}_{hh;xx}(t).\]

A brief description of the calculation strategy is as follows. From
~\eqref{eq:B ac dc}, we obtain\begin{equation} \dot{B}_{hh;xx}(t)
  =-\frac{2}{\hbar^2}\int_{-\infty}^{+\infty}\frac{d\omega}{2\pi}\gamma_{xh;hx}(\omega)S(\omega-\omega_{xh},t).\label{eq:Bhhxx}
\end{equation}
The controlling energy scale is the excess energy of the tunneling electron relative to the electrode Fermi energy, expressed by $\omega_{xh}$. We evaluate the above expression using
\eqref{eq:Gamma-greater-matrix-main-text}, \eqref{eq:r
  function}-\eqref{eq:sigma2}. Since the dipole field of an exciton is
weak and it does not affect the edge singularity by the Friedel sum
rule~\cite{mahanbook}, we set $[K_x,H_L]\approx0$, which also sets
$\mathbf{\bar{N}}_{xx}(t,0)\approx\mathbf{\bar{N}}$. The definition of
$\gamma_{xh;hx}(\omega,t';t)$ and this approximation together imply
that it is independent of the time arguments $t',t$, which we reflect
in the equation above by omitting the time arguments (see also the
discussion at the end of Sec.~\ref{sec:ct}). Taking only the surface
states as the final states in the charge transfer, we obtain
\begin{eqnarray}
  &&\gamma_{xh;hx}(\omega)=\label{eq:gamma_xhhx}%
  -i\sum_{\boldsymbol{k}\boldsymbol{k}'}T^{xh}_{s\boldk}T^{hx}_{s\boldk'}\times\\
  && \int_{0}^{+\infty}d\tau e^{-i\omega\tau}r_{xh;hx}(\tau)\times\nonumber\\
  &&\left\langle s\boldsymbol{k}\right|\mathbf{\bar{N}}
  \left[\mathbf{I}+\mathbf{\Sigma}^{(1)xx}_{xh;hx}(\tau)\mathbf{\bar{N}}\right]^{-1}
  e^{-iK_{x}\tau}\left|s\boldsymbol{k}'\right\rangle 
  ,\nonumber
\end{eqnarray}
where $s$ denotes the surface states, and
\begin{eqnarray*}
  r_{xh;hx}(\tau)&=&\frac{1}%
  {Z}\left\langle e^{iK_{h}\tau}e^{-iK_{x}\tau}e^{-\beta H_{L}}\right\rangle,\\
  \mathbf{\Sigma}^{(1)xx}_{xh;hx}(\tau)&=&e^{-iK_{x}\tau}e^{iK_{h}\tau}-\mathbf{I}.
\end{eqnarray*}
Note that \eqref{eq:gamma_xhhx} is different from the Fermi golden rule result in that it includes the effects of sudden switching of potential to all orders in perturbation theory, and therefore includes the effects of FES exactly. This results in dressing of all operators by the Coulomb potential of the QD. The above result is a perturbation expansion with the dressed hybridization as the small parameter. Eliminating the FES effects is equivalent to setting  $r_{xh;hx}(\tau)=1$, and $ \mathbf{\Sigma}^{(1)xx}_{xh;hx}(\tau)=0$. Substituting these two values into
\eqref{eq:gamma_xhhx}, and performing the integral over $\tau$, the Fermi golden rule expression without the FES effects emerges in the imaginary part of $\gamma_{xh;hx}$ representing the particle loss rate from the QD. 

To compute the integrand in~\eqref{eq:gamma_xhhx}, we represent $K_a$
in the plane-wave basis over the two-dimensional quasi-momentum within
the surface band. The circular symmetry of this band reduces the
problem significantly as the plane wave basis decomposes into a
product of angular momentum eigenfunctions, $e^{il\theta}$, and Bessel
functions $J_l(kr)$ for the total momentum $k$. In this basis, $K_a$
can be represented as a block-diagonal matrix with each block
corresponding to an angular momentum eigenvalue $l$, and given by
\begin{eqnarray*}
  \hbar K^{l}_{a;kk'}&=&\frac{\hbar^2k^2}{2m_L}\delta_{kk'}\\
  &+&\int_0^R dr\;r \hat{J}_{l}(kr)\hat{J}_{l}(k'r)\int dz |\psi_{L;s}(z)|^2 V_{aa}(r,z),
\end{eqnarray*}
where $m_L$ is the effective mass of the surface band, $V_{aa}(r,z)$
is the potential due to the QD in state $a$ (see \eqref{eq:Ka}),
$|\psi_{L;s}(z)|^2$ is the planar averaged probability density of the
surface state with negligible $k$-dependence (see
Fig. \ref{fig:electrode}, and Sec. \ref{sec:electrode-states}),
$\hat{J}_l(kr)$ are Bessel functions normalized to unity over $[0,R]$
such that $\hat{J}_l(kR)=0$, and $R$ is a cutoff radius set by
discretization of $k$ and is much larger than the screening length
within the two-dimensional surface electron gas. In addition,
$V_{aa}(r,z)$ is constructed in accordance with the discussion in the
previous section where the potential outside and inside the electrode
is given by \eqref{eq:Wqz} respectively, and
followed by Fourier transform back to real space.

We compute the resulting matrices $K^{l}_{a;kk'}$ over a discrete set
of $k$, and diagonalize them to compute their exponentials in
\eqref{eq:gamma_xhhx}, numerically exactly. Multiplying by
$T^{xh}_{sk}T^{hx}_{sk'}$ and summing over all $k,k'$ in the discrete set,
we numerically compute the integrand for a discrete set of $\tau$. By
experimenting with the discretization of $k$ and the total number of
angular momenta $l$, we obtained well-converged results by using $800$
$k$ points over the surface band shown in Fig. \ref{fig:electrode},
and setting maximum $l$ to 24.

The resulting integrand as a function of $\tau$ generally has a slowly
decaying tail that prevents a direct application of Fourier transform
to obtain $\gamma(\omega)$.  We follow the well-established methods to
handle this numerical
technicality~\cite{pawel-asym-qw-fes,Hawrylak,Mahan1980}, and then
Fourier transform the resulting expression to obtain
$\gamma(\omega)$. The remaining procedure to obtain $\dot{B}_{hh;xx}$
is straightforward.

\begin{figure}[hb]
  \includegraphics[width=3.3in]{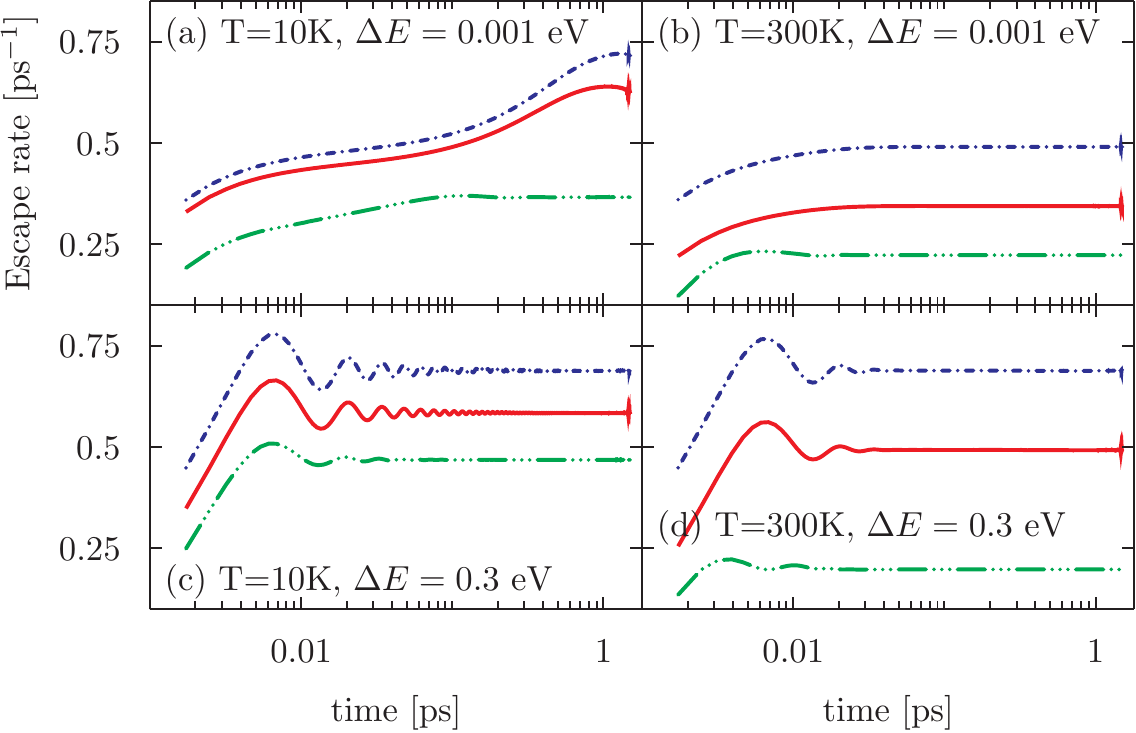}
  \caption{\label{fig:fes-tunneling}The escape rate of the electron in
    the presence of a hole with (solid red line) and without
    (dashed-dotted line) Coulomb coupling to the electrode.  The
    triple-dotted green line corresponds to the rate in the absence of
    the bulk substrate, but with interaction with the 2D surface gas
    retained. Note the suppression of tunneling due to the FES. Only
    the lowest electron level and all the hole levels correlated to it
    are included. The effective junction width is 6.25 \AA. Plots are
    shown for two different temperatures for the Fermi sea of surface
    states, and $\Delta E$ is the energy difference between the Fermi
    level and the lowest electron level. The oscillations at the
    rightmost end of each plot are due to the onset of aliasing as the
    time increases beyond the inverse of the largest energy
    differences imposed by the 2D momentum discretization of plane
    wave surface states. }
\end{figure}

In Fig.~\ref{fig:fes-tunneling}(a-d) we show $b_{x}(t)$ for two
different alignments between the energy level of the lowest electron
state and the Fermi level of the electrode. Results are also shown for
two different temperatures, and plots in each figure correspond to the
absence and presence of the Coulomb interaction between the QD and the
electrode. We see from the figure that the tunneling rate in the
presence of the interaction is always smaller than in the absence of
this coupling. This effect is the result of the FES, which in turn is
mainly dominated by the AOC function rather than the Mahan exciton
contribution. We verified this by comparing these results with
calculations in which the AOC is excluded. Decrease in the rate also
results from removal of the substrate because it eliminates screening
of the QD potential.

We observe in Fig.~\ref{fig:fes-tunneling} that the suppression in the
tunneling rate increases with temperature. This trend follows from the
Anderson-Yuval mapping~\citep{Ohtaka1990}, from which we expect the
AOC function to exponentially decaying for time $t>\hbar/k_BT$ and at
temperatures much smaller than the Fermi temperature of the electron
gas. Since the present calculation is precisely within this regime,
the suppression of rate increases with temperature. At temperatures
exceeding the Fermi temperature, the orthogonality catastrophe would
itself become exponentially suppressed. This regime could be reached
with an electrode made of lightly doped semiconductor in which the
Fermi energy can be an order of magnitude smaller. This is also the
kind of system studied in an experiment by Kleemans
\emph{et. al.}~\cite{Kleemans2010}.

In addition, note from Fig.~\ref{fig:fes-tunneling}(b-c) that the time
for $b_{x}(t)$ to reach its asymptotic value is much smaller than the
inverse rate implied by its magnitude. Thus the tunneling process may
be modeled accurately as Markovian with a constant rate given by
$b_{x}(t)$ for $t>1\mbox{ps}$. At temperature of 10K and below, as
shown in Fig.~\ref{fig:fes-tunneling}(a), non-Markov behavior may be
expected. The longer time for the approach to asymptotic value in this
regime is the result of sharp increase in the density of vacant
states.

We conclude that for a semi-infinite metallic electrode, the edge
singularity effect is small and quantitative, which is also in
agreement with recent \emph{ab-initio} work on the
topic~\citep{recent-Despoja2008}. On the other hand, since the effect
is proportional to the ratio of the scattering potential to the
bandwidth of the Fermi sea, a lightly doped semiconductor would be a
better system for its observation.

Let us briefly comment on how this crossover from the Markov to
non-Markov regime may be accessible for observation. Since it occurs
within the sub-picosecond timescale, the photoluminescence in this
regime is completely quenched. The crossover is therefore relevant
mainly to non-linear optical response of the system. A possible route
to accessing this in nonlinear optics is the reduction in bleaching of
exciton absorption line due to dissociation.  This bleaching can be
studied as a function of delay between pump pulses tuned to the
absorption frequency. The rise in absorption as a function of the
delay would then change from an exponential to a non-exponential
function as the temperature is lowered across the crossover, which
occurs at approximately 10K in the present model.
% ------------------------------------------------------------------------
% Figure: FES and tunneling
% ------------------------------------------------------------------------
% ------------------------------------------------------------------------

\subsection{Non-radiative exciton recombination}

In addition to dissociation, photoluminescence may also be quenched by
non-radiative recombination (NRR) in which the energy is transferred
to the electron-hole pair excitations in the electrode, rather than
being converted to a photon. In the present geometry, this process is
also \emph{faster} than the typical radiative recombination of
excitons in InAs, as we now discuss.

We first quantify the NRR of an exciton and its sensitivity to various
physical properties of the system. With reference to the full rate expression in~\eqref{eq:Massmatrix}, we focus on $M_{gg:xx}$ so $D_{gg}$ and $D_{xx}$ enter, both of which are simply unity.  Thus our theory describes this process
by the matrix element $ \dot{C}_{gg;xx}(t)$, which depends on
$\mathbf{d}_{cv}\cdot\mathbf{E}(\boldsymbol{r})$, where
$\mathbf{d}_{cv}$ is the conduction-valence band dipole matrix element
at the band edge. We set its value to 2.15 \AA~\cite{dcv-zunger}. The
vector $\mathbf{E}$ is the total electric field of the transition
density of the exciton (see Appendix~\ref{sec:matrix-elements}),
\[\mathbf{E}(\boldsymbol{r})=-\nabla \int
d\boldsymbol{s}V(\boldsymbol{r};\boldsymbol{s})
\varphi_n(\boldsymbol{s})\varphi_{n'}(\boldsymbol{s}),\]
where the potential $V$ also includes the electronic surface
polarization response of the QD.

%
% a c; a' c' g g; x x
\begin{equation}
  \dot{C}_{gg;xx}(t)=
  \frac{1}{\hbar^2}\sum_{\boldsymbol{q}}|V_{xg}(q)|^2 \int_{-\infty}^{+\infty}
  d\omega X(q,\omega)\dot{S}(\omega-\omega_{gx},t),
\end{equation}
where $X(\omega)$ is the susceptibility the computation of which we
described above in Sec.~\ref{sec:couplings}, and
\begin{equation}
  V_{xg}(q)=\mathbf{d}_{cv}\cdot%
  \int d^3\boldsymbol{r} e^{-i\boldsymbol{q}\cdot\boldsymbol{r}}%
  \mathbf{E}(\boldsymbol{r}).
\end{equation} 
We compute the integrand over a discrete set of $(q,\omega)$ over the
range shown in Figs.~\ref{fig:chi-2d}, and sum over this range to
compute $\dot{C}_{gg;xx}(t)$ at a discrete set of points $t$. The
range in $q$ is sufficient due to the fact that the transition density
of the exciton has a real space spread of approximately 4 nm. The
range in $\omega$ is justified from the exponentially suppressed
$X(q,\omega)$ above the acoustic plasmon line in
Fig.~\ref{fig:chi-2d}.

\begin{figure}[hb]
  \includegraphics[scale=1]{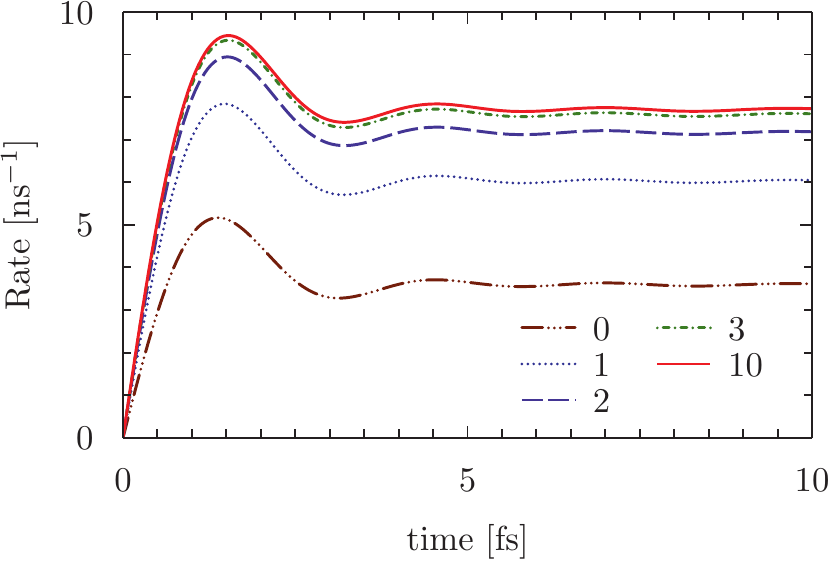}
  \caption{\label{fig:forster-vs-multipole}Forster rate calculated for
    increasing order $l$ (values indicated in legend) of the multipole
    expansion of the transition density of the exciton, including the
    electrostatic reaction field of the quantum dot. The oscillations
    result from the acoustic plasmon branch in the renormalized
    susceptibility of the surface electron gas.}
\end{figure}

In Fig.~\ref{fig:forster-vs-multipole}, we plot $\dot{C}_{gg;xx}(t)$
for various values of $l$ at which the multipole expansion of
$\mathbf{E}$ is truncated.  Each $l>0$ plot is therefore a correction
to the often used point dipole model, in which only the $l=0$ term of
the transition densities is employed. We see from the figure that the
converged solution is almost twice as large as $l=0$ case, and that
convergence occurs only beyond $l=3$.

As in the case of charge transfer, we may also model NRR as a
Markov process with a constant rate equal to the asymptotic value of
$\dot{C}_{gg;xx}(t)$. This is verified by plots of
$\dot{C}_{gg;xx}(t)$ in Fig.~\ref{fig:forster-vs-multipole}, which
show that after the initial appearance of an exciton state, the Fermi
sea responds at the ultrafast timescale of 10 fs, within which
$\dot{C}_{gg;xx}(t)$ settles to a constant value. This is much faster
than the rate implied by the asymptotic value.

\begin{figure}
  \includegraphics[width=3in]{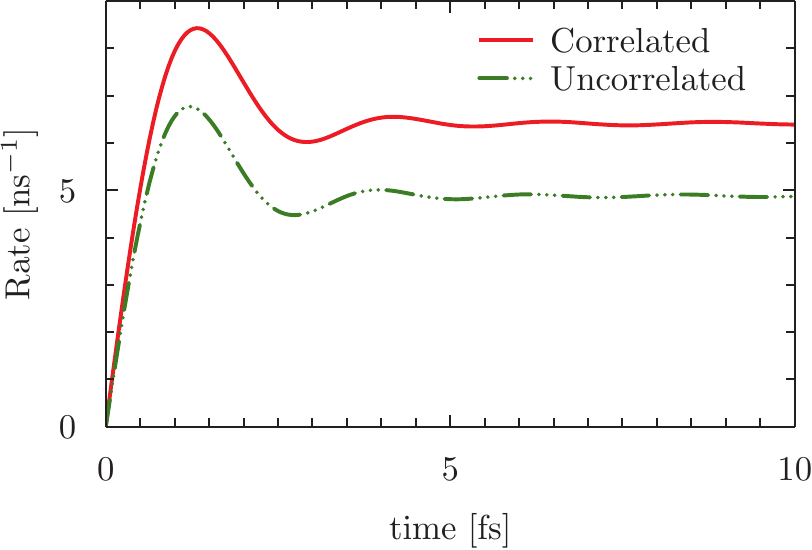}
  \caption{\label{fig:uncorr-corr} The Forster decay rate calculated
    for a product state and the fully correlated state for $l_z=0$. The slight
increase in the Forster rate for the correlated state is the
result of shift in the transition density as shown in
Fig.~\ref{fig:exciton-cd}.}
\end{figure}

\begin{figure}
%  \subfloat[]{\includegraphics[width=1.75in]{trans-plot-ex.pdf}}
%  % \begin{minipage}[t]{2in}
%  \subfloat[]{\includegraphics[width=1.75in]{trans-plot-eh.pdf}}
%  % \end{minipage}
\includegraphics[width=4in]{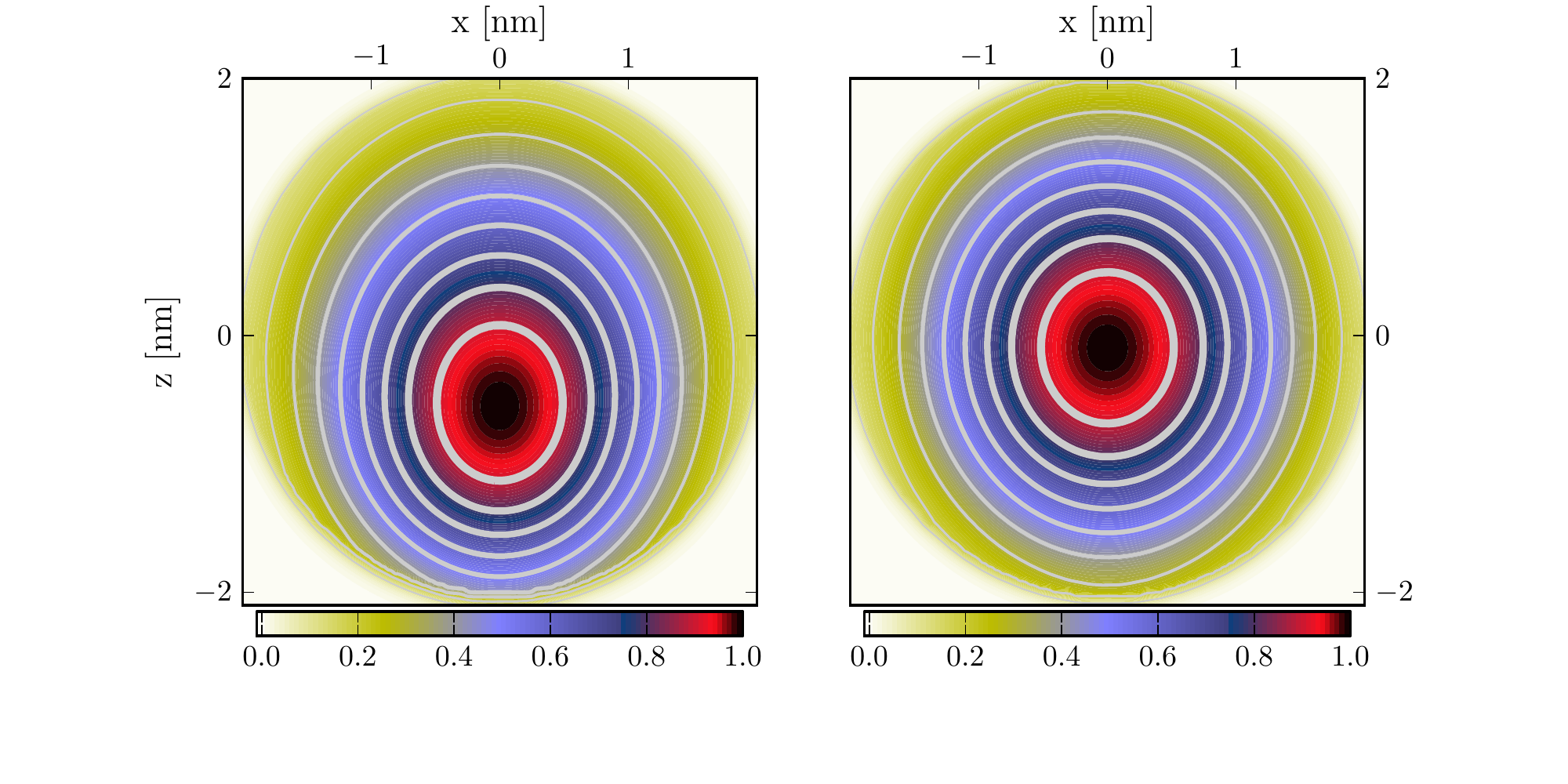}
  \caption{\label{fig:exciton-cd}Contours of transition density of
    the fully correlated exciton state (left) compared to the charge
    density of the product of lowest energy electron and hole
    wavefunctions (right). The thickness gray contour lines represents the contour
    level, ranging from 1 (maximum) to 0 (minimum).}
\end{figure}

We remark that while NRR is almost 100 times smaller than electron
tunneling rate in the figures shown, we expect from the exponential
suppression of tunneling with distance, as opposed to a power law
dependence of the NRR, that the non-radiative decay of an excited QD
would cross over from a dissociative to a non-dissociative channel as
its distance from the electrode is increased. In the present model,
the tunneling rate is reduced by a factor of 10 for every extra
1.1\AA\;of separation so that both regimes may be accessed for systems
with only sub-nanometer differences in the tunnel junction. Thus the
efficiency of current extraction in a photovoltaic device may be
heavily impacted by NRR, and a point dipole model with uncorrelated
electron-hole density suffices to estimate this impact within a factor of 2 as
can be seen from Fig.~\ref{fig:uncorr-corr}. The correlation between the
electron and the hole slightly increases the decay rate by 
shifting the transition density towards the metallic surface (see Fig.~\ref{fig:exciton-cd})

% ------------------------------------------------------------------------
% Figure: Forster decay vs. multipoles
% ------------------------------------------------------------------------
% ------------------------------------------------------------------------
% Figure: Forster decay vs. correlation
% ------------------------------------------------------------------------
% ------------------------------------------------------------------------
% Figure: Exciton charge density
% ------------------------------------------------------------------------
% ------------------------------------------------------------------------

\subsection{Hole cooling and tunneling}

Let us now turn to hole tunneling as well as cooling via Coulomb
driven energy transfer to pair excitations. The crossover between
charge and energy transfer as the dominant decay channel is also
relevant here. To show this we calculate the matrix elements
$\dot{B}_{gg;hh}(t)$ for tunneling, and $C_{hh;h'h'}(t)$ for
Coulomb-driven cooling. There is a symmetry restriction in the latter
so that transitions conserve the angular momentum along the surface
normal of the electrode.  Furthermore, due to our approximation
$[K_{x},H_L]\approx 0$ in Sec.~\ref{sec:ct}, the computation of
$\dot{B}_{gg;hh}(t)$ is obtained by replacing $x$ by $g$ in
$\dot{B}_{xx;hh}(t)$. Similarly, the computation of $C_{hh;h'h'}(t)$
closely parallels that of $C_{gg;xx}(t)$ described in
Sec.~\ref{sec:et}. Therefore, we have omitted a discussion of
computational details in this sub-section.

\begin{figure}
  \includegraphics{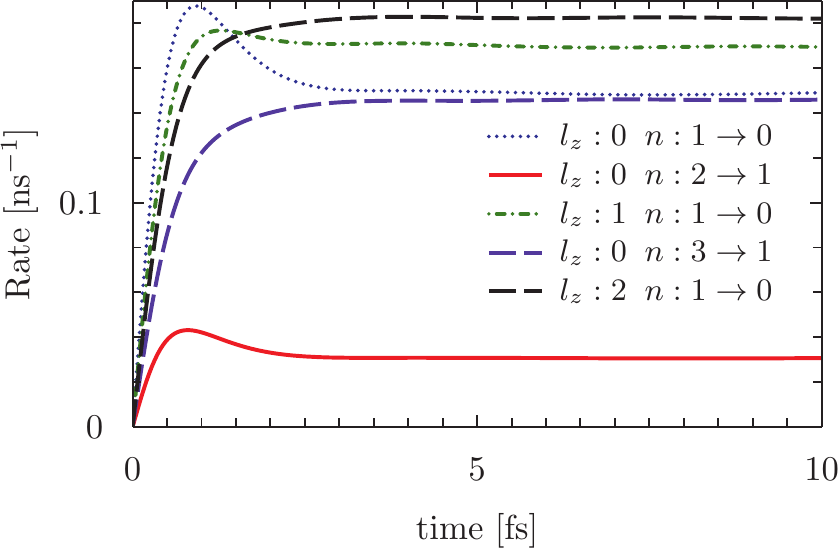}
  \caption{\label{fig:etf-hole} Cooling rates of hot holes in which
    energy is transferred to the electrode via the Coulomb coupling to
    its dynamical dielectric function. By symmetry the transfer
    conserves $l_z$, as indicated along each curve where the change in
    the principal quantum number $n$ is also indicated. The static
    screening by bulk substrate is included both in the response of
    the surface electron gas, and as an image correction to the field
    of the quantum dot.}
\end{figure}

\begin{figure}
  \includegraphics{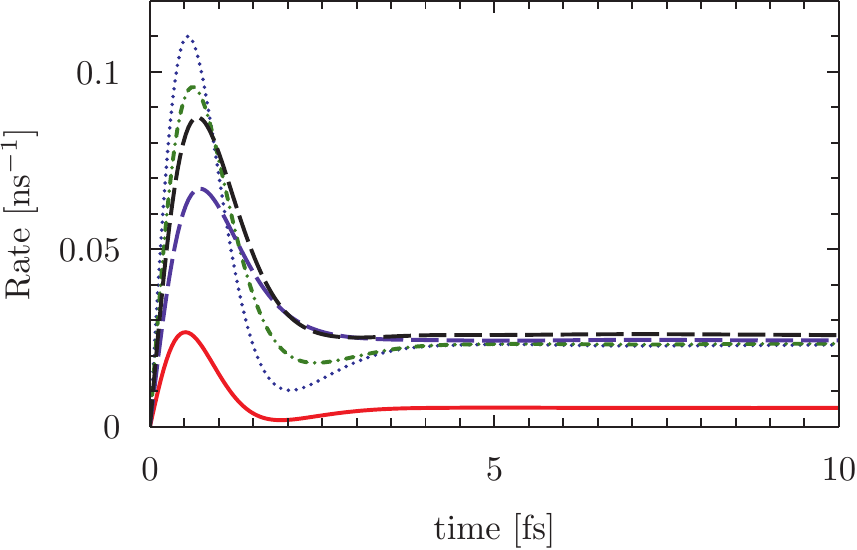}
  \caption{\label{fig:etf-hole-mixed} Cooling rates for the same
    states as in Fig.~\ref{fig:etf-hole} but without the substrate
    induced screening of the electron-electron interaction in the
    surface electron gas. The reduction in the cooling rate is almost
    by a factor of 8, and is mainly due to the shifting of the plasmon
    branch up in energy or farther away from the part of loss function
    that couples to the transition densities. Line styles are as in
    Fig.~\ref{fig:etf-hole}. }
\end{figure}

We plot the results of our calculations for energy transfer in the
presence of a semi-infinite bulk electrode in Fig.~\ref{fig:etf-hole}.
To understand the dynamical effects of bulk screening, we repeated
this calculation by keeping the static image potential of the QD, but
removed the screening of electron-electron interaction within the 2D
gas. Thus we capture only the dynamical effects of the changes in the
plasmon branch, and plot the resulting rates in
Fig.~\ref{fig:etf-hole-mixed} (the underlying loss function for this
calculation is shown in left panel of Fig.~\ref{fig:chi-2d}). The
reduction in the rates compared to the screened case can be understood
by observing that only the points in proximity of $q\approx 0.5/nm$
and $E=0.2$ eV couple to the QD potential. This region, which lies
entirely inside the incoherent pair excitation regime is farther from
the plasmon branch in the absence than in the presence of the
substrate screening. The resulting loss function is therefore smaller
in the unscreened than in screened case. Thus the \emph{dynamical}
effect of screening is to enhance the rate.

\begin{figure}
  \includegraphics{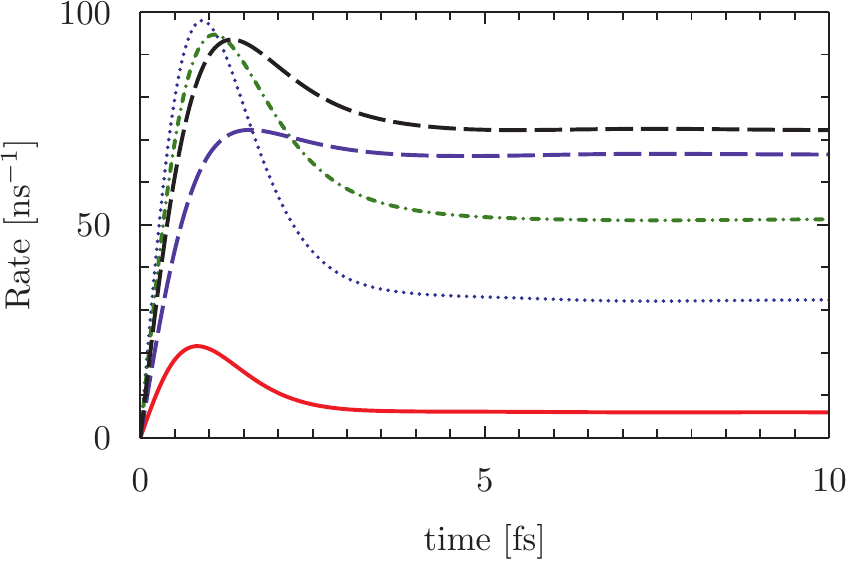}
  \caption{\label{fig:etf-hole-no-screen} Cooling rates for the same
    states as in Fig.~\ref{fig:etf-hole} but after completely
    excluding the effects of the bulk substrate. The almost 200-fold
    increase in the rate is due to the removal of image cancellation
    due to the static substrate response. A low dielectric substrate
    supporting a conducting thin film, such as metal-on-oxide, would
    correspond to this case. Line styles are as in
    Fig.~\ref{fig:etf-hole}.}
\end{figure}

However, we emphasize that the largest quantitative effect of bulk
screening is to \emph{suppress} this rate via the instantaneous image
potential within the plane of the surface electron gas. In
Fig.~\ref{fig:etf-hole-no-screen} we plot the rates when the image
potential is completely removed, which corresponds to the lack of any
substrate. Due to the fact that the coupling is proportional to the
square of the QD potential, we see a substantial increase, by
approximately a factor of 250, in the cooling rate.

\begin{figure}
  \includegraphics{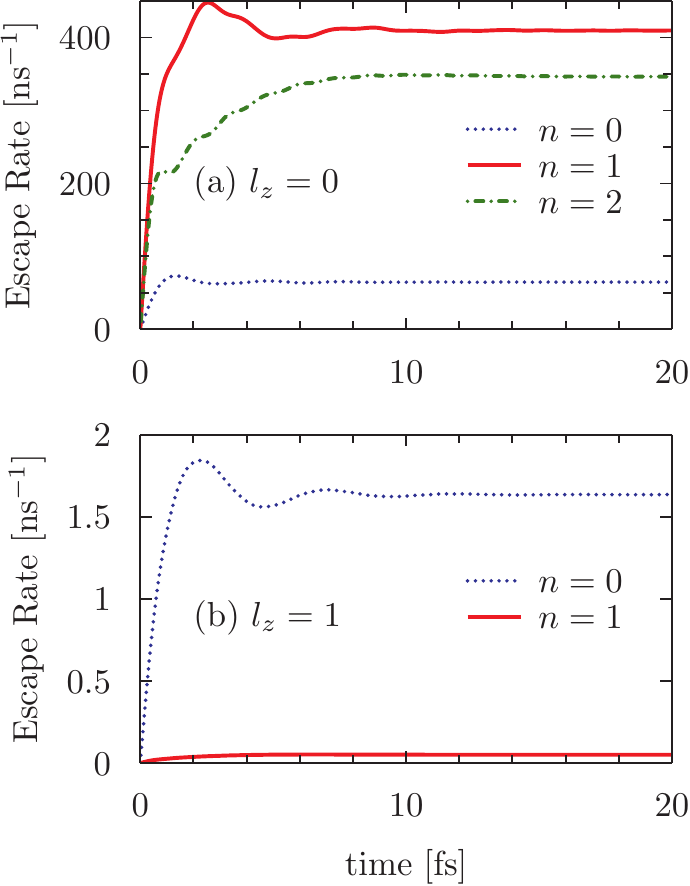}
  \caption{\label{fig:hole-tunnel} Rate of tunneling into a hole level
    of the QD with a single hole. Plots are shown for lowest energy
    states (principal quantum number $n$) at two different values of
    $l_z$.  The charge distribution of the $l_z=1$ hole state is
    minimal near the tunnel junction, and therefore, it has a much
    smaller escape rate than the $l_z=0$ state.}
\end{figure}

We now consider these results in light of the tunneling rates of
holes, plotted in Fig.~\ref{fig:hole-tunnel}. In the presence of the
substrate, tunneling is much faster for $s$ ($l=0$) hole states, but
approaches the energy transfer rates for $p$ ($l=1$) states. The much
smaller tunneling rate for the $p$ state is expected from its spatial
profile which contains a node in the tunnel junction. In the absence of the bulk, the
energy and charge transfer rates for all states lie within factors of
two to four and the two processes can thus compete. We expect this to
hold true even when we account for the effect of substrate removal on
tunneling.

To see why, we note that the 2D gas also generates an image potential
away from its own plane, which would have the same effect on the
junction potential as the image potential of a metallic substrate. The
formation of this image charge may be treated as instantaneous for the
purpose of tunneling, because as we have seen in the results above,
the plasmon oscillations are in the regime of 10-20 fs, and the
dynamical rates settle to constant values beyond this timescale.  Thus
the required charge re-arrangement in the 2D gas is still much faster
than the tunneling rates. Also, as shown in
Fig~\ref{fig:fes-tunneling}, the substrate removal \emph{reduces}
tunneling rates due to the FES, but this is still within a factor of
2. The reduced dimensionality of the electrode thus only lowers the
screening of the external field within the plane of the 2D gas, and
has only a small effect on the charge transfer.  We conclude that
tuning the substrate allows energy transfer to be controlled almost
independently of charge transfer.

% ------------------------------------------------------------------------
% Figure: Hole Cooling
% ------------------------------------------------------------------------

% ------------------------------------------------------------------------
% Figure: Hole tunneling
% ------------------------------------------------------------------------

\section{Discussion}\label{sec:discussion}

In the previous sections, we have derived a microscopic theory and an
effective model of dynamics to obtain a self-contained framework for
studying surface coupled quantum dots. In this section, we discuss how
the model can be applied to analyze a realistic experimental scenario,
and also to analyze the experimental data. We also discuss how various
physical processes, neglected here for brevity, can be described
within our framework without any further modifications of the theory.

As an experiment would generally involve a collection of quantum dots
spread over a region possibly much larger than their size, the charge
and energy transfer rates must be averaged across the collection, and
the spatial dependence of the electrode density of states. Thus both
types of rates must be computed for a distribution of tunnel junction
widths and changes in density of surface states across the collection
of QDs. Similarly, while the optical wavelength is much larger than
the size of the QD, the total optical response must also account for
spatial changes in the phase of the waves across the region containing
the QDs. This could be taken into account by including the factor
$e^{i\boldsymbol{k}\cdot\boldsymbol{r}_j}$ in the optical field, where
$\boldsymbol{k}$ is the wave-vector of the field and
$\boldsymbol{r}_j$ points to the center of of the $j^{th}$ dot.

Quantum dot arrays may be designed to exploit the vast range of
timescale of tunneling and the qualitative changes in exciton decay
pathways, which we have demonstrated above, in order to control the
transport of energy and charge across these arrays. Our theory can be
extended straightforwardly to address these systems. The Hilbert space
of QD states may be expanded to include electronic states of the
entire array, with appropriate expansion in the size of the ``system
density matrix'' used in our dynamical model. The Hamiltonian, $H_D$
would then involve both the intra-dot energy levels and inter-dot
couplings, for example in a tight-binding form.  The coupling to
electrodes may then acquire additional dependence on QD locations, but
they could still be calculated using the same methodology as outlined
in this paper. Population dynamics of the density matrix of this dot
array, computed in the same manner as for a single QD above, would
then provide the information necessary for energy and charge transport
studies.

In a similar vein, the QD density matrix may also be extended to
include bi-exciton, tri-excitons, and even higher order charge
complexes. The fundamental structure of our theory is a set of levels
for the QD, and their dynamical coupling to the electrode. From our
semi-analytical model of the surface polarization in the surface
coupled QD system, the correlation energy of all higher-particle
states may be determined by expanding them in product of single
particle basis (though it could be impractical beyond the biexcitons).
The calculation of hybridization and Coulomb interaction with the
electrode then follows the same path as was used for excitons. The
implications of inter-particle correlations on their Coulomb coupling
to the electrode surface, and the energy transfer rates may be studied
within this as was done above for exciton decay by NRR. We recall that
the correlations become important when the monopole moment of the
charge distribution of a state vanishes, and therefore does not set
the energy scale for Coulomb coupling far above all the multipole
contributions.

We have neglected vibrational levels of the QD in formulating the
present theory. At the simplest level, the electron-phonon coupling
within the QD provides an extra energy transfer channel for the
electronic system. The discretization of vibrational levels would
generally result in a non-Markovian time-dependent rate of energy
exchange, which may be accounted for via a model spectral density of
phonons in the QD. This would result in an extra energy transfer
matrix defined in the same way as $C_{ac;a'c'}(t)$ in~\pref{eq:dC/dt}
above, the spectral function of the susceptibility in that formula
replaced by the model spectral density of phonons. However, this
treatment would still neglect the coherent dynamics of the electronic
and the lattice system. The most extensive model would be a coupled
system of electronic, photonic, and vibronic density matrices. This
would allow a full study of phonons, excitons, polaritons, and
photoluminescence.

\section{Conclusion}\label{sec:conclusion}

In summary, we have presented a theoretical and computational
framework to model the charge kinetics of optically excited quantum
dots on surfaces. We have started from a microscopic construction of
the eigenstates of the sub-systems, and the couplings between them,
and employ them in an effective model of dynamics restricted to a
small subset of these states. The model is ideally suited to explore
various questions regarding the effect of the surface on charge
extraction, exciton lifetime, and photoluminescence quenching in these
systems.  We have also developed a the dynamical theory taking into
account the effects of the FES in the electrode response.

We illustrated the use of our theory by applying it to an InAs quantum
dot above an Au[111] surface to which it is coupled via the Shockley
surface states. Both the charge and energy transfer processes are
essentially Markovian in this system, but complex dynamics may result
from their interplay.  We found that the FES can lower the exciton
dissociation, but does so only by a factor of 2 for a Fermi sea that
has a bandwidth much larger than the screened Coulomb coupling.

We also discussed the effects of multipole moments of excitons on the
rate of energy transfer to the electron-hole pair excitations in the
surface electron gas. The electron-hole correlation and its interplay
with the image potential of the electrode has significant quantitative
effects on the oscillator strength for energy loss to the plasma
excitations of the electron gas. This non-radiative decay of excitons
was found to be about 100 times slower than the dissociative rate for
the material and geometry used in the calculations. However, with
exponential scaling of the dissociative rate with the barrier width
and height, and for larger QDs exhibiting greater correlation effects,
competition between the two decay pathways can be expected. This would
also yield rich dynamics.

Such a Coulomb driven process can also act as a cooling mechanism for
a charged QD, and we have discussed in particular the cooling of
holes. The energy transfer is very sensitive to the excitation
spectrum at the electrode surface, and in particular the coupling of
the QD to the acoustic surface plasmons. The size of the QD, and the
energy level spacing, control the strength of its dynamical coupling
via this mechanism, and we find that the shifting and broadening of
the plasmon mode due to screening by the bulk yields a large
(approximately 200-fold) reduction in the cooling rate. By arguing
that this screening does not affect the formation of image potentials
in the junction, we claim that screening properties of the substrate
can sensitively tune this system between energy transfer and charge
transfer modes of operations.

Our work can be applied to model and analyze experiments on colloidal
quantum dots near semiconductor and metallic surfaces, and epitaxial
quantum dots in multiple quantum wells. We have emphasized the vast
range of timescales that can exist in the dynamics of these systems,
and point to the various cross-overs in the dominant decay pathways
for exciton states as well as the charge kinetics. In a subsequent
paper, we will develop the theory of linear and non-linear optical
response of these systems. Thus the excitation process itself will be
studied dynamically in the presence of charge kinetics described here.

Another interesting application of this model that we have briefly
discussed is the exploration of spatially dependent coupling between a
quantum dot array and an electrode to control transport physics in
these systems. Finally, extensions of the present theory are necessary
to describe coupling to the driven vibrational modes of the quantum
dot, and dynamics of photon degrees of freedom for exploring the
competition between charge and energy exchange with photoluminescence
and phonons. These extensions will be considered in future
publications.

% ------------------------------------------------------------------------
% Appendices
% ------------------------------------------------------------------------
\appendix
% ------------------------------------------------------------------------
% Appendix: Matrix-elements
% ------------------------------------------------------------------------

\begin{widetext}

  \section{Matrix elements}\label{sec:matrix-elements}

  \subsection{Hybridization}

  \newcommand{\expkappa}{} %{e^{-\kappa_{\nu\boldk}z}}
  For QD states that are exponentially suppressed at the electrode
  surface, the hybridization may be defined by adding and subtracting
  $H_L$ to the total Hamiltonian. The decomposition has the advantage
  of ease of evaluation in terms of potential energy differences
  outside the electrode. Thus for the \emph{single} electron or hole QD state we write
  \begin{eqnarray}
    T^{ge}_{\nu\boldsymbol{k}}	&=&	\int
    d\boldsymbol{r}u^*_{\boldsymbol{k}\nu}(\boldsymbol{r})e^{-i\boldk\cdot\boldr}
    \left[H(\boldsymbol{r})-H_{L}(\boldsymbol{r})\right]\varphi_{e}(\boldsymbol{r}
    ),\label{eq:hybridization-gen-e}\\
    T^{hg}_{\nu\boldsymbol{k}}	&=&	\int
    d\boldsymbol{r}u^*_{\boldsymbol{k}\nu}(\boldsymbol{r})e^{-i\boldk\cdot\boldr}
    \left[H(\boldsymbol{r})-H_{L}(\boldsymbol{r})\right]\varphi_{h}(\boldsymbol{r}
    ),\label{eq:hybridization-gen-h}\\
    T^{eg}_{\nu\boldsymbol{k}}&=&T^{ge*}_{\nu\boldsymbol{k}}\\
    T^{gh}_{\nu\boldsymbol{k}}&=&T^{hg*}_{\nu\boldsymbol{k}}
  \end{eqnarray}
Since $\varphi_h$ represents a single electron
orbital in the valence band, such that $\varphi_h^*$ is the state of the hole, all of the above matrix elements are defined with respect to an electron transfer.
  Here we have shifted the origin so that $z=0$ lies at the image
  plane of the electrode. The integrand vanishes on the electrode side
  ($z<0$), and we simplify the above formula within the effective mass
  approximation with a uniform mass, $m$ inside the QD. Since the
  kinetic energy operator acting on $\phi_a$ yields $E_a$, and the
  function $u_{\nu\boldk}(\boldr)$ varies on a much faster scale than
  the rest of the functions under the integral, we obtain
  \begin{equation}
    T^{ga}_{\nu\boldsymbol{k}}	=\int d\boldsymbol{r} \langle
    u^*_{\boldsymbol{k}\nu}(z)\rangle
    e^{-i\boldk\cdot\boldr}U_a(\boldr)\varphi_{a}(\boldsymbol{r}).\label{eq:T2}
  \end{equation}
  Here we have replaced $u_{\nu\boldk}(\boldr)$ by its average,
  $\langle u^*_{\boldsymbol{k}\nu}(z)\rangle$, over the planar [111]
  unit cell of Au lattice. Outside the electrode, $\langle
  u^*_{\boldsymbol{k}\nu}(z)\rangle$ decays exponentially, but may
  have additional dependence on $z$ depending on the pseudopotential
  in the DFT calculation. The potential energy $U_{a}$ in~\pref{eq:T2},
  which depends on the QD state, is defined as,
  \begin{eqnarray*}
U_{a}(\boldsymbol{r})&=&V(\boldsymbol{r})-U_{L}(\boldsymbol{r}
   )+\left(1-\frac{m}{m_{0}}\right)\left[E_{\alpha}-V(\boldsymbol{r})\right]
    \Theta(\boldr\in\Omega_D).
  \end{eqnarray*}
  Here the pseudopotential $U_L$ for the electrode can be extracted
  from the DFT calculation, the planar average of which in our
  calculations is shown in Fig.~\ref{fig:electrode}. The function
  $\Theta(\boldr\in\Omega_D)$ is unity inside the QD and zero outside.
  In the presence of cylindrical symmetry, we may write the above
  integral in terms of a Bessel transform of order equal to the
  angular momentum quantum number, $l$, of the QD state,
  \begin{eqnarray*}
    T^{ag}_{\nu\boldk}(z) &=& (-1)^{l/2}\int dr r U_\alpha(r,z)\varphi_{a}(r,z)J_l(kr),
  \end{eqnarray*}
  and write,
  \begin{eqnarray}
    T^{ag}_{\nu\boldk}&=&\int dz\left\langle u^*_{\boldk\nu}(z)\right\rangle
    T^{ag}_{\nu\boldk}(z).\label{eq:hybridization}
  \end{eqnarray}
 
Finally, we introduce matrix elements defining the charge transfer processes that form or dissociate an exciton state. Using the coefficients, $\Phi_{eh;x}$, for the expansion of the exciton state in terms of the electron-hole product states, we write the matrix elements for charge transfer involving an exciton state in terms of the above matrix elements connecting the ground and single particle states,
\begin{eqnarray}
T^{hx}_{\nu\boldsymbol{k}} &=& \sum_{e}\Phi_{he;x}T^{ge}_{\nu\boldsymbol{k}},\\
T^{ex}_{\nu\boldsymbol{k}} &=& \sum_{h}\Phi_{he;x}T^{gh}_{\nu\boldsymbol{k}},\\
T^{xh}_{\nu\boldsymbol{k}} &=& T^{hx*}_{\nu\boldsymbol{k}},\\
T^{xe}_{\nu\boldsymbol{k}} &=& T^{ex*}_{\nu\boldsymbol{k}}.
\end{eqnarray}

\subsection{Coulomb interaction}
We write the matrix elements in \pref{eq:Vhat nn' def} by separating
the classical contribution,
\begin{eqnarray}
  V_{n,\nu\boldsymbol{k};n',\nu'\boldsymbol{k}'} & = & W_{n,\nu\boldsymbol{k};n',
    \nu'\boldsymbol{k}'}-\delta_{\nu\nu'}\delta_{\boldsymbol{k},\boldsymbol{k}'}W_{
    nn'}^{(img)}.
  \label{eq:V and W}
\end{eqnarray}
In any expectation value, the subtracted term multiplies the sum over
occupied states, and therefore produces $W_{nn'}^{(img)}$ so long as
the trace of the statistical matrix for the electrode is normalized to
unity. The matrix elements in \eqref{eq:V and W} are given by,
\begin{eqnarray}
  W_{n,\nu\boldsymbol{k};n',\nu'\boldsymbol{k}'}\label{eq:eff coulomb mat elem-app}
  & = & \frac{1}{2}\int d\boldsymbol{r}W_{nn'}(\boldsymbol{r})e^{-
    i(\boldsymbol{k}-\boldsymbol{k}')\cdot\boldsymbol{r}}\\
  &  & \times
  u_{\nu\boldsymbol{k}}^{*}(\boldsymbol{r})u_{\nu\boldsymbol{k}'}(\boldsymbol{r}),
  \nonumber 
\end{eqnarray}
and
\begin{eqnarray}
  W_{nn'}^{(img)} & = & \frac{1}{2}\int d\boldsymbol{s}\int
  d\boldsymbol{r}W_{nn'}(\boldsymbol{r})\label{eq:image pot mat
    elem}\varrho_{img}(\boldsymbol{r};\boldsymbol{s}),\nonumber 
\end{eqnarray}
where $\varrho_{img}(\boldsymbol{r};\boldsymbol{s})$ is the charge
distribution producing an image potential due to a point charge
located at $\boldsymbol{s}$ outside the surface. In the above
equations, $W_{nn'}$ is the potential due of the "transitions density"
of the two QD states, $\varphi_n(\boldr)\varphi_{n'}(\boldr)$. We
determine this potential from our semi-analytical
solution~\citep{virk-apl} for the electrostatic potential, which
yields the coefficients of multipole moments, $Q_{lm}(\boldsymbol{s})$
as a function of $\boldsymbol{s}$. Following the calculation of
$Q_{lm}$ we calculate the potential of the transition density as
follows. When $n$ and $n'$ refer to the same bulk band, we exploit the
fast scale of Bloch envelope functions and set their overlap to unity,
and obtain
\begin{equation*}
  W_{nn'}(\boldsymbol{r})=\sum_{l,m}\frac{1}{2\varepsilon_{0}a}\int
  d\boldsymbol{s}\frac{Q_{lm}(\boldsymbol{s})}{2l+1}\frac{a^{l}}{r^{l+1}}Y_{lm}
  (\hat{\mathbf{r}})\varphi_{n}^{*}(\boldsymbol{s})\varphi_{n'}(\boldsymbol{s}),
\end{equation*}
When $n$ and $n'$ belong to different bulk bands, the underlying Bloch
envelope functions become orthogonal. In this case, we must expand the
Coulomb potential over the bulk unit cell of the QD. Let
$\boldsymbol{x}$ vary over the unit cell,
\[V(\boldr-\boldsymbol{s}-\boldsymbol{x})=V(\boldr-\boldsymbol{s})-\boldsymbol{x
}\cdot\nabla V(\boldr-\boldsymbol{s}).\] When this is averaged over
the macroscopic envelope functions $\varphi_n(\boldsymbol{s})$, we
replace $W_{nn'}(\boldr)$ by
\[W_{nn'}(\boldr)+\mathbf{d}_{nn'}\cdot\nabla W_{nn'}(\boldr) ,\]
where $\mathbf{d}_{nn'}$ is the dipole moment between the Bloch states
between bands $n$ and $n'$. This is used in calculating energy
transfer for exciton states in which case the zeroth order term in the
field vanishes. When $\mathbf{d}_{nn'}$ also vanishes, higher order
derivatives of the potential must be coupled to the multiple moments
of the microscopic Bloch functions.
Replacing the overlap of Bloch envelope functions by their planar
average,
\[O_{\nu\boldk,\nu\boldk '}(z)=\int d\boldr_\parallel
u^*_{\nu\boldk}(\boldr)u_{\nu\boldk '}(\boldr),\] we obtain,
\begin{eqnarray}
  &&W_{n,\nu\boldsymbol{k};n',\nu'\boldsymbol{k}'}\\\label{eq:eff coulomb mat
    elem}
  &=& \frac{1}{2}\int dz O_{\nu\boldk,\nu\boldk '}(z)\int
  d\boldsymbol{r}_\parallel e^{i(\boldk
    '-\boldk)\cdot\boldr_\parallel}W_{nn'}(\boldsymbol{r}).\nonumber
\end{eqnarray}

% ------------------------------------------------------------------------
% Appendix: Mathematical
% ------------------------------------------------------------------------
\section{Charge and energy transfer rates}\label{sec:math}

In this appendix, we collect the necessary mathematical steps to
derive the energy and charge transfer rates as defined in \pref{eq:B
  ac dc} and \pref{eq:dC/dt}. We make use of the equivalence of the
following two expressions. Let,
\begin{eqnarray}
  f(t) & = & -i\int_{0}^{t}d\tau X(\tau)e^{i\omega_{0}\tau},\label{eq:f(t)}
\end{eqnarray}
where $X(\tau)$ is a causal function, and the factor $i$ is for later
convenience. The Fourier transform of $X(\tau)$, which is complex,
\[
X(\omega)=X_{R}(\omega)+iX_{I}(\omega),
\]
satisfies the Kramers-Kronig relations~\citep{mahanbook}. Then, we may
write $f(t)$ entirely in terms of $X_{I}(\omega)$ as,
\begin{eqnarray}
  f(t) & = & \frac{1}{2}\int
  d\omega\mbox{Spec}\left[X(\omega)\right]S(\omega,t),\label{eq:Basic 
    Relation}
\end{eqnarray}
where we have defined the \emph{spectral function}
$\mbox{Spec}\left[X(\omega)\right]=-2X_{I}(\omega)$, and
\begin{eqnarray*}
  S(\omega,t) & = &
  -\frac{\sin[(\omega-\omega_{0})t]}{\omega-\omega_{0}}+\frac{2i\sin^{2}
    [(\omega-\omega_{0})\frac{t}{2}]}{\omega-\omega_{0}}.
\end{eqnarray*}
It is useful to note that the limiting behavior of the function $S$
is,
\begin{eqnarray*}
  \lim_{t\rightarrow\infty}S(\omega,t) & = & -\pi\delta(\omega)t,\\
  \lim_{t\rightarrow0}S(\omega,t) & = & -t+i\frac{\omega t^2}{2}.
\end{eqnarray*}

\subsection{Charge transfer}

The charge transfer matrix can be described by the functions defined
in \eqref{eq:Lambda-Gamma}, which are explicitly equal to two of the
four correlation functions in the definition of $B_{ac;a'c'}(t)$
in~\eqref{eq:B ac dc}. To handle all these correlations, and to derive
this reduction to two functions explicitly, we define here two
additional functions,
\begin{eqnarray*}
  \Lambda^{-ca}_{c'b;bc}(t'',t';t) & = & -i\Theta(t'-t'')\left\langle
    \mathcal{T}^{c'b}(t'')\mathcal{T}^{bc}(t')\mathcal{T}_0^{ca}(t)\mathfrak{R}
  \right\rangle, \nonumber \\
  \Gamma^{-}_{c'c;aa'}(t'',t';t) & = & -i\Theta(t'-t'')\left\langle
    \mathcal{T}^{c'c}(t'')\mathcal{T}_0^{ca}(t)\mathcal{T}^{aa'}(t')\mathfrak{R}
  \right\rangle.
\end{eqnarray*}
It follows from the above definitions that
\begin{eqnarray}
  \Gamma_{c'c;aa'}^{-}(t'',t';t)& =
  &-\left[\Gamma_{a'a;cc'}(t'',t';t)\right]^{*},\label{eq:rel-G}\\
  \Lambda^{-ca}_{c'b;bc}(t'',t';t)& =
  &-\left[\Lambda^{ac}_{cb;bc'}(t'',t';t)\right]^{*}\label{eq:rel-L},
\end{eqnarray}
Let us introduce the Fourier transforms of the correlation functions
as,
\begin{eqnarray*}
  \Gamma_{c'c;aa'}(\omega,t';t) & = &
  \int_{-\infty}^{+\infty}dt''\Gamma_{c'c;aa'}(t'',t';t)e^{i\omega t''}.
\end{eqnarray*}
The above definitions of the correlation functions and their
relationship,\eqref{eq:rel-G} and \eqref{eq:rel-L} imply that
\begin{eqnarray*}
  \mbox{Spec}\left[\Gamma_{aa';c'c}(\omega,t';t)\right] & = &
  \mbox{Spec}\left[\Gamma^{-}_{cc';a'a}(-\omega,t';t)\right],\\
  \mbox{Spec}\left[\Lambda^{ca}_{aa';c'c}(\omega,t';t)\right] & = &
  \mbox{Spec}\left[\Lambda^{-ac}_{cc';a'a}(-\omega,t';t)\right].
\end{eqnarray*}
We now use the basic relation \eqref{eq:Basic Relation}, and exploit
the fact that only the $\mbox{Spec} [\Gamma(\omega,t';t)]$ and
$\mbox{Spec} [\Lambda(\omega,t';t)]$ are necessary in the final
expressions. Thus performing the required permutations of states in
the subscript, and setting
\begin{eqnarray*}
  \gamma_{aa';cc'}(\omega,t';t)&=&\mbox{Spec}\left[\Gamma_{aa';cc'}(\omega,
    t';t)\right],\\
  \lambda_{aa';cc'}(\omega,t';t)&=&\mbox{Spec}\left[\Lambda_{aa';cc'}(\omega,
    t';t)\right],
\end{eqnarray*}
we obtain~\pref{eq:B ac dc} in the main text. We now turn to the
evaluation of these spectral functions.

The functions $\Gamma$ and $\Lambda$ can be expressed in terms of
single particle Green functions of the electrode, within the
interaction picture discussed in the main body. Here we explicitly
develop the expressions for $\Gamma(\omega,t';t)$, since the
derivation of $\Lambda(\omega,t';t)$ follows the same form.  These Green functions are
physically different depending on whether they describe propagation of
the system under addition or removal of a particle. Following
conventional notation we use superscript `` $^>$'' to describe
propagation under addition of an electron to the electrode, and ``$^<$'' for the removal of an electron. Thus letting
$n_c$ be the \emph{electron} occupation of the QD in state $c$, we define
\begin{equation}\label{eq:Gamma-cases}
  \Gamma_{c'c;aa'}(t'',t';t)=
  \begin{cases}\Gamma^>_{c'c;aa'}(t'',t';t) & n_c'=n_c+1,n_a'=n_a-1, \\
    \Gamma^<_{c'c;aa'}(t'',t';t) & n_c'=n_c-1,n_a'=n_a+1,\end{cases}
\end{equation}
where $\Gamma_{c'c;aa'}^{>}(t'',t';t)$ describes the tunneling of an electron out of the quantum dot and is given by~\eqref{eq:Gamma-green-func} in the main text, while
\begin{eqnarray}
  \Gamma_{c'c;aa'}^{<}(t'',t';t)
  &=&
  \frac{-i}{Z}\Theta(t'-t'')\sum_{\nu\nu'\boldsymbol{k}\boldsymbol{k}'}T^{cc'}_{\nu\boldk}T^{aa'}_{\nu'\boldk'}
  \left\langle
    e^{iK_{c'}t'}c_{\nu\boldsymbol{k}}^{\dagger}e^{-iK_{c}(t'-t)}e^{iK_{a}(t''-t)}c_{\nu'\boldsymbol{k}'}e^{-iK_{a'}t''}e^{-\beta
      H_{L}}\right\rangle,\label{eq:Gamma-green-func-less}
\end{eqnarray}
describes the tunneling from the electrode to the quantum dot. We also
define $\Lambda^\lessgtr$ in a similar fashion to correspond to the
ordering of $c^\dagger$ and $c$. Note that the superscripts on the hybridization matrix elements, \emph{e.g.} $T^{cc'}_{\nu\boldk}T^{aa'}_{\nu\boldk}$ match the pairs of subscripts on the correlation function, and these pairs of indices are constrained by the type of correlation function, ``lesser'' or ``greater'', in accordance with~\eqref{eq:Gamma-cases}.

In the case of a quadratic Hamiltonian, this formula can be further
simplified using two mathematical relations which can be proven using expressions derived by Hirsch~\citep{hirsch}.  In particular, our $H_L$ and $K$ operators are all quadratic.  The Hirsch approach allows us to connect the required full traces to the evaluation of an expression based on those operators in a single particle basis.  In the following, the left hand side gives the ratio of two thermal averages while the right hand side is a specific matrix element in a single particle basis of the enclosed series of operators represented by matrices in that basis set:
{\small
  \begin{eqnarray}
    &&\frac{\left\langle
        e^{iA_{n}x}e^{iA_{m}y}c_{i}e^{iA_{2}}c_{j}^{\dagger}e^{iA_{3}}e^{-\beta
          H_L}\right
      \rangle }{\left\langle e^{iA_{n}x}e^{iA_{m}y}e^{iA_{2}}e^{iA_{3}}e^{-\beta
          H_L}\right\rangle }\nonumber\\
    &=&	\left\langle\nu\boldsymbol{k}
      \left|\mathbf{\bar{N}}_{mn}(-y,-x)\left[\mathbf{I}+\left(e^{-iA_{m}y}e^{-iA_n
              x}e^{-iA_{3}}e^{-
              iA_{2}}-\mathbf{I}\right)\mathbf{\bar{N}}_{mn}(-y,-x)\right]^{-1}e^{-iA_{m}y}e^{-iA_n
          x}e^{-iA_{3}}\right|\nu'\boldsymbol{k}'\right
    \rangle \nonumber\\
    & & \label{eq:hirsch-1}\\
%
    % second one
%
    &&\frac{\left\langle
        e^{iA_{3}}c_{i}^{\dagger}e^{iA_{2}}c_{j}e^{iA_{n}x}e^{iA_{m}y}e^{-\beta
          H_{L}}\right
      \rangle }{\left\langle e^{iA_{3}}e^{iA_{2}}e^{iA_{n}x}e^{iA_{m}y}e^{-\beta
          H_{L}}\right\rangle }\nonumber\\
    &=&\left\langle\nu\boldsymbol{k}
      \left|\mathbf{N}_{nm}(x,y)\left[\mathbf{I}+\left(e^{iA_nx}e^{iA_my}e^{iA_{3}}e^
            {iA_{2}}-\mathbf{I}
          \right)\mathbf{N}_{nm}(x,y)\right]^{-1}e^{iA_nx}e^{iA_my}e^{iA_{3}}
      \right|\nu'\boldsymbol{k}'\right\rangle \label{eq:hirsch-2}.
  \end{eqnarray}
}%
In the second of the two preceding equations,
$\mathbf{N}_{nm}(x,y)=\mathbf{I}-\mathbf{\bar{N}}_{nm}(x,y)$, and in
the the first equation,
\[\mathbf{\bar{N}}_{mn}(y,x)=e^{iK_{m}y}e^{iK_{n}x}\left[1+e^{-\beta H_L}\right]^{-1}e^{-iK_{n}x}e^{-iK_{m}y},
\] where the central part is given by~\eqref{eq:W0} in the main text. To apply expressions to (\ref{eq:hirsch-1} and \ref{eq:hirsch-2}), we
substitute,
\begin{eqnarray*}
  e^{iA_nx}e^{iA_my}&=&e^{iK_{c'}t'}=e^{iK_{c'}t}e^{iK_{c'}(t'-t)},\\
  e^{iA_2}&=&e^{-iK_c(t'-t)}e^{iK_a(t''-t)},\\
  e^{iA_3}&=&e^{-iK_{a'}t''}.
\end{eqnarray*}
It is also convenient to define a matrix functions,
\begin{eqnarray*}
  \boldsymbol{\Sigma}^{(1)bb'}_{c'c;aa'}(t',t'';t)
  &=&e^{-iK_{c'}t'}e^{iK_{a'}t''}e^{-iK_{a}t''}\mathcal{T}_0^{bb'}(t)e^{iK_{c}t'}-
  \mathbf{I},\nonumber\\
  \boldsymbol{\Sigma}^{(2)bb'}_{c'c;aa'}(t',t'';t)
  &=&e^{-iK_{c'}t'}\mathcal{T}_0^{bb'}(t)e^{iK_{a'}t''}e^{-iK_{a}t''}e^{iK_{c}t'}-
  \mathbf{I},\nonumber
\end{eqnarray*}
and a function describing the AOC,
\begin{eqnarray}
  r_{c'c;aa'}(t'',t',t)\label{eq:r function-app}
  & = & \left\langle
    e^{iK_{c'}t'}e^{-iK_{c}(t'-t)}e^{iK_{a}(t''-t)}e^{-iK_{a'}t''}e^{-\beta H_L}
  \right\rangle .\nonumber 
\end{eqnarray}
To apply these expressions to $\Lambda^{>ca}_{ab;ba'}$, we substitute,
\begin{eqnarray*}
  e^{iA_nx}e^{iA_my}&=&e^{iK_{c}t}e^{iK_{a}(t'-t)}\\
  e^{iA_2}&=&e^{-iK_{b}t'}e^{iK_bt''}\\
  e^{iA_3}&=&e^{-iK_{a'}t''}.
\end{eqnarray*}

With these definitions, we obtain the following set of functions to
describe electron and hole tunneling between the QD and the electrode,
%\begin{widetext}
{\small
\begin{eqnarray}
&
&\Gamma_{c'c;aa'}^{>}(t'',t';t)=-i\Theta(t'-t'')r_{c'c;aa'}(t'',t',t)\label{
eq:Gamma-greater-matrix}\\
 & & \times\sum_{\nu\nu'\boldsymbol{k}\boldsymbol{k}'}%
 T^{c'c}_{\nu\boldk}T^{aa'*}_{\nu'\boldk'}%
\left\langle \nu\boldsymbol{k}\right|\mathbf{\bar{N}}_{c'c'}(-t',0)
\left[\mathbf{I}+\boldsymbol{\Sigma}^{(1)ca}_{c'c;aa'}(t',t'';t)\mathbf{\bar{N}}_{c'c'
}(-t',0)\right]^{-1}\left[e^{-
iK_{c'}t'}e^{iK_{a'}t''}\right]\left|\nu'\boldsymbol{k}'\right\rangle ,\nonumber \\
&&\Gamma_{c'c;aa'}^{<}(t'',t';t)=i\Theta(t'-t'')r_{c'c;a'a}(t'',t',t)\label{eq:Gamma-lesser-matrix}\\
&&\sum_{\boldsymbol{k}
\boldsymbol{k}'}T^{aa'}_{\nu\boldk}T^{c'c}_{\nu'\boldk'}\left\langle \nu\boldsymbol{k}
\right|\mathbf{N}_{c'c'}(t'',0)\left[\mathbf{I}+\boldsymbol{\Sigma}^{(1)ca}_{
a'a;cc'}(t'',t';t)
\mathbf{N}_{c'c'}(t'',0)\right]^{-1}\left[e^{-iK_{a'}t''}e^{iK_{c'}t'}\right]
\left|
\nu'\boldsymbol{k}'\right\rangle,\nonumber
\end{eqnarray}}
and for $\Lambda$
{\small
  \begin{eqnarray}
    &&\Lambda^{>ca}_{ab;ba'}(t'-t'',t';t)=-i\Theta(t'-t'')r_{ca;ba'}(t,t'',t')\label
    {eq:Lambda-greater-matrix}\\
    && \times\sum_{\nu\nu'\boldsymbol{k}\boldsymbol{k}'}%
    T^{ab}_{\nu\boldk}T^{ba'}_{\nu'\boldk'}%
    \left\langle\nu \boldsymbol{k}\right|\mathbf{\bar{N}}_{ac}(t-t',-t)
    \left[\mathbf{I}+\boldsymbol{\Sigma}^{(2)ac}_{ab;ba'}(t',t'';t)\mathbf{\bar{N}}_{ac}
      (t-t',-t)\right]^{-1}\left[e^{-
        iK_{c'}t'}e^{iK_{a'}t''}\right]\left|\boldsymbol{k}'\nu'\right\rangle ,\nonumber \\
    \nonumber\\
    &&\Lambda^{<ca}_{ab;ba'}(t'-t'',t';t)=-i\Theta(t'-t'')r_{ca;ba'}(t,t'',t')\label
    {eq:Lambda-lesser-matrix}\\
    &  & \times\sum_{\nu\nu'\boldsymbol{k}\boldsymbol{k}'}%
    T^{ba'}_{\nu\boldk}T^{ab}_{\nu'\boldk'}%
    \left\langle\nu \boldsymbol{k}\right|\mathbf{N}_{ca}(t-t',t)
    \left[\mathbf{I}+\boldsymbol{\Sigma}^{(2)ac}_{ab;ba'}(t',t'';t)\mathbf{N}_{ca}
      (t-t',t)\right]^{-1}\left[e^{iK_{c}t}e^{iK_{a}(t-t')}e^{-iK_{a'}t''}\right]
    \left|\nu'\boldsymbol{k}'\right\rangle .\nonumber 
  \end{eqnarray}}
% \end{widetext}

%  \begin{eqnarray}
%    &&\Lambda^{>ca}_{ab;ba'}(t'-t'',t';t)=-i\Theta(t'-t'')r_{ca;ba'}(t,t'',t')\label
%    {eq:Lambda-greater-matrix}\\
%    && \times\sum_{\nu\nu'\boldsymbol{k}\boldsymbol{k}'}%
%    T_{a;\nu\boldk}T^*_{b;\nu'\boldk'}%
%    \left\langle\nu \boldsymbol{k}\right|\mathbf{\bar{N}}_{ac}(t-t',-t)
%    \left[\mathbf{I}+\boldsymbol{\Sigma}^{(2)ac}_{ab;ba'}(t',t'';t)\mathbf{\bar{N}}_{ac}
%      (t-t',-t)\right]^{-1}\left[e^{-
%        iK_{c'}t'}e^{iK_{a'}t''}\right]\left|\boldsymbol{k}'\nu'\right\rangle ,\nonumber \\
%%%
%%
%    \nonumber\\
%    &&\Lambda^{<ca}_{ab;ba'}(t'-t'',t';t)=-i\Theta(t'-t'')r_{ca;ba'}(t,t'',t')\label
%    {eq:Lambda-lesser-matrix}\\
%    &  & \times\sum_{\nu\nu'\boldsymbol{k}\boldsymbol{k}'}%
%    T_{a';\nu\boldk}T^*_{b;\nu'\boldk'}%
%    \left\langle\nu \boldsymbol{k}\right|\mathbf{N}_{ca}(t-t',t)
%    \left[\mathbf{I}+\boldsymbol{\Sigma}^{(2)ac}_{ab;ba'}(t',t'';t)\mathbf{N}_{ca}
%      (t-t',t)\right]^{-1}\left[e^{iK_{c}t}e^{iK_{a}(t-t')}e^{-iK_{a'}t''}\right]
%    \left|\nu'\boldsymbol{k}'\right\rangle .\nonumber 
%  \end{eqnarray}}

\subsection{Energy transfer}\label{sec:energytransfer}

\subsubsection{First order term}

Since we are only dealing with the Coulomb term in this appendix, we will omit the subscript $C$ used in the main text for functions related to this interaction.
From the definitions of the superoperators, the first order term takes
the form shown in~\eqref{eq:Pmaintext}  in the main text. We now impose the condition that the subscripts of $\hat{V}_{ab}$
correspond to states with the same net charge, and that only the
monopole terms are significant in the operators $K_a$. Thus we replace
$K_a$ and $K_a'$ by their average, and define the time evolution of
the electrode density operator under the potential $(V_a+V_a')/2$ as
\[\mathfrak{R}^{(aa')}(t)=e^{i\frac{t}{2}(K_a+K_{a'})}\mathfrak{R}e^{-i\frac{t}{
    2}(K_a+K_{a'})},\]
% S
The corrections to this are small, and quantified at the end of this
discussion. The first order term now becomes, {\small
  \begin{eqnarray*}
    \frac{d}{dt}P_{ac;a'c'}(t) &=&
    -iD_{ac}(t)D_{a'c'}^{-1}(t)\left[\frac{d}{dt}D^{-1}_{ac}(t)\int_0^t d\tau
      \sum_{\boldk,\boldk',\nu\nu'}\right.\\
    &&\left.\delta_{cc'}V_{a,\nu\boldk;a'\nu'\boldk'}\langle
      c^{\dagger}_{\nu\boldk} c_{\nu\boldk'} \mathfrak{R}^{(aa')}(\tau)\rangle
      -\delta_{aa'}V_{c',\nu\boldk;c,\nu'\boldk'}\langle c^{\dagger}_{\nu\boldk}
      c_{\nu\boldk'} \mathfrak{R}^{(cc')}(\tau)\rangle\right]
  \end{eqnarray*}
}
Clearly, each term under the sum is the expectation value of the
potential energy as the electrode state evolves under the average
potentials of the states that coupled by the off-diagonal Coulomb
interaction. Thus by using our definition of the Hartree energy matrix in~\eqref{eq:hartree-energy},
\begin{eqnarray}
  \frac{d}{dt}P_{ac;a'c'}(t)&=&-i\delta_{cc'}D_{a'c}^{-1}(t)\left(\Delta_{aa'}(t)+\frac{d\log{D_{ac}(t)}}{dt
    }\int_0^t\Delta_ {aa'}(\tau)\right)\nonumber\\
  &&+i\delta_{aa'}D_{ac'}^{-1}(t)\left(\Delta_{c'c}(t)+\frac{d\log{D_{ac}(t)}}{dt
    }\int_0^t\Delta_ {c'c}(\tau)\right)\label{eq:first order app}
\end{eqnarray}

\newcommand{\avk}{\ensuremath{\bar{K}^{aa'}}} To obtain the
corrections beyond the above, we let $\bar{K}^{aa'}=(K_a+K_{a'})/2$,
and $\delta W_{aa'}=(\hat{V}_{aa}-\hat{V}_{a'a'})/2$, and write
\begin{eqnarray*}
  \left\langle e^{-iK_a t}\hat{V}_{aa'}e^{iK_{a'}t}\mathfrak{R}\right\rangle&=&
  \left\langle \hat{V}_{aa'}e^{i(\avk-\delta W_{aa'})t}\mathfrak{R}e^{-i(\avk+\delta W_{aa'})t}\right\rangle\\
  &=&\left\langle \hat{V}_{aa'}e^{it \avk}U_{aa'}(t)\mathfrak{R}U^\dagger_{aa'}(t)e^{-it\avk}\right\rangle,
\end{eqnarray*}
where
\[U_{aa'}(t) = 1 - i\int_0^t dt' e^{-it \avk}\delta W_{aa'}(t') e^{it
  \avk} - \int_0^t\int_0^{t'} e^{-it' \avk}\delta W_{aa'}(t')
e^{i(t'-t'') \avk}\delta W_{aa'}(t'') e^{it'' \avk}+\ldots.\]

From this result, we see that the leading correction to the first
order term, as approximated by \eqref{eq:first order app} is given by
a potential $\delta W_{aa'}$, which is equal to the difference between
two electrostatic potentials of equal charge.

\subsubsection{Second order term}
We derive the expression for one of the terms in the energy transfer
matrix, while the remaining three terms can be computed from it. Thus
let,
\begin{eqnarray*}
  F_{ac;a'c'}(\tau;t) & = & -e^{i(\omega_{aa'}+\omega_{c'c})t}\times\\
  &  & \int_{0}^{t}d\tau\left\langle
    \Delta\hat{V}_{c'c}(t-\tau)\Delta\hat{V}_{aa'}(t)\mathfrak{R}
  \right\rangle e^{-i\omega_{c'c}\tau}.
\end{eqnarray*}
We define
\begin{eqnarray*}
  \chi_{ab;b'a'}^{+}(\tau;t) & = & -i\Theta(\tau)\left\langle
    \Delta\hat{V}_{c'c}(t-\tau)\Delta
    \hat{V}_{aa'}(t)\mathfrak{R}\right\rangle \\
  \chi_{ab;b'a'}^{-}(\tau;t) & = & -i\Theta(\tau)\left\langle
    \Delta\hat{V}_{c'c}(t)\Delta\hat{V}_{aa'}(t-
    \tau)\mathfrak{R}\right\rangle .
\end{eqnarray*}
These two functions also satisfy the relation
\begin{eqnarray*}
  \left(\chi_{c'c;aa'}^{+}(\tau;t)\right)^{*} & = & -\chi_{a'a;c'c}^{-}(\tau;t)\\
  \Rightarrow\mbox{Spec}\left[\chi_{c'c;aa'}^{+}(\omega;t)\right] & = &
  \mbox{Spec}
  \left[\chi_{c'c;aa'}^{-}(-\omega;t)\right].
\end{eqnarray*}

Using the expressions for $\Delta V$, and implying summation over
repeated indices, we obtain,
\begin{eqnarray*}
  \chi_{c'c;aa'}^{+}(\omega;t) & = &
  V_{c',\nu\boldsymbol{k};c,\nu'\boldsymbol{k}'}V_{a,\nu
    \boldsymbol{p};a',\nu'\boldsymbol{p}'}\times\\
  &  & \left[\left\langle
      c_{\nu\boldsymbol{k}}^{\dagger}(t-\tau)c_{\nu'\boldsymbol{k}'}(t-
      \tau)c_{\mu\boldsymbol{p}}^{\dagger}(t)c_{\mu'\boldsymbol{p}'}(t)\right\rangle
  \right.\\
  &  & \left.-\left\langle
      c_{\nu\boldsymbol{k}}^{\dagger}(t-\tau)c_{\nu'\boldsymbol{k}'}(t-\tau)\right
    \rangle \left\langle
      c_{\mu\boldsymbol{p}}^{\dagger}(t)c_{\mu'\boldsymbol{p}'}(t)\right\rangle 
  \right].
\end{eqnarray*}
In most circumstances, we may neglect defects in the electrode
surfaces and assume a uniform electron gas. Within this approximation,
the correlation functions must conserve momentum, and the planar
Fourier transform of the QD potential also becomes a function only of
the difference between the two momenta
\begin{eqnarray*}
  V_{c',\nu\boldsymbol{k};c,\nu'\boldsymbol{k}'}=V_{c'\nu';c\nu}(\boldsymbol{k}
  -\boldsymbol{k}') 
  & = & \int
  V_{c'c}(\boldsymbol{r})e^{-i(\boldsymbol{k}-\boldsymbol{k}')\cdot\boldsymbol{r}}
  d
  \boldsymbol{r}.
\end{eqnarray*}
In this case the number of momenta in the summation reduces, and we
obtain
\begin{eqnarray*}
  \chi_{c'c;aa'}^{+}(\omega;t) & = & \sum_{\boldsymbol{q}}V_{c',\nu;c,\nu'}(-
  \boldsymbol{q})X_{\nu\nu';\mu\mu'}(\boldsymbol{q},\omega)V_{a,\mu;a',\mu'}
  (\boldsymbol{q}).
\end{eqnarray*}
Here the function, $X_{\nu\nu';\mu\mu'}(\boldsymbol{q},\omega)$, is
the density density correlation function, generalized to include
inter-subband transitions,
\begin{eqnarray*}
  &&X_{\nu\nu';\mu\mu'}(\boldsymbol{q},\omega) = \int_0^\infty d\tau\;e^{i\omega
    \tau}\times\\
  &&\left[\left\langle c_{\nu\boldsymbol{k}-\boldsymbol{q}}^{\dagger}(t-
      \tau)c_{\nu'\boldsymbol{k}}(t-\tau)c_{\mu\boldsymbol{p}+\boldsymbol{q}}^{\dagger
      }
      (t)c_{\mu'\boldsymbol{p}}(t)\right\rangle \right.\\
  && \left.-\left\langle
      c_{\nu\boldsymbol{k}}^{\dagger}(t-\tau)c_{\nu'\boldsymbol{k}-
        \boldsymbol{q}}(t-\tau)\right\rangle \left\langle
      c_{\mu\boldsymbol{p}+\boldsymbol{q}}^{\dagger}
      (t)c_{\mu'\boldsymbol{p}}(t)\right\rangle \right].
\end{eqnarray*}
The calculation of density-density correlation can be found in most
textbooks on solid state physics~\citep{mahanbook}. Substitution of
these calculations into the above formula and setting $\chi =
\mbox{Spec}[\chi^+]$ yields formula~\pref{eq:dC/dt} in the main body.

\end{widetext}
\clearpage

\begin{acknowledgements}
  This work is part of the Center for Re-Defining Photovoltaic
  Efficiency Through Molecule Scale Control, an Energy Frontier
  Research Center funded by the U.S. Department of Energy (DOE),
  Office of Science, Office of Basic Energy Sciences under award
  no. DE-SC0001085 and the research was carried out in part at the
  Center for Functional Nanomaterials, Brookhaven National Laboratory,
  contract no. DE-AC02-98CH10886.  KSV also acknowledges partial
  support by the Natural Sciences and Engineering Council of Canada.
\end{acknowledgements}
\bibliographystyle{apsrev4-1} \bibliography{BibliographyCitedOnly}

% ------------------------------------------------------------------------
% Figure: Wavefunctions electron and hole
% ------------------------------------------------------------------------
% \begin{figure*}
%   \begin{minipage}[t]{4in}
%     \includegraphics[width=4in]{electron-wf-contour.pdf}
%   \end{minipage}\\
%   \begin{minipage}[t]{4in}
%     \includegraphics[width=4in]{hole-wf-contour.pdf}
%   \end{minipage}
%   \caption{\label{fig:eh-wavefunctions}Three lowest energy
%   wavefunctions of electron (top) and hole (bottom) levels}
% \end{figure*}
% ------------------------------------------------------------------------

% ------------------------------------------------------------------------
% Figure: Charge
% ------------------------------------------------------------------------
% \begin{figure}
%   \includegraphics[width=8cm]{total-charge.pdf}
%   \caption{Total charge}
% \end{figure}

\end{document}